\documentclass{article}

\PassOptionsToPackage{numbers, compress}{natbib}
\usepackage[main, preprint]{neurips_2026}
\makeatletter
\renewcommand{\@noticestring}{\@trackname}
\makeatother

\usepackage[utf8]{inputenc} %
\usepackage[T1]{fontenc}    %
\usepackage{hyperref}       %
\usepackage{url}            %
\usepackage{booktabs}       %
\usepackage{array}          %
\usepackage{amsmath,amssymb,mathtools}        %
\usepackage{amsfonts}       %
\usepackage{amsthm}         %
\usepackage{algorithm}
\usepackage{algpseudocode}
\usepackage{nicefrac}       %
\usepackage{microtype}      %
\usepackage{xcolor}         %
\usepackage{graphicx}

\newtheorem{theorem}{Theorem}
\newtheorem{proposition}{Proposition}
\newtheorem{lemma}{Lemma}
\newtheorem{corollary}{Corollary}
\theoremstyle{definition}

\usepackage{enumitem}
\providecommand{\Tr}{\operatorname{Tr}}
\providecommand{\diag}{\operatorname{diag}}
\providecommand{\R}{\mathbb{R}}
\providecommand{\one}{\mathbf{1}}
\providecommand{\Top}{\operatorname{Top}}
\providecommand{\polar}{\operatorname{polar}}
\providecommand{\LR}{\operatorname{LR}}
\providecommand{\inner}[2]{\ensuremath{\left\langle #1,\,#2\right\rangle}}
\providecommand{\sym}{\operatorname{sym}}
\newcommand{\tablesetup}{\setlength{\tabcolsep}{3pt}\renewcommand{\arraystretch}{1.05}}

\title{Orthogonal Sparse Subgraph Alignment for Structure-Function Coupling in Brain Networks}

\author{%
  \textbf{Haonan Gao$^{1}$ \quad Daeyoung Ham$^{2}$ \quad Yifei Zhang$^{1}$ \quad Xinyuan Tian$^{1}$} \\
  \textbf{Shengxian Ding$^{1}$ \quad Yize Zhao$^{1,}$\setcounter{footnote}{1}\thanks{Corresponding authors: Yize Zhao (\texttt{yize.zhao@yale.edu}) and Tianxi Li (\texttt{tianxili@umn.edu}).} \quad Tianxi Li$^{3,}$\footnotemark[2]\setcounter{footnote}{0}} \\[6pt]
  \normalfont $^{1}$Department of Biostatistics, Yale School of Public Health, Yale University \\
  \normalfont $^{2}$Department of Statistics and Data Science, University of Texas at San Antonio \\
  \normalfont $^{3}$School of Statistics, University of Minnesota, Twin Cities
}

\begin{document}

\maketitle

\begin{abstract}
Characterizing the relationship between brain anatomical wiring and functional coordination remains a fundamental challenge in computational neuroscience. This difficulty primarily arises because structure-function coupling in the human brain is spatially heterogeneous, often localized to subnetworks, and inherently difficult to align across disparate representational modalities. To address this methodological gap, we introduce \textbf{O}rthogonal \textbf{S}parse \textbf{S}ubgraph \textbf{A}lignment (\textsc{Ossa}), a principled and interpretable framework designed to identify compact structural cliques that exhibit maximal concordance with functional connectivity (FC) across subjects. Within each FC-derived community, \textsc{Ossa} optimizes an alignment objective over an orthogonal transformation, which absorbs coordinate mismatch between the two connectivity modalities, and a sparse node-selection vector, which identifies the induced structural connectivity (SC) subgraph. 
Furthermore, for canonical \textsc{Ossa} within a fixed population community, we establish local statistical guarantees under the required assumptions. With high probability, an empirical local maximizer exists in a specified neighborhood of the population target, and every local maximizer in that neighborhood exactly recovers the active support and satisfies a nonasymptotic error bound for the aligned functional projector.
Extensive empirical evaluations on simulated data, alongside analyses of the Adolescent Brain Cognitive Development (ABCD) baseline and longitudinal cohorts, demonstrate that \textsc{Ossa} successfully recovers aligned structural cliques under latent cross-modal rotations. Ultimately, the proposed methodology significantly improves out-of-sample SC--FC alignment over size-matched baselines and identifies compact structural supports with longitudinal consistency in node selection, while multi-support diagnostics suggest a dominant aligned core within the estimated FC communities.
\end{abstract}

\section{Introduction}
Understanding how anatomical wiring governs coordinated functional activity in the brain remains a fundamental problem in computational neuroscience. This objective naturally motivates the study of two complementary representations of the brain connectome. Specifically, structural connectivity (SC), inferred from diffusion imaging, charts the physical pathways along which regions communicate, whereas functional connectivity (FC), derived from correlated activity fluctuations, captures the statistical dependence between regional signals \citep{friston1994functional,sporns2013structure}. Characterizing the relationship between SC and FC has emerged as a salient analytical task: it provides a principled framework to investigate how anatomy constrains function, how individual variability manifests across modalities, and which subnetworks offer the most robust structural support for observed functional coordination \citep{honey2009predicting,fotiadis2024structure}.

Meanwhile, the SC-FC relationship is \textbf{intrinsically indirect and spatially heterogeneous} \citep{suarez2020linking,fotiadis2024structure}. An extensive body of literature demonstrates that functional connectivity is only partially constrained by direct anatomical pathways, and the cross-modal agreement varies systematically across cortical hierarchies, individuals, and developmental stages \citep{honey2009predicting,vazquezrodriguez2019gradients,baum2020development,zamani2022local,gao2026gradient}. Specifically, while certain functional communities are anchored by a dense set of anatomical connections, others rely on diffuse or polysynaptic interactions \citep{yeo2011organization,goni2014resting,seguin2020network,zamani2022local}. Consequently, the most pronounced cross-modal signals are concentrated within localized subnetworks rather than distributed uniformly across the connectome \citep{vazquezrodriguez2019gradients,zamani2022local,gao2026gradient}. Existing analytical approaches often summarize SC--FC coupling using a single connectome-wide correlation or region-wise scalar scores \citep{honey2009predicting,vazquezrodriguez2019gradients,baum2020development}. Global correlations obscure spatial variation, whereas region-wise scores characterize such variation without directly identifying a compact node set whose induced structural subgraph maximizes cross-modal concordance. To move beyond global and region-wise scalar summaries, we therefore seek a framework that identifies structural subnetworks with maximal SC--FC concordance within functional communities.

Extensive prior research has sought to characterize the relationship between structural and functional connectivity (SC-FC). Global correlation approaches \citep{honey2009predicting,suarez2020linking,fotiadis2024structure} reduce the connectome to a single summary statistic, thereby failing to identify localized subnetworks. Predictive SC-to-FC models \citep{honey2009predicting,abdelnour2014network,suarez2020linking} aim to explain functional connectivity from structural connectivity, while graph-theoretic summaries \citep{bullmore2009complex,rubinov2010complex} characterize network topology. These formulations do not directly target the joint estimation of a cross-modal alignment transformation and a sparse node set defining an induced structural subgraph. Subnetwork inference \citep{zalesky2010network}, functional parcellation \citep{yeo2011organization,craddock2012whole}, and regional SC--FC analysis \citep{zamani2022local} provide complementary tools for characterizing localized network organization. Here, SC and FC share anatomical node labels, and alignment relates latent functional geometry to a selected structural subgraph while preserving the anatomical identities of its nodes. In practice, leading functional gradients are often rotated or otherwise transformed relative to the structural connectome \citep{margulies2016situating,vazquezrodriguez2019gradients,vosdewael2020brainspace,gao2026gradient}. Consequently, a localized search that ignores this mismatch tends to recover the structurally densest subset rather than the subnetwork genuinely aligned with functional organization.

Our primary objective in this paper is to address this fundamental challenge: given paired SC and FC matrices across subjects, we aim to identify a sparse, interpretable set of brain regions whose induced structural subgraph is maximally concordant with the dominant functional organization. Specifically, the proposed methodology is designed to overcome the two primary limitations of existing SC-FC analyses. First, the relevant signal manifests at the subnetwork level. Concordance is an intrinsic property of a node set rather than of individual nodes. Consequently, a valid estimation procedure must jointly model the connection patterns of node sets instead of evaluating individual regions sequentially. Second, as FC and SC are typically embedded in disparate coordinate systems, the two modalities may differ by a nontrivial transformation that a purely localized search cannot adequately capture. Therefore, a rigorous methodological solution must explicitly account for this cross-modal mismatch to localize the target subnetwork, while simultaneously ensuring the statistical interpretability of the identified brain regions.

\begin{figure}[!t]
    \centering
    \includegraphics[width=\linewidth]{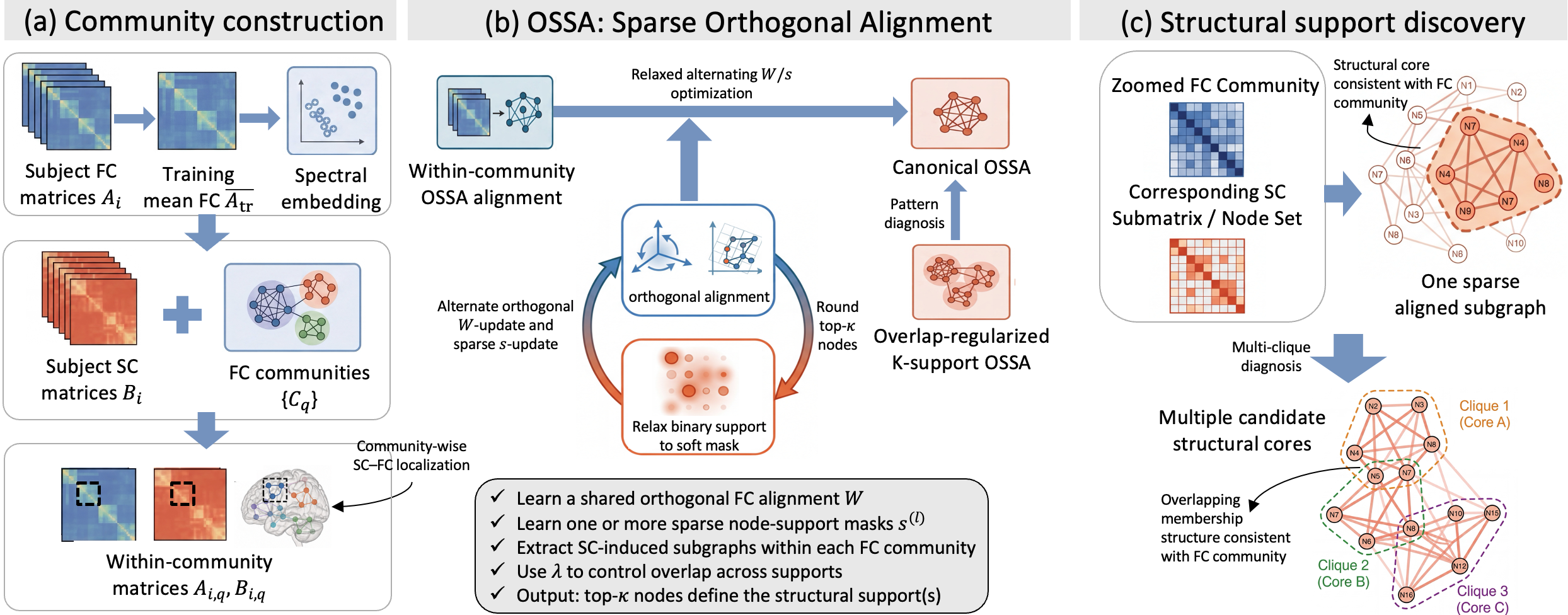}
    \caption{Overview of the \textsc{Ossa}. \textbf{(a)} SC and FC are restricted to functional communities obtained from a spectral clustering of population-averaged FC. \textbf{(b)} Within a community, \textsc{Ossa} jointly learns an orthogonal FC alignment $W$ and $K$ sparse node-support masks. \textbf{(c)} Rounding each mask gives an induced SC subgraph. An overlap penalty controls whether the $K$ supports collapse or separate.}
    \label{fig:overview}
\vspace{-0.2cm}
\end{figure}

To address this methodological gap, we introduce \textbf{O}rthogonal \textbf{S}parse \textbf{S}ubgraph \textbf{A}lignment (\textbf{\textsc{Ossa}}), a novel framework that integrates these dual requirements into a unified variational problem. Specifically, \textsc{Ossa} couples an orthogonal transformation component with a sparse subgraph selection component (Figure~\ref{fig:overview}). The method jointly estimates an orthogonal transformation of the functional connectivity and a sparse node-selection vector, thereby inducing a structural subgraph that maximizes cross-modal concordance. The transformation component serves as a structure-aware alignment mechanism, absorbing the global coordinate mismatch between modalities while preserving the intrinsic geometry of the functional data. Concurrently, the sparse selection component directly encodes the combinatorial subgraph structure within the continuous optimization space. Together, these components extend the classical orthogonal Procrustes problem \citep{schonemann1966generalized} to a setting where the alignment target is itself an unknown, induced subgraph of the structural connectome. More generally, \textsc{Ossa} addresses the joint alignment and localization of paired similarity representations on a common index set when both the transformation and the informative subset are unknown. Related settings include comparing learned representations with localized shared information and aligning embedding spaces with only a subset of reliable anchors.

To the best of our knowledge, \textsc{Ossa} represents the first principled statistical formulation for the cross-modal subnetwork identification problem in SC-FC analysis. The problem can be solved by a computationally efficient algorithm that seamlessly integrates the orthogonal alignment and sparse selection components. Furthermore, we establish theoretical guarantees for canonical \textsc{Ossa} within a fixed population community. Under the required conditions, a local maximizer exists near the population target, and every local maximizer in the specified neighborhood achieves exact support recovery and the finite-sample aligned functional projector error bound with high probability.
The empirical utility of the proposed methodology is evaluated through extensive simulation experiments, which highlight its superior accuracy and robustness compared to conventional baseline methods. Finally, we apply \textsc{Ossa} to analyze multi-modal brain connectome data from the Adolescent Brain Cognitive Development (ABCD) study. Our analysis successfully extracts highly interpretable structural modules that tightly couple with functional organization, yielding novel scientific insights into the anatomical constraints of the developing human brain. The identified supports also remain highly consistent across the baseline and Year-2 follow-up releases, indicating that the dominant aligned cores are stable under longitudinal neurodevelopmental variation.

\section{Orthogonal sparse subgraph alignment}
\label{sec:model}

Consider a study comprising $n$ subjects, indexed by $i \in \{1, \ldots, n\}$. For each subject $i$, we observe a paired set of $P \times P$ symmetric matrices, $A_i$ and $B_i$, representing the functional connectivity (FC) and structural connectivity (SC) networks, respectively, defined on a common atlas of $P$ brain regions. We assume that $A_i$ and $B_i$ have been preprocessed according to standard pipelines in brain connectome studies.

\paragraph{Communities.} Let $\{\mathcal{C}_q\}_{q=1}^Q$ denote a partition of the atlas nodes derived from the training FC data. Each set $\mathcal{C}_q$ represents a module of brain regions, hereafter referred to as a functional community, which is assumed to be shared across all subjects. Such community structures are commonly studied in the context of functional connectivity patterns \citep{garcia2018applications,chen2022harmonizing}. Let $p_q = |\mathcal{C}_q|$ denote the size of community $q$. Under the assumption of a multiple degree-corrected block structure, the algorithm proposed by \citep{agterberg2025joint} exactly identifies the communities $\mathcal{C}_q$ with high probability. Consequently, we condition our subsequent analysis on the successful recovery of these communities, and we formulate the \textsc{Ossa} objective within each FC-derived community. In practice, the community detection algorithm of \cite{agterberg2025joint} is applied prior to \textsc{Ossa}. Define $A_{i,q}$ and $B_{i,q}$ as the principal submatrices of $A_i$ and $B_i$, respectively, induced by the node set $\mathcal{C}_q$. For notational conciseness, when the context restricts focus to a single community, we suppress the subscript $q$ and overload the notation, denoting the submatrices as $A_i$ and $B_i$, and the community size as $p$.

Fixing each community, our target is to recover a shared subset of brain regions $\mathcal V\subseteq\{1,\ldots,p\}$, where $|\mathcal V|\le \kappa$, such that the $\mathcal V$-induced SC subgraph has maximal agreement with the FC geometry across subjects. This $\mathcal V$-induced SC subgraph thus serves as a structural backbone for the FC patterns.

\subsection{The canonical \textsc{Ossa}}
\label{subsec:single-ossa}
Let $O(p)$ be the set of all $p\times p$ orthogonal matrices, and write \(S=\operatorname{diag}(s)\). Fixing one community, the \textsc{Ossa} problem is defined as the following optimization problem
\begin{equation}
\label{eq:ossa-hard}
\begin{aligned}
\max_{W,s}\quad
&J(W,s)=
\sum_{i=1}^n
\Tr\!\left[
\left(W^\top A_i W\right)
\left(SB_iS\right)
\right]\\
\text{s.t.}\quad
&W^\top W=I_p,\qquad
s\in\{0,1\}^p,\qquad
\one^\top s=\kappa.
\end{aligned}
\end{equation}
where $\Tr$ is the trace operator. Two parameters are introduced in the \textsc{Ossa} problem. The selection vector $s$ identifies a subset of nodes from the SC, such that $S B_i S$ represents the induced structural subgraph of $B_i$. Concurrently, the rotation matrix $W$ captures a shared cross-modal transformation mapping. The objective function $J(W,s)$ explicitly evaluates the alignment between the transformed FC matrix and the corresponding induced SC subgraph. Because the induced matrix $S B_i S$ is generally expected to form a locally dense structural module within the selected node set, it can be conceptualized as a structural clique of the SC network, a topological connectivity pattern widely documented in empirical connectome studies \citep{reimann2017cliques, sizemore2018cliques, vangimalla2022communities}.  Both components are indispensable to this formulation: without the orthogonal transformation $W$, the model would impose an exact coordinate-level comparability between FC and SC, a restrictive assumption rarely supported by empirical connectome data. Conversely, without the sparse selection vector $s$, the objective degenerates into a global alignment metric, failing to achieve precise anatomical localization. As \textsc{Ossa} is formulated within individual functional communities, its integration with functional community detection establishes a hierarchical refinement of the cross-modal SC-FC relationship.

Directly optimizing the objective in \eqref{eq:ossa-hard} is NP-hard due to the combinatorial nature of the binary constraints. To circumvent these difficulties, we employ a continuous relaxation of the domain for $s$, formulating the following surrogate optimization problem.
\begin{equation}
\label{eq:ossa-relaxed}
\max_{W\in O(p),\;s\in\mathcal S_\kappa} J(W,s),
\qquad
\mathcal S_\kappa=\{s\in[0,1]^p \colon \one^\top s\le \kappa\}.
\end{equation}
\begin{algorithm}[t]
\caption{Canonical \textsc{Ossa}}
\label{alg:single-ossa}
\begin{algorithmic}[1]
\Require Symmetric matrices $\{A_i,B_i\}_{i=1}^n$ in one community; budget $\kappa$; initial steps $\eta_W,\eta_s$; backtracking factor $\beta\in(0,1)$
\Ensure Alignment $W$, relaxed support $s$, and hard support $\widehat{\mathcal V}$
\State Initialize $W\leftarrow I_p$ or a random orthogonal matrix; initialize $s\in\mathcal S_\kappa$
\Repeat
    \State $S\leftarrow\diag(s)$ and $M_i\leftarrow S B_i S$ for $i=1,\ldots,n$
    \State $G_W\leftarrow 2\sum_{i=1}^n A_i W M_i$ and $W^+(\eta)\leftarrow\Pi_{O(p)}(W+\eta G_W)$
    \State Choose the largest $\eta\in\{\eta_W\beta^m:m\ge0\}$ satisfying
    \Statex \hskip\algorithmicindent $J(W^+(\eta),s)\ge J(W,s)$, and set $W\leftarrow W^+(\eta)$
    \State $D \leftarrow \sym\{\sum_{i=1}^n (W^\top A_i W) \circ B_i\}$
    \State $g_s\leftarrow 2Ds$ and $s^+(\eta)\leftarrow\Pi_{\mathcal S_\kappa}(s+\eta g_s)$
    \State Choose the largest $\eta\in\{\eta_s\beta^m:m\ge0\}$ satisfying
    \Statex \hskip\algorithmicindent $J(W,s^+(\eta))\ge J(W,s)$, and set $s\leftarrow s^+(\eta)$
\Until{the objective and iterate changes satisfy the stopping tolerances}
\State $\widehat{\mathcal V}\leftarrow\Top_\kappa(s)$
\State \Return $W$, $s$, $\widehat{\mathcal V}$
\end{algorithmic}
\end{algorithm}

Upon solving the relaxed formulation in \eqref{eq:ossa-relaxed}, the final induced subnetwork is recovered via a hard thresholding procedure, $\widehat{\mathcal{V}} = \Top_\kappa(\widehat{s})$, where $\Top_\kappa$ extracts the indices corresponding to the $\kappa$ largest entries. This continuous relaxation preserves the natural interpretation of $s$ as a soft subgraph-membership indicator while simultaneously rendering the optimization computationally tractable. We solve \eqref{eq:ossa-relaxed} by a projected block-coordinate ascent procedure, outlined in Algorithm~\ref{alg:single-ossa}.

\paragraph{Theoretical properties.}
We study \eqref{eq:ossa-relaxed} within a fixed population community.
The matrices restricted to that community satisfy the sampling conditions
in Assumptions~\ref{ass:iid-symmetry}--\ref{ass:sub-exponential-tail}.
Define
\begin{equation}
\label{eq:empirical-objective}
\bar J_n(W,s)=\frac{J(W,s)}{n},
\qquad
\mathcal J(W,s)=\mathbb E\bar J_n(W,s).
\end{equation}
For $V\subseteq\{1,\ldots,p\}$, let $\one_V$ be its indicator vector and
$P_V=\diag(\one_V)$. Fix $V^\star$ with $|V^\star|=\kappa$ and write
$u^\star=\one_{V^\star}$.
Assumption~\ref{ass:population-target} specifies a compact smooth embedded
submanifold $\mathcal W^\star\subset O(p)$ and a rank-$\kappa$ orthogonal
projector $\Pi^\star$ such that
$\mathcal J(U,u^\star)=J^\star$,
$\operatorname{grad}_W\mathcal J(U,u^\star)=0$, and
$UP_{V^\star}U^\top=\Pi^\star$ for every $U\in\mathcal W^\star$, where $\operatorname{grad}_W$ denotes the gradient for the Frobenius metric on $O(p)$.
We let \(J^\star\) denote the common value of the population alignment objective \(\mathcal J(U,u^\star)\) for \(U\in\mathcal W^\star\). 
Under the curvature condition in Assumption~\ref{ass:normal-curvature}, each such alignment is a local maximizer of \(\mathcal J(\,\cdot\,,u^\star)\).
The matrices in \(\mathcal W^\star\) are therefore stationary alignments whose columns indexed by \(V^\star\) span the same subspace.

Theorem~\ref{thm:canonical-ossa} establishes the existence of local maximizers satisfying \eqref{eq:main-localization} and proves $\widehat s=u^\star$ and the bound on $\|\widehat W P_{V^\star}\widehat W^\top-\Pi^\star\|_F$ in \eqref{eq:main-local-conclusions} for every such maximizer.
\begin{theorem}\label{thm:canonical-ossa}
Suppose Assumptions~\ref{ass:iid-symmetry}--\ref{ass:support-margin}
hold.  
In addition, let $x>0$ and $0<r<r_0$ satisfy Assumption~\ref{ass:local-separation}.
With probability at least $1-2e^{-x}$, \eqref{eq:ossa-relaxed} has a
local maximizer $(\widehat W,\widehat s)$ in a neighborhood of $(\mathcal W^\star,u^\star)$ satisfying
\begin{equation}
\label{eq:main-local-conclusions}
\widehat s=u^\star,
\qquad
\|\widehat W P_{V^\star}\widehat W^\top-\Pi^\star\|_F
\le
\frac{8(\kappa-1)\sqrt{\kappa}\,\eta_n(x)}{\alpha_n}.
\end{equation}
Here $\eta_n(x)$, and $\alpha_n$ are specified in Section~\ref{app:proof}.
\end{theorem}
\noindent Proof is given in Section~\ref{app:proof}. 
In Theorem~\ref{thm:canonical-ossa}, \(\eta_n(x)\) denote a high-probability bound on the operator-norm differences between sample and population averages of FC matrices weighted by SC edge values, uniformly over all edges. 
The constant \(\alpha_n\) characterizes the lower bound of the magnitude of the negative curvature of the population objective in directions normal to the target alignment set $\mathcal W^\star$. 
The radius \(r_0\) limits the size of the neighborhood in which the local recovery analysis applies.
For every local maximizer covered by Theorem~\ref{thm:canonical-ossa}, the first conclusion of the support recovery yields \(\widehat V=\operatorname{Top}_\kappa(\widehat s)=V^\star\). Consequently, \(\widehat W P_{\widehat V}\widehat W^\top=\widehat W P_{V^\star}\widehat W^\top\) is the orthogonal projector onto the functional subspace spanned by the columns of \(\widehat W\) indexed by the recovered SC support. 
The two conclusions therefore address both components of OSSA: identifying the structural node set and recovering its aligned functional subspace. 
Along any sequence satisfying the assumptions of Theorem~\ref{thm:canonical-ossa} with \(x=x_n\to\infty\), these estimated projectors are consistent in probability for \(\Pi^\star\) if the right-hand side of \eqref{eq:main-local-conclusions} tends to zero.
These guarantees concern local maximizers in the specified neighborhood. We do not claim that Algorithm~\ref{alg:single-ossa}, including its multi-restart selection rule below, reaches this neighborhood or returns one of these maximizers. The monotonicity of the objective values alone does not ensure this; see Appendix~\ref{app:monotonicity} and Appendix~\ref{app:broader-impact}.

\subsection{Extension to multiple cliques}
\label{subsec:multi-ossa}
The canonical \textsc{Ossa} framework identifies a single dominant clique within each functional community. This configuration is practically well-justified, as communities in functional connectivity networks are typically already restricted in scale (comprising approximately 10-30 nodes). Further partitioning these modules into multiple cliques is rarely necessary from a dimensionality reduction perspective. Indeed, as demonstrated in our analysis of the ABCD dataset (Section~\ref{subsec:real}), empirical evidence strongly corroborates the prevalence of a single-clique structural topology within these localized functional modules. Nevertheless, for broader applicability, the \textsc{Ossa} formulation can be naturally extended to extract multiple cliques per community. Specifically, we generalize the method by introducing $K$ to denote the predefined number of target cliques to be identified within a given community, as \textsc{Ossa}-$K$. Define
\[
s^{(\ell)}\in\mathcal S_\kappa,
\qquad
S^{(\ell)}=\diag(s^{(\ell)}),
\qquad
\mathcal S_\kappa=\{s\in[0,1]^p \colon \one^\top s\le\kappa\}, \quad 1\le \ell \le K.
\]
The $K$-clique \textsc{Ossa} problem solves
\begin{equation}
\label{eq:multi-ossa-opt}
\begin{aligned}
\max_{W\in O(p),\;s^{(1)},\ldots,s^{(K)}\in\mathcal S_\kappa} \sum_{\ell=1}^K\sum_{i=1}^n
\Tr\!\left[
\left(W^\top A_i W\right)
\left(S^{(\ell)}B_iS^{(\ell)}\right)
\right]
-
\lambda\sum_{\ell<\ell'}\inner{s^{(\ell)}}{s^{(\ell')}}.
\end{aligned}
\end{equation}
This objective is a direct generalization of the canonical formulation, incorporating an explicit regularization term to penalize redundant or highly overlapping node memberships. Crucially, the orthogonal transformation matrix $W$ is shared across all structural subgraphs. This shared rotation ensures that all candidate subnetworks are evaluated within a unified functional coordinate system, thereby preventing clique-specific transformations from artificially absorbing the underlying separation structure. We employ a similar projected block-coordinate ascent procedure. The complete optimization routine is detailed as Algorithm~\ref{alg:multi-ossa} in Appendix~\ref{app:multi-algo}.

\paragraph{Remark.}
We employ the $K$-clique extension primarily as a diagnostic tool. If multiple distinct cliques adequately explain the functional connectivity, the estimated supports should diverge at moderate values of the regularization parameter $\lambda$, while maintaining robust cross-modal alignment on held-out data. Conversely, if a given FC community contains only a single aligned clique, the supports would collapse into a unified structure at small values of $\lambda$. When artificially forced apart by increased penalization, the extraneous supports may capture generic correlation structures but will cease to remain competitive under the primary alignment objective.

\section{Simulation experiments}
\label{sec:experiments}
We first evaluate \textsc{Ossa} in simulations with known target supports. These experiments isolate support recovery, the role of the orthogonal block, and whether the $K$-support extension separates repeated recovery from distinct cores. All simulation experiments use fixed train/validation/test splits. We use the $W=I_p$ ablation, \textsc{Ossa}-\textsc{noW}, to isolate the learned rotation. Uncertainty is reported as standard error over simulation replicates.

\subsection{Evaluation of the canonical \textsc{Ossa}}
\label{subsec:single-sim}
\paragraph{Experiment setup.}
We first test recovery of one planted clique under FC-SC coordinate mismatch. Each replicate contains $n=30$ subjects in one community of size $p=40$, with a planted support $\mathcal V^\star$ of size $\kappa^\star=10$. For each subject $i$,
\[
B_i = C_i + \zeta_i^{B} + \sigma_B E_i^{B},
\qquad
A_i = W^\star C_i {W^\star}^\top + \zeta_i^{A} + \sigma_A E_i^{A},
\]
with all matrices symmetrized and diagonal entries set to zero. The shared signal is $C_i=\LR_3(\mathcal V^\star,1.5)$, where $\LR_r(D,c)$ denotes a symmetric mixed-sign block supported on $D\times D$, generated as $UU^\top\circ\Sigma$ with entries $U_{ab}\overset{\mathrm{i.i.d.}}{\sim}\mathcal N(0,c^2/r)$ and a symmetric random sign matrix $\Sigma$. The additional terms $\zeta_i^A,\zeta_i^B$ collect subject-level background and modality-specific nuisance variation, their exact construction is given in Appendix~\ref{app:simu-setup}. These terms are included to stress-test recovery under extra single-modality variation, not as part of the \textsc{Ossa} model itself. We control mismatch by $W^\star=\polar((1-\alpha)I_p+\alpha Q)$, where $Q\sim\mathrm{Haar}(O(p))$ and $\polar(X)=UV^\top$ for $X=U\Sigma V^\top$ is the orthogonal factor in the polar decomposition. Thus $\alpha=0$ gives matched coordinates, and $\alpha=1$ gives a random rotation. 

We compare \textsc{Ossa} with size-matched baselines that select $\kappa$ nodes by simple marginal or graph summaries: (1) Marginal degree: \textsc{SC-Degree} and \textsc{FC-Degree} take the top-$\kappa$ nodes by row-sum of $\overline{|B|}$ and $\overline{|A|}$, respectively, where the average is over subjects. (2) Edgewise modality fusion: \textsc{Joint-SCxFC} takes the top-$\kappa$ nodes by row-sum of $\overline{|A_i \circ B_i|}$. (3) SC graph summary: \textsc{Communicability}~\citep{estrada2008communicability} takes the top-$\kappa$ nodes by column-sum of $\exp(\overline{B})$. (4) \textsc{Random} samples a uniform $\kappa$-node subset as a control floor. We also include two \textsc{Ossa} controls: \textsc{Ossa}-\textsc{noW} fixes $W=I_p$, and \textsc{Ossa}-\textsc{oracle} fixes $W=W^\star$ as an upper bound. All \textsc{Ossa} variants use eight restarts, and we keep the largest training objective. We report Jaccard similarity with the planted support and exact recovery rate, averaged over 10 independent replicates for each $(\alpha,\sigma,n)$ setting. Detailed setup is defined in appendix~\ref{app:simu-setup}.

\paragraph{Results.}
Among the tested methods, \textsc{Ossa} remains accurate under latent FC-SC rotation (Table~\ref{tab:single-summary}; full rotation and noise sweeps are reported in Appendix~\ref{app:single-sweeps}). In the full sweep, fixed-coordinate methods work at $\alpha=0$ but fail as rotation increases. At full rotation, \textsc{Ossa} reaches Jaccard $0.861\pm0.042$ and exactly recovers $\mathcal V^\star$ in $40\%$ of replicates, compared with $0.205\pm0.045$ for \textsc{Ossa}-\textsc{noW}. The oracle run with $W=W^\star$ is nearly perfect, suggesting that the remaining gap comes from estimating $W$. \textsc{Ossa} also remains stable at $\sigma=0.5$ ($0.827\pm0.043$) and reaches $0.85$-$0.89$ for $n\ge30$, where baselines show little gain with larger $n$.

\begin{table}[t]
\caption{Canonical \textsc{Ossa} clique support recovery on three diagnostic slices (mean $\pm$ s.e.). \textsc{Exact} reports $\Pr(\widehat{\mathcal V} = \mathcal V^\star)$ at $\alpha = 1$. \textsc{Ossa}-\textsc{oracle} is reported separately as an upper bound.}
\label{tab:single-summary}
\centering
\tablesetup
\begin{tabular}{lcccc}
\toprule
Method & \multicolumn{2}{c}{Rot.\ $\alpha{=}1$} & Noise $\sigma{=}0.5$ & $n{=}80$ ($\alpha{=}1$) \\
\cmidrule(lr){2-3}
 & Jaccard & Exact & Jaccard & Jaccard \\
\midrule
\textsc{Ossa}            & \textbf{0.861\,$\pm$\,0.042} & 0.40 & \textbf{0.827\,$\pm$\,0.043} & \textbf{0.894\,$\pm$\,0.038} \\
\textsc{Random}          & 0.162\,$\pm$\,0.028 & 0.00 & 0.154\,$\pm$\,0.025 & 0.219\,$\pm$\,0.031 \\
\textsc{SC-Degree}       & 0.203\,$\pm$\,0.027 & 0.00 & 0.207\,$\pm$\,0.034 & 0.174\,$\pm$\,0.024 \\
\textsc{FC-Degree}       & 0.037\,$\pm$\,0.012 & 0.00 & 0.050\,$\pm$\,0.018 & 0.045\,$\pm$\,0.019 \\
\textsc{Joint-SCxFC}     & 0.157\,$\pm$\,0.031 & 0.00 & 0.190\,$\pm$\,0.028 & 0.263\,$\pm$\,0.028 \\
\textsc{Communicability} & 0.179\,$\pm$\,0.035 & 0.00 & 0.186\,$\pm$\,0.035 & 0.162\,$\pm$\,0.027 \\
\textsc{Ossa}-\textsc{noW}        & 0.205\,$\pm$\,0.045 & 0.00 & 0.199\,$\pm$\,0.033 & 0.240\,$\pm$\,0.040 \\
\midrule
\textsc{Ossa}-\textsc{oracle}     & 1.000\,$\pm$\,0.000 & 1.00 & 1.000\,$\pm$\,0.000 & 0.982\,$\pm$\,0.018 \\
\bottomrule
\end{tabular}
\vspace{-0.2em}
\end{table}

\subsection{Multi-clique simulation}
\label{subsec:multi-sim}
We further evaluate \textsc{Ossa}-$K$ in a multi-clique setting. We use one community ($p=50$, $n=30$), a shared latent rotation $\alpha=0.6$, and noise $\sigma=0.4$, with either two disjoint supports, two supports sharing four nodes, or three disjoint supports. We compare against random, SC-spectral, communicability, and \textsc{Sequential-Ossa} $K$-support baselines. \textsc{Sequential-Ossa} is a greedy baseline that fits canonical \textsc{Ossa}, masks the selected nodes, and refits. Details of the generator, baselines, $\lambda$ grid, and full results are in Appendices~\ref{app:multi-sim-details}-\ref{app:lambda-traj}.

Table~\ref{tab:multi-summary} summarizes the diagnostic outcome. \textsc{Ossa}-$K$ recovers all three regimes. In the overlap case, it estimates $3.75\pm0.56$ shared nodes for a planted overlap of four. The strongest baseline, \textsc{Sequential-Ossa}, works mainly in disjoint regimes and degrades under planted overlap because its masking step enforces disjoint supports. Thus the shared-$W$ objective in \eqref{eq:multi-ossa-opt} recovers multi-clique structure while allowing controlled overlap.

\begin{table}[t]
\caption{Multi-support simulation summary (mean $\pm$ s.e.). The \textsc{Ossa}-$K$ and best-baseline columns report best-match Jaccard. Fitted overlap is the mean pairwise support overlap. The full baseline table is in Appendix~\ref{app:multi-sim-details}.}
\label{tab:multi-summary}
\centering
\tablesetup
\begin{tabular}{lcccc}
\toprule
Regime & Planted ovl. & \textsc{Ossa}-$K$ Jacc. & Best base. Jacc. (\textsc{Seq.-Ossa}) & \textsc{Ossa}-$K$ ovl. \\
\midrule
Disjoint $K=2$      & 0 & $\mathbf{0.945 \pm 0.023}$ &  $0.710 \pm 0.040$ & $0.00 \pm 0.00$ \\
Overlap $r=4,K=2$  & 4 & $\mathbf{0.910 \pm 0.066}$ &  $0.448 \pm 0.026$ & $\mathbf{3.75 \pm 0.56}$ \\
Disjoint $K=3$      & 0 & $\mathbf{0.975 \pm 0.016}$ &  $0.744 \pm 0.063$ & $0.00 \pm 0.00$ \\
\bottomrule
\end{tabular}
\vspace{-0.2em}
\end{table}

\subsection{Robustness}
\label{subsec:robust-sim}
We run three robustness checks for choices that are usually unknown in practice, full results are in Appendix~\ref{app:robustness}. \textbf{(i) Sparsity-budget misspecification} (Appendix~\ref{app:budget}): When $\kappa$ is swept from $4$ to $16$ around $\kappa^\star=10$ at $\alpha=0.6$ and $\sigma=0.4$, \textsc{Ossa} stays above all non-\textsc{Ossa} baselines. Smaller budgets favor precision and larger budgets favor recall. \textbf{(ii) Selection stability} (Appendix~\ref{app:stability}): Bootstrap selection frequency is higher on planted than non-planted nodes ($0.46$ vs.\ $0.18$), and per-restart pairwise Jaccard has median $0.43$. 
The observed variation in selected supports across initializations indicates sensitivity to initialization in the joint nonconvex optimization problem. We therefore use multiple restarts in the main experiments and keep the run with the largest training objective.
\textbf{(iii) Continuous $\lambda$-trajectory} (Appendix~\ref{app:lambda-traj}): the dense $\lambda$ grid spans four decades, and fitted overlap decreases from collapse, through the planted value, to full disjointness. These checks motivate the fixed-budget, multi-restart, and $\lambda$-diagnostic choices used for ABCD.

\section{Cross-modal analysis of adolescent brain connectome}
\label{subsec:real}
We next evaluate \textsc{Ossa} on Adolescent Brain Cognitive Development (ABCD) study \citep{casey2018abcd,hagler2019image}, a large multi-site longitudinal neuroimaging study of adolescent brain development. We use ABCD structural and functional connectomes on the 68-region Desikan-Killiany atlas \citep{desikan2006automated}. The main evaluation uses the baseline release: FC is signed Pearson correlation and SC is normalized streamline density. We symmetrize both modalities, ignore diagonals, Fisher-$z$ transform FC, clip negative FC values, and apply $\log(1+\cdot)$ to SC. A cohort-median $5\sigma$ QC rule on off-diagonal means removes 10 subjects, leaving $n_{\mathrm{ABCD}}=2{,}350$. We use deterministic 60/20/20 train/validation/test splits over 10 seeds. Community count, sparsity budget, and baseline-side rotations are selected using only training/validation data. The test set is evaluated once after these choices are fixed.

\begin{figure*}[t]
  \centering
  \includegraphics[width=\textwidth]{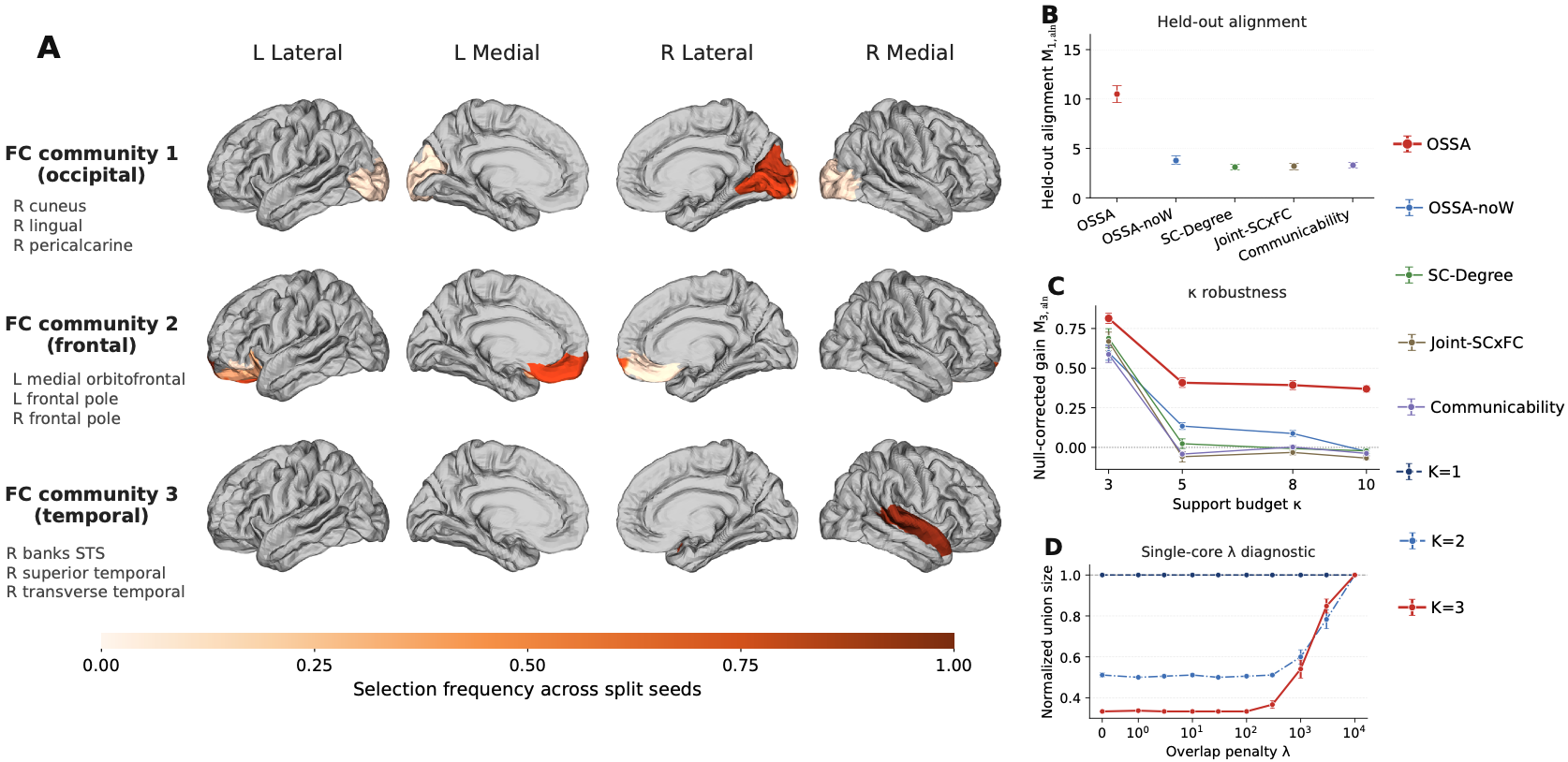}
  \caption{ABCD localization and diagnostics.
  (A) \textsc{Ossa} selection frequency for the three FC communities at $\kappa=3$. Labels denote consensus supports.
  (B) Mean $\pm$ s.e.\ of held-out rotation-aware alignment $M_{1,\mathrm{aln}}$ for the five methods shown (Table~\ref{tab:abcd-main}).
  (C) Budget robustness: \textsc{Ossa} maintains positive SC-degree-quartile-matched gain $M_{3,\mathrm{aln}}$.
  (D) $K$-support $\lambda$ diagnostic: supports collapse at small $\lambda$ and separate only under large penalties.}
  \label{fig:abcd-main}
\vspace{-0.2cm}
\end{figure*}

\paragraph{Community partition and budget.}
Network cross-validation offers an alternative framework for selecting the number of communities under network models \citep{chen2018network,li2020network}.
Across all 10 split seeds, we select $Q=3$ communities by jointly considering eigengap \citep{vonluxburg2007tutorial}, Newman-Girvan modularity \citep{newman2004finding}, and silhouette \citep{rousseeuw1987silhouettes}, with partitions stable up to label permutation. This stability is assessed under the same community-construction procedure; the identities of the regions selected by \textsc{Ossa} remain conditional on that procedure (see Appendix~\ref{app:abcd-community-sensitivity}). The training-only diagnostic curves are reported in Appendix~\ref{app:abcd-community}. The pooled community sizes are $(30,23,15)$. We set $\kappa=3$ in each community, the smallest nontrivial induced SC subgraph after zeroing SC diagonals: $\kappa=1$ gives zero objective, and $\kappa=2$ selects one edge. This choice is also supported by the baseline validation curves: averaged across split seeds, alignment per selected node, $J_{\mathrm{val}}(\kappa)/\kappa$, is highest at $\kappa=3$ among the tested budgets $\kappa\ge3$ in each community. We also check $\kappa\in\{3,5,8,10\}$ to test sensitivity to this scale (Appendix~\ref{app:abcd-kappa-robust}).

\paragraph{Baselines and metrics.}
We compare \textsc{Ossa} with fourteen size-matched support-selection baselines and the \textsc{Ossa}-\textsc{noW} ablation. (1) \emph{Marginal degree} (\textsc{SC-Degree}, \textsc{FC-Degree}) rank nodes by row sums of the training means of $|B_i|$ and $|A_i|$, respectively. (2) \emph{Cross-modal fusion} (\textsc{Joint-SCxFC}) ranks row sums of the averaged $|A_i\circ B_i|$. (3) \emph{SC graph summary} (\textsc{Communicability}~\citep{estrada2008communicability}) uses column sums of $\exp(\bar B)$. (4) \textsc{SDI} \citep{preti2019decoupling} ranks nodes by the high/low-frequency FC energy ratio on the normalized SC Laplacian. (5) \emph{Deep learning} (\textsc{BrainGNN}~\citep{li2021braingnn},  \textsc{BrainNetCNN}~\citep{kawahara2017brainnetcnn}) both provide learned node saliencies, with hyperparameters selected by validation loss under the same compute budget as one \textsc{Ossa} run. (6) \textsc{Random} samples a uniform $\kappa$-node support. The ablation \textsc{Ossa}-\textsc{noW} runs Algorithm~\ref{alg:single-ossa} with $W=I_p$ fixed. We additionally evaluate nonnegative matrix factorization (NMF) \citep{lee1999learning}, sparse partial least squares (PLS) \citep{chun2010sparse}, path efficiency, mean first passage time, FAQ graph matching \citep{vogelstein2015fast}, and SC-to-FC ridge prediction as size-matched node-selection baselines. For fairness, each baseline support is refit with its own rotation. For a selected support $\widehat{\mathcal V}_{m,q}$ from method $m$ in community $q$, we hold $s=\mathbf 1_{\widehat{\mathcal V}_{m,q}}$ fixed (protocol detail in Appendix~\ref{app:abcd-baseline-W}) and fit a method-specific $\widehat W_{m,q}$ by running the \textsc{Ossa} $W$-block on training subjects. 
Thus the evaluator is the same across methods and only the support changes. Our primary metric is the out-of-sample \textsc{Ossa} objective evaluated on held-out test data, given by,
\[
M_{1,\mathrm{aln}}(\widehat{\mathcal V}, W)
= \frac{1}{n_{\text{te}}}\sum_{i=1}^{n_{\text{te}}}
\Tr\!\left[\bigl(W^\top A_i^{\text{te}} W\bigr)\bigl(S B_i^{\text{te}} S\bigr)\right],
\qquad S=\diag(\one_{\widehat{\mathcal V}}).
\]
Let $\mathrm{vec}_{>}\!(\cdot)$ extract strictly upper-triangular entries and let $\rho(\cdot,\cdot)$ denote Pearson correlation. The scale-free, rotation-aware concordance is
\[
M_{2,\mathrm{aln}}(\widehat{\mathcal V}, W) = \frac{1}{n_{\text{te}}}\sum_i
\rho\ \!\bigl(\mathrm{vec}_{>}\!\bigl((W^\top A_i^{\text{te}} W)[\widehat{\mathcal V}, \widehat{\mathcal V}]\bigr),\, \mathrm{vec}_{>}\!(B_i^{\text{te}}[\widehat{\mathcal V}, \widehat{\mathcal V}])\bigr),
\]
with $M_{2,\mathrm{raw}}=M_{2,\mathrm{aln}}(\widehat{\mathcal V},I_p)$ as a rotation-unaware control. To test whether a support merely selects structurally dense nodes, we use an SC-degree-quartile-matched null:
\[
M_{3,\mathrm{aln}}(\widehat{\mathcal V}, W) = M_{2,\mathrm{aln}}(\widehat{\mathcal V}, W) - \frac{1}{N}\sum_{r=1}^N M_{2,\mathrm{aln}}(\widehat{\mathcal V}_r, W_r),
\]
where each $\widehat{\mathcal V}^{(r)}$ matches the SC-degree-quartile profile of $\widehat{\mathcal V}$ within the same community and $\widehat W^{(r)}$ is refit by the same training-side $W$-block. We use $N=1000$ null supports and report the one-sided permutation $p$-value as the fraction whose $M_{2,\mathrm{aln}}$ exceeds the observed value.

\paragraph{Canonical \textsc{Ossa} evaluation.}
\label{subsec:abcd-single}
At $\kappa=3$, \textsc{Ossa} outperforms all size-matched baselines on held-out alignment (Table~\ref{tab:abcd-main}; Figure~\ref{fig:abcd-main}b). Across $30$ seed-community units, \textsc{Ossa} attains $M_{1,\mathrm{aln}}=10.55\pm0.85$; the closest competitors are NMF ($9.96\pm0.82$) and sparse PLS ($9.22\pm1.10$) (Tables~\ref{tab:abcd-main} and~\ref{tab:abcd-additional-baselines}). \textsc{Ossa}'s improvement over \textsc{Ossa}-\textsc{noW} ($3.82\pm0.42$), which fixes $W=I_p$, points to the contribution of the learned orthogonal block. \textsc{Ossa} achieves higher held-out alignment in all $30$ paired seed-community comparisons against each baseline in Table~\ref{tab:abcd-main}. These comparisons are descriptive, because the repeated splits reuse subjects from the same cohort. For held-out subjects from one split, \textsc{Ossa}'s mean $M_{1,\mathrm{aln}}$ exceeds NMF's by $0.468$ (paired subject-bootstrap 95\% CI: $[0.400,0.534]$; Appendix~\ref{app:abcd-baseline-W}). The recovered supports are compact and stable across split seeds. After aligning community labels to the pooled $Q=3$ partition, the consensus supports localize to right occipital, fronto-orbitofrontal/frontal-pole, and right temporal regions, respectively (Figure~\ref{fig:abcd-main}a; see Table~\ref{tab:abcd-consensus-sc} for SC edge weights). Scale-free metrics suggest that the gain is not just a structural-density effect. In raw coordinates, \textsc{SC-Degree} slightly exceeds \textsc{Ossa} on $M_{2,\mathrm{raw}}$ ($+0.674$ vs. $+0.635$), consistent with its preference for dense SC nodes. After each support is evaluated with its own fitted rotation, \textsc{Ossa} leads on $M_{2,\mathrm{aln}}$ ($+0.821$ vs. $+0.695$) and on the SC-degree-quartile-matched gain $M_{3,\mathrm{aln}}$ ($+0.81$ vs. next-best $+0.69$). These aligned metrics quantify the concordance of the selected structural subgraph with the latent functional geometry represented by the fitted transformation (see Appendix~\ref{app:abcd-alignment-diagnostics} for full empirical diagnostics); the selected SC nodes retain their anatomical identities. The $\kappa$ sweep in Figure~\ref{fig:abcd-main}c keeps positive null-corrected gain for $\kappa\in\{3,5,8,10\}$, full values are in Appendix~\ref{app:abcd-kappa-robust}.

\begin{table}[t]
\centering
\caption{ABCD canonical \textsc{Ossa} evaluation, mean $\pm$ s.e.\ across $30$ (seed,
community) units, $\kappa = 3$. $M_{3,\mathrm{aln}}$ gain is the difference
between the method's $M_{2,\mathrm{aln}}$ and the mean of $N=1000$
SC-degree-quartile-matched random nulls. $M_3\,p$ is mean (median) over the
$30$ units.}
 \label{tab:abcd-main}
\tablesetup
\begin{tabular}{lccccc}
\toprule
Method & $M_{1,\mathrm{aln}}$ & $M_{2,\mathrm{raw}}$ & $M_{2,\mathrm{aln}}$ &
$M_{3,\mathrm{aln}}$ gain & $M_3$ $p$-value \\
\midrule
\textsc{Ossa}     & $\mathbf{10.55 \pm 0.85}$ & $+0.635$ & $\mathbf{+0.821}$ & $\mathbf{+0.81}$ & $\mathbf{0.080\,(0.059)}$ \\
\textsc{Ossa}-\textsc{noW}        & $3.82 \pm 0.42$  & $+0.605$ & $+0.605$ & $+0.60$ & $0.209\,(0.180)$ \\
\textsc{Communicability} & $3.31 \pm 0.30$  & $+0.572$ & $+0.594$ & $+0.59$ & $0.190\,(0.139)$ \\
\textsc{Joint-SCxFC}     & $3.19 \pm 0.32$  & $+0.653$ & $+0.675$ & $+0.67$ & $0.147\,(0.059)$ \\
\textsc{SC-Degree}       & $3.14 \pm 0.32$  & $\mathbf{+0.674}$ & $+0.695$ & $+0.69$ & $0.138\,(0.028)$ \\
\textsc{FC-Degree}       & $2.72 \pm 0.32$  & $+0.066$ & $+0.066$ & $+0.06$ & $0.433\,(0.598)$ \\
\textsc{BrainNetCNN}     & $0.56 \pm 0.22$  & $+0.192$ & $+0.283$ & $+0.28$ & $0.345\,(0.390)$ \\
\textsc{BrainGNN}        & $0.49 \pm 0.08$  & $+0.336$ & $+0.427$ & $+0.42$ & $0.275\,(0.196)$ \\
\textsc{SDI}             & $0.31 \pm 0.04$  & $+0.299$ & $+0.389$ & $+0.38$ & $0.286\,(0.343)$ \\
\textsc{Random}          & $0.18 \pm 0.04$  & $+0.397$ & $+0.386$ & $+0.38$ & $0.300\,(0.218)$ \\
\bottomrule
\end{tabular}
\vspace{-0.2em}
\end{table}

\paragraph{Structural and alignment interpretation.}
The occipital and temporal supports include primary visual and auditory regions, consistent with stronger SC--FC correspondence in unimodal cortex \citep{vazquezrodriguez2019gradients}. Table~\ref{tab:abcd-consensus-sc} reports subject-averaged $\log(1+x)$ SC edge weights and mean-weight ratios to SC-degree-matched nulls.

\begin{table}[t]
\centering
\caption{SC connectivity of the consensus supports. Ratios use SC-degree-matched null means.}
\label{tab:abcd-consensus-sc}
\tablesetup
\begin{tabular}{lccc}
\toprule
Community & Three SC edge weights & Mean & SC/null ratio \\
\midrule
Occipital & $1.75,\ 2.62,\ 3.46$ & $2.612$ & $2.09\times$ \\
Frontal & $2.34,\ 3.11,\ 3.42$ & $2.958$ & $1.84\times$ \\
Temporal & $0.42,\ 1.24,\ 2.37$ & $1.343$ & $5.95\times$ \\
\bottomrule
\end{tabular}
\vspace{-0.2em}
\end{table}

For $\widehat\Pi=\widehat W P_{\widehat V}\widehat W^\top$, the community-mean normalized squared distance from $P_{\widehat V}$ is $0.54$--$0.67$, versus $0.80$--$0.90$ for dimension-matched Haar references. Loading participation ratios indicate $5$--$10$ effective FC regions versus $10$--$19$ under Haar. These characterize latent functional geometry; SC node identities remain unchanged. The diagnostics alone do not establish biological specificity (Appendix~\ref{app:abcd-alignment-diagnostics}).

\paragraph{Multi-clique sensitivity (\(\lambda\)-diagnostic).}
At $\kappa=3$, the ABCD path shows collapse for $\lambda\le 10^2$ (pairwise Jaccard $0.98$, $|U|/\kappa\approx1$) and separation only under large penalties. Forced separation at $\lambda=10^4$ reduces the primary held-out objective from $10.42$ to $6.42$ ($K=2$) and $4.60$ ($K=3$). Although \(M_{2,\mathrm{aln}}\) and \(M_{3,\mathrm{aln}}\) gain increase at high \(\lambda\), these appear to reflect secondary correlation motifs rather than additional high-\(M_1\) cores. These results support using \(K>1\) as a diagnostic rather than the default estimator. Full trajectories and baseline-extension checks are reported in Appendices~\ref{app:abcd-multi}-\ref{app:abcd-multisupport-baselines}.

\paragraph{Second-visit replication.}
We apply the unchanged baseline protocol and QC to the ABCD Year-2 release, yielding $n_{\mathrm{ABCD2}}=1{,}897$ subjects (Appendix~\ref{app:abcd-secondvisit}). The three-community structure is preserved, with cross-seed ARI $0.99\pm0.01$ and pooled sizes $(31,23,14)$, close to baseline $(30,23,15)$. The canonical ranking is also preserved: \textsc{Ossa} attains $M_{1,\mathrm{aln}}=10.16\pm0.87$, compared with $3.23\pm0.30$ for \textsc{Ossa}-\textsc{noW} and at most $3.19$ for classical baselines. \textsc{Ossa} leads each Table~\ref{tab:abcd-secondvisit-main} baseline in held-out alignment across all $30$ paired seed-community comparisons; these are descriptive because splits reuse follow-up subjects. The $\lambda$ diagnostic gives the same single-core pattern: collapse at small $\lambda$ and forced separation only at large $\lambda$. \textsc{Ossa}'s cross-visit support replication score (CVSR), measured by support-set Jaccard, is $0.928\pm0.033$, with mean overlap $2.86/3$ nodes. Graph-summary baselines also replicate well, whereas deep-learning baselines replicate less well. CVSR supports longitudinal stability, which alone does not establish biological specificity. Appendix~\ref{app:abcd-secondvisit} gives per-method results and the follow-up $\lambda$ sweep.

\section{Conclusion and discussion}
\label{sec:conclusion}
\textsc{Ossa} jointly localizes structural subnetworks and aligns them with latent functional geometry. In ABCD, the selected supports are compact, stable, and longitudinally reproducible; the $\lambda$ path suggests a dominant aligned core within each FC community. 
Our theoretical guarantees cover canonical \textsc{Ossa} within a fixed population community.
Future work includes establishing that the current algorithm reaches the required neighborhood with sufficiently small stationarity residuals, or developing an algorithm with these guarantees, as well as extending the statistical guarantees to the multi-clique setting.
Empirically, the analyses are restricted to the Desikan--Killiany parcellation and the ABCD cohort; the identities of the selected regions are conditional on the functional-community construction and the sparsity budget. Rigorous replication across parcellations, acquisition pipelines, and clinical cohorts is needed before assigning definitive biological significance to these supports.

\begin{ack}
This work was supported by the National Institutes of Health (NIH) under grants R01AG081413, R01EB034720, and R01AG068191 (to Y. Zhao). T. Li was also supported in part by the National Science Foundation (NSF) under grant DMS-2515367.
We thank the anonymous reviewers for their constructive feedback, which substantially improved the paper.
\end{ack}

\newpage
\appendix
 
\section{Optimization details and derivations}
\label{app:optimization}
 
This appendix gives the first-order quantities used in Algorithms~\ref{alg:single-ossa} and~\ref{alg:multi-ossa}. All derivations assume the matrices have been symmetrized. In implementation we use $X\leftarrow (X+X^\top)/2$ for numerical stability and ignore diagonal entries when evaluating inter-regional connectivity.
 
\subsection{Gradient for the canonical \textsc{Ossa} clique W-block}
\label{app:w-gradient}
 
Fix $s$ and define $M_i=SB_iS$. For one subject,
\[
f_i(W)=\Tr(W^\top A_i W M_i).
\]
Taking differentials,
\begin{align*}
df_i
&=\Tr(dW^\top A_i W M_i)+\Tr(W^\top A_i dW M_i)\\
&=\Tr\!\left((A_i W M_i)^\top dW\right)
  +\Tr\!\left((A_i^\top W M_i^\top)^\top dW\right).
\end{align*}
Therefore
\[
\nabla_W f_i(W)=A_i W M_i + A_i^\top W M_i^\top.
\]
When $A_i$ and $M_i$ are symmetric, this simplifies to
\begin{equation}
\label{eq:app-gradW-single}
\nabla_W J(W,s)=2\sum_{i=1}^n A_i W(SB_iS).
\end{equation}
A Euclidean ascent proposal $\widetilde W=W+\eta\nabla_WJ$ is mapped back to the orthogonal group by the polar projection \citep{higham1986computing}
\begin{equation}
\label{eq:app-orth-proj}
\Pi_{O(p)}(\widetilde W)=UV^\top,
\qquad
\widetilde W=U\Sigma V^\top,
\end{equation}
which is the closest orthogonal matrix to $\widetilde W$ in Frobenius norm.
 
\subsection{Reduction and gradient for the canonical \textsc{Ossa} clique s-block}
\label{app:s-gradient}
 
Fix $W$ and set $Q_i=W^\top A_iW$. Since $S=\diag(s)$,
\[
(SB_iS)_{ab}=s_a(B_i)_{ab}s_b,
\qquad
SB_iS=(ss^\top)\circ B_i.
\]
Using symmetry of $Q_i$ and $B_i$,
\begin{align*}
\Tr(Q_iSB_iS)
&=\sum_{a,b}(Q_i)_{ab}(B_i)_{ab}s_as_b
=s^\top(Q_i\circ B_i)s.
\end{align*}
Thus, with
\[
D(W)=\sum_{i=1}^n(Q_i\circ B_i),
\]
the fixed-$W$ objective is $s^\top D(W)s$, and
\begin{equation}
\label{eq:app-grads-single}
\nabla_s J(W,s)=2D(W)s.
\end{equation}
 
\subsection{Projection onto the capped simplex}
\label{app:capped-simplex}
 
The support update requires Euclidean projection \citep{wang2015projection,duchi2008efficient} onto
\[
\mathcal S_\kappa
=
\{s\in\mathbb R^p:0\le s\le 1,\ \mathbf 1^\top s\le\kappa\}.
\]
For \(y\in\mathbb R^p\), define \(z_j(\tau)=\min\{1,\max\{0,y_j-\tau\}\}\) for \(j=1,\ldots,p\). If \(\sum_j z_j(0)\le\kappa\), then \(\Pi_{\mathcal S_\kappa}(y)=z(0)\). Otherwise, \(\Pi_{\mathcal S_\kappa}(y)=z(\tau)\), where \(\tau>0\) is chosen so that \(\sum_j z_j(\tau)=\kappa\). This scalar equation is monotone in \(\tau\) and can be solved by bisection or a sorting-based capped-simplex projection method.
 
\subsection{Multi-clique gradients}
\label{app:multi-gradients}
 
For the multi-clique objective \eqref{eq:multi-ossa-opt}, fix $\{s^{(\ell)}\}_{\ell=1}^K$ and define
\[
M_i=\sum_{\ell=1}^K S^{(\ell)}B_iS^{(\ell)}.
\]
The overlap penalty is constant in $W$, so the same calculation as Appendix~\ref{app:w-gradient} gives
\begin{equation}
\label{eq:app-gradW-multi}
\nabla_W J_K(W,\{s^{(\ell)}\})=2\sum_{i=1}^n A_i W M_i.
\end{equation}
 
For fixed $W$, the objective is
\[
J_K(\{s^{(\ell)}\}\mid W)
=
\sum_{\ell=1}^K {s^{(\ell)}}^\top D(W)s^{(\ell)}
-
\lambda\sum_{\ell<\ell'}\inner{s^{(\ell)}}{s^{(\ell')}}.
\]
Holding all other supports fixed, the gradient with respect to $s^{(\ell)}$ is
\begin{equation}
\label{eq:app-grads-multi}
\nabla_{s^{(\ell)}}J_K
=
2D(W)s^{(\ell)}-
\lambda\sum_{r\ne\ell}s^{(r)}.
\end{equation}
Each support is projected onto its own capped simplex $\mathcal S_\kappa$.
 
\subsection{Monotonicity and nonconvexity}
\label{app:monotonicity}
 
Algorithms~\ref{alg:single-ossa} and~\ref{alg:multi-ossa} use backtracking steps that accept only updates that do not decrease the objective. Since $O(p)\times\mathcal S_\kappa$ and $O(p)\times\mathcal S_\kappa^K$ are compact, the corresponding objective sequences are bounded and therefore converge in value. The optimization problems remain nonconvex because of the orthogonality constraint and the quadratic support variables. We therefore treat the algorithms as first-order local solvers. In all experiments, we use multiple random restarts and retain the run with the largest training objective. 
This monotonicity and boundedness imply convergence of the objective values, but do not by themselves establish convergence of the iterates, vanishing stationarity residuals, or convergence to a global maximizer. 
The objective and iterate change tolerances are numerical stopping criteria and do not by themselves imply the residual bounds required in theory below.

\section{Technical details}\label{app:proof}
\newcounter{condition}
\renewcommand{\thecondition}{A\arabic{condition}}
\newenvironment{condition}
  {\refstepcounter{condition}\par\medskip\noindent\textbf{Assumption~\thecondition.}\ }
  {\par\medskip}
For $W\in O(p)$, define
\begin{equation}
\label{eq:Dn-D0}
D_n(W)
=
\frac{1}{n}\sum_{i=1}^n
\left\{(W^\top A_iW)\circ B_i\right\},
\qquad
D_0(W)=\mathbb E D_n(W).
\end{equation}
Then
\begin{equation}
\label{eq:quadratic-objectives}
\bar J_n(W,s)=s^\top D_n(W)s,
\qquad
\mathcal J(W,s)=s^\top D_0(W)s.
\end{equation}
For a vector $v$, write $\|v\|_1=\sum_j|v_j|$,
$\|v\|_2=(\sum_jv_j^2)^{1/2}$, and
$\|v\|_\infty=\max_j|v_j|$.  For a matrix $M$, let
$\|M\|_F=\{\Tr(M^\top M)\}^{1/2}$ and
$\|M\|_{\mathrm{op}}=\sup_{\|v\|_2=1}\|Mv\|_2$.
For a real random variable $Z$ and $r\in\{1,2\}$, define
\begin{equation}
\label{eq:orlicz-norm-definition}
\|Z\|_{\psi_r}
=
\inf\left\{t>0:\mathbb E\exp\left(\frac{|Z|^r}{t^r}\right)\le2\right\}.
\end{equation}
The cases $r=1$ and $r=2$ are the sub-exponential and
sub-Gaussian norms, respectively \citep{wainwright2019high}.

For matrices $X,Y$ of the same size, write
$\inner{X}{Y}=\Tr(X^\top Y)$.  We equip $O(p)$ with the
Riemannian metric induced by this Frobenius inner product.
For $W,U$ in the same connected component, define
\begin{equation}
\label{eq:pointwise-geodesic-distance}
d_O(W,U)
=
\inf_{\substack{\gamma:[0,1]\to O(p)\ \text{piecewise }C^1\\
                 \gamma(0)=U,\ \gamma(1)=W}}
\int_0^1\|\dot\gamma(t)\|_F\,dt.
\end{equation}
Set $d_O(W,U)=+\infty$ for points in different components.
For every nonempty $\mathcal W\subset O(p)$, define
\begin{equation}
\label{eq:alignment-distances}
d_O(W,\mathcal W)
=
\inf_{U\in\mathcal W}d_O(W,U),
\qquad
d_V(W,\mathcal W)
=
\inf_{U\in\mathcal W}\|(W-U)P_V\|_F.
\end{equation}

Differentiating $W^\top W=I_p$ along a differentiable curve gives
\begin{equation}
\label{eq:orthogonal-tangent-space}
T_WO(p)
=
\left\{
\xi\in\R^{p\times p}:W^\top\xi+\xi^\top W=0
\right\}
=
\left\{
W\Omega:\Omega^\top=-\Omega
\right\}.
\end{equation}
For $U\in O(p)$ and $\xi\in T_UO(p)$, the Riemannian exponential
for this metric is
\begin{equation}
\label{eq:orthogonal-exponential}
\operatorname{Exp}_U(\xi)=U\exp(U^\top\xi),
\qquad
\exp(M)=\sum_{k=0}^{\infty}\frac{M^k}{k!}.
\end{equation}
The curve $t\mapsto\operatorname{Exp}_U(t\xi)$ is the geodesic
with initial position $U$ and initial velocity $\xi$; see
\citet[Section~2.2.2, Eq.~(2.14)]{edelman1998geometry}.
For $Z\in\R^{p\times p}$, note that
\begin{equation}
\label{eq:tangent-projection-decomposition}
Z
=
\left\{Z-W\sym(W^\top Z)\right\}
+
W\sym(W^\top Z),
\end{equation}
where $\sym(M)=(M+M^\top)/2$.  The first term in
\eqref{eq:tangent-projection-decomposition} belongs to
$T_WO(p)$, the second belongs to its orthogonal
complement, and the two terms are orthogonal.  Hence
\begin{equation}
\label{eq:tangent-projection}
\mathcal P_W(Z)
=
Z-W\sym(W^\top Z)
\end{equation}
is the orthogonal projection onto $T_WO(p)$ (under the stated inner product) and satisfies
\begin{equation}
\label{eq:tangent-projection-contraction}
\|\mathcal P_W(Z)\|_F\le\|Z\|_F.
\end{equation}
We remark that the disaplys \eqref{eq:orthogonal-tangent-space} and
\eqref{eq:tangent-projection} are the orthogonal-group specializations of
\citet[Examples~3.5.3 and~3.6.2]{absil2008optimization}; see also
\citet[Section~2.2.1]{edelman1998geometry}.

For a differentiable function $F$ on $O(p)$,
$\operatorname{grad}_W F$ is the unique tangent vector satisfying
\begin{equation}
\inner{\operatorname{grad}_W F}{\xi}
=
DF(W)[\xi],
\qquad
\xi\in T_WO(p).
\end{equation}
If $\nabla_WF$ is the Euclidean gradient of any differentiable
ambient extension of $F$, then
\begin{equation}
\label{eq:riemannian-gradient-projection}
\operatorname{grad}_W F
=
\mathcal P_W(\nabla_WF)
=
\nabla_WF-W\sym(W^\top\nabla_WF).
\end{equation}
The projected expression is independent of the chosen extension.
For twice continuously differentiable $F$, the Riemannian Hessian
is given by
\begin{equation}
\label{eq:riemannian-hessian-definition}
\operatorname{Hess}_{WW}F(W)[\xi]
=
\mathcal P_W\left\{
\left.\frac{d}{dt}\operatorname{grad}_W F(\gamma(t))\right|_{t=0}
\right\},
\end{equation}
where $\gamma$ is any smooth curve in $O(p)$ with
$\gamma(0)=W$ and $\dot\gamma(0)=\xi$.
This is the Riemannian covariant derivative of the gradient, which is further illustrated in Section 5 of \citet{absil2008optimization}.

We use the following conditions at the target $u^\star$.

\begin{condition}
\label{ass:iid-symmetry}
The pairs $(A_i,B_i)$, $i=1,\ldots,n$, are independent and
identically distributed,
$A_i=A_i^\top$, $B_i=B_i^\top$, and
$\diag(B_i)=0$ almost surely.  Moreover,
$2\le\kappa<p$.
\end{condition}

\begin{condition}
\label{ass:sub-exponential-tail}
There exists $\nu_n>0$ such that, for every
$a,b\in\{1,\ldots,p\}$ and every pair of unit vectors
$x,y\in\R^p$,
\begin{equation}
\label{eq:sub-exponential-tail}
\|B_{i,ab}x^\top A_i y\|_{\psi_1}\le\nu_n.
\end{equation}
\end{condition}

\begin{condition}
\label{ass:population-target}
There are a nonempty compact smooth embedded submanifold
$\mathcal W^\star\subset O(p)$ and $J^\star\in\R$ such that
\begin{equation}
\label{eq:population-maximizer}
\mathcal J(U,u^\star)=J^\star,
\qquad
\operatorname{grad}_W\mathcal J(U,u^\star)=0,
\qquad
U\in\mathcal W^\star.
\end{equation}
There is also a rank-$\kappa$ projector $\Pi^\star$ satisfying
\begin{equation}
\label{eq:target-projector}
UP_{V^\star}U^\top=\Pi^\star,
\qquad
U\in\mathcal W^\star.
\end{equation}
The set contains every unselected-column completion: if
$U\in\mathcal W^\star$, $R\in O(p)$, and
$RP_{V^\star}=P_{V^\star}$, then
$UR\in\mathcal W^\star$.
\end{condition}

\begin{condition}
\label{ass:normal-curvature}
There exists $\alpha_n>0$ such that, for every
$U\in\mathcal W^\star$ and every
$\xi\in T_UO(p)$ orthogonal to $T_U\mathcal W^\star$,
\begin{equation}
\label{eq:normal-curvature}
-\inner{\xi}{
\operatorname{Hess}_{WW}\mathcal J(U,u^\star)[\xi]}
\ge
\alpha_n\|\xi\|_F^2.
\end{equation}
\end{condition}

\begin{condition}
\label{ass:support-margin}
There exists $m_n>0$ such that
\begin{equation}
\label{eq:support-margin}
\inf_{U\in\mathcal W^\star}
\left[
\min_{a\in V^\star}\{2D_0(U)u^\star\}_a
-
\max\left\{
0,\max_{b\notin V^\star}\{2D_0(U)u^\star\}_b
\right\}
\right]
\ge
m_n.
\end{equation}
\end{condition}

The diagonal restriction in
\ref{ass:iid-symmetry} applies only to $B_i$.  In particular,
\begin{equation}
\label{eq:D-zero-diagonal}
\{D_n(W)\}_{aa}
=
\{D_0(W)\}_{aa}
=
0,
\qquad
a=1,\ldots,p,
\end{equation}
whereas a zero diagonal of $A_i$ would not generally be preserved
by the transformation $W^\top A_iW$.

The completion requirement in \ref{ass:population-target} ensures
that the distance to $\mathcal W^\star$ is finite in both connected
components of $O(p)$.  Write $e_j$ for the $j$th standard basis
vector of $\R^p$.  Choose $j\notin V^\star$ and set
\begin{equation}
R_j=I_p-2e_je_j^\top.
\end{equation}
Then
$R_jP_{V^\star}=P_{V^\star}R_j=P_{V^\star}$ and
$\det(R_j)=-1$.  Hence, for every
$U\in\mathcal W^\star$,
\begin{equation}
\label{eq:completion-reflection-projector}
(UR_j)P_{V^\star}=UP_{V^\star},
\qquad
(UR_j)P_{V^\star}(UR_j)^\top
=
\Pi^\star,
\end{equation}
but $\det(UR_j)=-\det(U)$.  Thus $\mathcal W^\star$ intersects both
components.  For every $W\in O(p)$,
\begin{equation}
d_O(W,\mathcal W^\star)
=
\inf_{\substack{U\in\mathcal W^\star\\
\det(U)=\det(W)}}d_O(W,U),
\end{equation}
and the infimum is finite and attained.  If $W_1$ and $W_2$ lie in
the same component, the triangle inequality gives
\begin{equation}
\label{eq:set-distance-lipschitz}
\left|
d_O(W_1,\mathcal W^\star)
-
d_O(W_2,\mathcal W^\star)
\right|
\le
d_O(W_1,W_2).
\end{equation}
Consequently, $W\mapsto d_O(W,\mathcal W^\star)$ is continuous on
each component and hence on $O(p)$, because the two components are
both open and closed.

Condition \ref{ass:normal-curvature} permits directions tangent
to $\mathcal W^\star$ to remain unidentified and requires curvature
only in directions normal to the stationary manifold
$\mathcal W^\star$.  Condition \ref{ass:support-margin} is the
corresponding strict first-order requirement for the support block.
To make the support requirement explicit, fix
$U\in\mathcal W^\star$ and write
\begin{equation}
\label{eq:population-support-score}
g(U)
=
\nabla_s\mathcal J(U,u^\star)
=
2D_0(U)u^\star.
\end{equation}
Equations \eqref{eq:D-zero-diagonal} and
\eqref{eq:population-support-score} imply
\begin{align}
g_a(U)
&=
2\sum_{\ell\in V^\star\setminus\{a\}}
\{D_0(U)\}_{a\ell},
&&
a\in V^\star,
\\
g_b(U)
&=
2\sum_{\ell\in V^\star}
\{D_0(U)\}_{b\ell},
&&
b\notin V^\star,
\end{align}
where
\begin{equation}
\{D_0(U)\}_{a\ell}
=
\mathbb E\!\left[
B_{i,a\ell}(Ue_a)^\top A_i(Ue_\ell)
\right].
\end{equation}
Thus \eqref{eq:support-margin} compares every selected (active) coordinate's
aggregate within-support SC--FC alignment with both zero and every
outside (inactive) coordinate's aggregate alignment.

A vector $d$ is a feasible first-order direction from $u^\star$ if
and only if
\begin{equation}
\label{eq:feasible-support-directions}
d_a\le0\quad(a\in V^\star),
\qquad
d_b\ge0\quad(b\notin V^\star),
\qquad
\one^\top d\le0.
\end{equation}
The directions $-e_a$ and $e_b-e_a$ remove active mass and
replace active mass by inactive mass, respectively; their
directional derivatives are
$-g_a(U)$ and $g_b(U)-g_a(U)$.  The comparison with zero in
\eqref{eq:support-margin} is therefore essential since \eqref{eq:ossa-relaxed} permits active mass to be removed without being reallocated.

More generally, for $d$ satisfying
\eqref{eq:feasible-support-directions}, set
\begin{equation}
A_d
=
-\sum_{a\in V^\star}d_a,
\qquad
B_d
=
\sum_{b\notin V^\star}d_b,
\qquad
c(U)
=
\max\left\{0,\max_{b\notin V^\star}g_b(U)\right\}.
\end{equation}
Then $0\le B_d\le A_d$, and
\eqref{eq:support-margin} implies
\begin{align}
\inner{g(U)}{d}
&=
-\sum_{a\in V^\star}g_a(U)(-d_a)
+
\sum_{b\notin V^\star}g_b(U)d_b
\notag\\
&\le
-\{c(U)+m_n\}A_d+c(U)B_d
\le
-m_nA_d
\le
-\frac{m_n}{2}\|d\|_1.
\end{align}
Hence the directional derivative is uniformly negative along every
nonzero feasible direction. 

For $x>0$, define
\begin{equation}
\label{eq:concentration-rate}
R_n(x)
=
2\log p+p\log 9+x,
\qquad
\eta_n(x)
=
C\nu_n
\left\{
\sqrt{\frac{R_n(x)}{n}}
+
\frac{R_n(x)}{n}
\right\},
\end{equation}
where $C>0$ is a universal constant.  For
$a,b\in\{1,\ldots,p\}$, let
\begin{equation}
\label{eq:H-ab}
H_{ab}
=
\frac{1}{n}\sum_{i=1}^nB_{i,ab}A_i
-
\mathbb E(B_{i,ab}A_i),
\end{equation}
and define
\begin{equation}
\label{eq:good-event}
\mathcal G_x
=
\left\{
\max_{a,b}\|H_{ab}\|_{\mathrm{op}}\le\eta_n(x)
\right\}.
\end{equation}

Lemma~\ref{lem:operator-concentration} bounds $\max_{a,b}\|H_{ab}\|_{\mathrm{op}}$ and gives the probability of $\mathcal G_x$.

\begin{lemma}
\label{lem:operator-concentration}
Under \ref{ass:iid-symmetry}--\ref{ass:sub-exponential-tail},
the universal constant in \eqref{eq:concentration-rate} can be
chosen so that
\begin{equation}
\label{eq:good-event-probability}
\mathbb P(\mathcal G_x)\ge1-2e^{-x}.
\end{equation}
\end{lemma}

\begin{proof}
Fix $a,b$ and a unit vector $z\in\R^p$.  The variables
\begin{equation}
Y_i
=
B_{i,ab}z^\top A_i z
-
\mathbb E(B_{i,ab}z^\top A_i z)
\end{equation}
are independent and centered.  By
\eqref{eq:sub-exponential-tail}, Jensen's inequality, and the
triangle inequality for the $\psi_1$ norm,
$\|Y_i\|_{\psi_1}\le C_0\nu_n$ for a universal constant $C_0$.
The sub-exponential Bernstein inequality
\citep{vershynin2018highdimensional} therefore yields
\begin{equation}
\mathbb P\!\left(
|z^\top H_{ab}z|>t
\right)
\le
2\exp\!\left[
-c n
\min\left\{
\frac{t^2}{\nu_n^2},
\frac{t}{\nu_n}
\right\}
\right],
\qquad
t>0,
\end{equation}
for a universal constant $c>0$.

Let $\mathcal N$ be a $1/4$-net of the unit sphere in $\R^p$ with
$|\mathcal N|\le9^p$.  Because $H_{ab}$ is symmetric,
\begin{equation}
\|H_{ab}\|_{\mathrm{op}}
\le
2\max_{z\in\mathcal N}|z^\top H_{ab}z|.
\end{equation}
Indeed, for every unit vector $v$, choose $z\in\mathcal N$ such
that $\|v-z\|_2\le1/4$.  Then
\begin{equation}
|v^\top H_{ab}v|
\le
|z^\top H_{ab}z|
+
\frac{1}{2}\|H_{ab}\|_{\mathrm{op}}.
\end{equation}
A union bound over at most $p^2 9^p$ pairs and net points yields
\begin{align}
\mathbb P\!\left(
\max_{a,b}\|H_{ab}\|_{\mathrm{op}}>2t
\right)
\le
2\exp\!\left[
2\log p+p\log9
-c n
\min\left\{
\frac{t^2}{\nu_n^2},
\frac{t}{\nu_n}
\right\}
\right].
\label{eq:operator-union-bound}
\end{align}
Taking
$t=C_1\nu_n\{\sqrt{R_n(x)/n}+R_n(x)/n\}$ with $C_1$
sufficiently large makes the exponent in
\eqref{eq:operator-union-bound} at most $-x$.  Absorbing the fixed
constants into $C$ in \eqref{eq:concentration-rate} proves
\eqref{eq:good-event-probability}.
\end{proof}

Define
\begin{equation}
\label{eq:M0}
M_0
=
\max_{a,b}
\left\|
\mathbb E(B_{i,ab}A_i)
\right\|_{\mathrm{op}}.
\end{equation}
This quantity is finite under
\ref{ass:sub-exponential-tail}.  Indeed, symmetry and
\eqref{eq:sub-exponential-tail} imply
\begin{equation}
\left\|
\mathbb E(B_{i,ab}A_i)
\right\|_{\mathrm{op}}
=
\sup_{\|z\|_2=1}
\left|
\mathbb E(B_{i,ab}z^\top A_i z)
\right|
\le
C_0\nu_n.
\end{equation}
The possible dependence of $M_0$ on $n$ and $p$ is suppressed.

Lemma~\ref{lem:uniform-stochastic} bounds $(\bar J_n-\mathcal J)(W,s)$ and $\nabla_s(\bar J_n-\mathcal J)(W,s)$ uniformly on $O(p)\times\mathcal S_\kappa$ using $\mathcal G_x$.

\begin{lemma}
\label{lem:uniform-stochastic}
Under \ref{ass:iid-symmetry}--\ref{ass:sub-exponential-tail},
on $\mathcal G_x$, uniformly over
$W\in O(p)$ and $s\in\mathcal S_\kappa$,
\begin{align}
&\textnormal{(i)}\quad
\left|(\bar J_n-\mathcal J)(W,s)\right|
\le
\kappa^2\eta_n(x),
\label{eq:uniform-objective-deviation}
\\
&\textnormal{(ii)}\quad
\left\|
\nabla_s(\bar J_n-\mathcal J)(W,s)
\right\|_\infty
\le
2\kappa\eta_n(x).
\label{eq:stochastic-support-gradient}
\end{align}
\end{lemma}

\begin{proof}
Write $w_a$ for column $a$ of $W$.  The symmetry in
\ref{ass:iid-symmetry} gives
$H_{ab}=H_{ba}=H_{ab}^\top$.  Expanding $\bar J_n$ and \eqref{eq:Dn-D0} yields
\begin{equation}
(\bar J_n-\mathcal J)(W,s)
=
\sum_{a,b=1}^p
s_as_b w_a^\top H_{ab}w_b.
\end{equation}
Since $s\ge0$, $\one^\top s\le\kappa$, and
$\|w_a\|_2=1$, the event \eqref{eq:good-event} implies
\begin{align}
\left|
(\bar J_n-\mathcal J)(W,s)
\right|\le
\eta_n(x)\sum_{a,b=1}^p s_as_b=
\eta_n(x)(\one^\top s)^2
\le
\kappa^2\eta_n(x),
\end{align}
which is \eqref{eq:uniform-objective-deviation}.  The $a$th
coordinate of the support-gradient difference is
\begin{equation}
\label{eq:support-gradient-coordinate}
2\sum_{b=1}^p
s_b w_a^\top H_{ab}w_b.
\end{equation}
Equation \eqref{eq:good-event} bounds the absolute value of
\eqref{eq:support-gradient-coordinate} by
$2\eta_n(x)\one^\top s\le2\kappa\eta_n(x)$, proving
\eqref{eq:stochastic-support-gradient}.
\end{proof}

Lemma~\ref{lem:uniform-W-stochastic} bounds $\operatorname{grad}_W(\bar J_n-\mathcal J)(W,u^\star)$ and the change in $\operatorname{grad}_W\bar J_n(W,s)$ when $s$ is replaced by $u^\star$.

\begin{lemma}
\label{lem:uniform-W-stochastic}
Under \ref{ass:iid-symmetry}--\ref{ass:sub-exponential-tail},
on $\mathcal G_x$, uniformly over
$W\in O(p)$ and $s\in\mathcal S_\kappa$,
\begin{align}
&\textnormal{(i)}\quad
\left\|
\operatorname{grad}_W
(\bar J_n-\mathcal J)(W,u^\star)
\right\|_F
\le
2(\kappa-1)\sqrt{\kappa}\,\eta_n(x),
\label{eq:stochastic-W-gradient}\\
&\textnormal{(ii)}\quad
\left\|
\operatorname{grad}_W\bar J_n(W,s)
-
\operatorname{grad}_W\bar J_n(W,u^\star)
\right\|_F
\leq
2(\kappa+\sqrt{\kappa})
\{M_0+\eta_n(x)\}
\|s-u^\star\|_1.
\label{eq:empirical-W-lipschitz}
\end{align}
\end{lemma}

\begin{proof}
Write $w_a$ for column $a$ of $W$.
The Euclidean $W$-gradient of the stochastic difference at
$u^\star$ has column $a$
\begin{equation}
\label{eq:W-gradient-column}
2u_a^\star
\sum_{b\in V^\star}H_{ab}w_b.
\end{equation}
Equations \ref{ass:iid-symmetry} and \eqref{eq:H-ab} give
$H_{aa}=0$.  Hence each of the at most $\kappa$ possibly nonzero columns in
\eqref{eq:W-gradient-column} has norm at most
$2(\kappa-1)\eta_n(x)$, and the Euclidean gradient has Frobenius
norm at most
$2(\kappa-1)\sqrt{\kappa}\,\eta_n(x)$.  Applying
\eqref{eq:tangent-projection-contraction} proves
\eqref{eq:stochastic-W-gradient}.

For \eqref{eq:empirical-W-lipschitz}, let
\begin{equation}
\widehat M_{ab}
=
\frac{1}{n}\sum_{i=1}^nB_{i,ab}A_i,
\qquad
h=s-u^\star.
\end{equation}
Equations \eqref{eq:good-event} and \eqref{eq:M0} imply
\begin{equation}
\label{eq:Mhat-bound}
\max_{a,b}\|\widehat M_{ab}\|_{\mathrm{op}}
\le
M_0+\eta_n(x).
\end{equation}
The difference between column $a$ of the two Euclidean gradients is
\begin{equation}
\label{eq:W-gradient-difference-column}
2h_a\sum_{b=1}^p s_b\widehat M_{ab}w_b
+
2u_a^\star\sum_{b=1}^p h_b\widehat M_{ab}w_b.
\end{equation}
By \eqref{eq:Mhat-bound}, the matrices formed by the first and
second terms in \eqref{eq:W-gradient-difference-column} have
Frobenius norms at most
\begin{equation}
2\kappa\{M_0+\eta_n(x)\}\|h\|_2
\qquad\text{and}\qquad
2\sqrt{\kappa}\{M_0+\eta_n(x)\}\|h\|_1,
\end{equation}
respectively.  Since $\|h\|_2\le\|h\|_1$, the Euclidean-gradient
difference is bounded by the right-hand side of
\eqref{eq:empirical-W-lipschitz}.  Both gradients are evaluated at
the same $W$, so \eqref{eq:riemannian-gradient-projection} applies
the same linear map $\mathcal P_W$ to both.  Equation
\eqref{eq:tangent-projection-contraction} therefore proves
\eqref{eq:empirical-W-lipschitz} for the Riemannian gradients.
\end{proof}

\subsection{Bounds near $\mathcal W^\star\times\{u^\star\}$}

Lemma~\ref{lem:local-population} relates the distance to $\mathcal W^\star$ to the population gradient norm and objective loss through \eqref{eq:local-gradient-growth} and \eqref{eq:local-value-growth} below.

\begin{lemma}
\label{lem:local-population}
Under Assumptions~\ref{ass:iid-symmetry}--\ref{ass:normal-curvature},
there exists $\rho_{2,n}>0$ such that, for every $W\in O(p)$,
$d_O(W,\mathcal W^\star)\le\rho_{2,n}$ implies
\begin{align}
&\textnormal{(i)}\quad
\label{eq:local-gradient-growth}
\left\|
\operatorname{grad}_W\mathcal J(W,u^\star)
\right\|_F
\ge
\frac{\alpha_n}{2}
d_O(W,\mathcal W^\star),
\\
&\textnormal{(ii)}\quad
\label{eq:local-value-growth}
J^\star-\mathcal J(W,u^\star)
\ge
\frac{\alpha_n}{4}
d_O(W,\mathcal W^\star)^2.
\end{align}
\end{lemma}

\begin{proof}
Equation \eqref{eq:sub-exponential-tail} implies that every
coefficient defining $\mathcal J$ is finite.  Since
\eqref{eq:quadratic-objectives} is a finite polynomial in the
entries of $W$ and $s$, the map
$(W,s)\mapsto\mathcal J(W,s)$ is smooth.
By \citet[Theorem~5.25]{lee2018introduction}, the tubular-neighborhood
theorem applied within each connected component of $O(p)$ gives a
radius $\rho_{1,n}>0$ on which the normal exponential map around
$\mathcal W^\star$ is one-to-one.  By
\citet[Theorem~6.40]{lee2018introduction}, the radius can be chosen
so that every $W$ satisfying
$d_O(W,\mathcal W^\star)<\rho_{1,n}$ has the unique representation
\begin{equation}
\label{eq:tubular-representation}
\begin{aligned}
W
&=
\operatorname{Exp}_{U(W)}(\xi_W),\\
U(W)
&\in
\mathcal W^\star,\\
\xi_W
&\in
T_{U(W)}O(p),
\qquad
\xi_W\perp T_{U(W)}\mathcal W^\star,\\
\|\xi_W\|_F
&=
d_O(W,\mathcal W^\star).
\end{aligned}
\end{equation}
The same metric tubular-neighborhood result also gives the converse
distance identity
\begin{equation}
\label{eq:tubular-normal-distance}
d_O\!\left(
\operatorname{Exp}_U(\zeta),\mathcal W^\star
\right)
=
\|\zeta\|_F
\end{equation}
for every $U\in\mathcal W^\star$ and every
$\zeta\in T_UO(p)$ satisfying $\zeta\perp T_U\mathcal W^\star$ and
$\|\zeta\|_F<\rho_{1,n}$.
The two componentwise radii can be replaced by their minimum because
$\mathcal W^\star$ intersects both components by
\eqref{eq:completion-reflection-projector}.

We next extend the curvature in
\eqref{eq:normal-curvature} uniformly to nearby points.  Consider
\begin{equation}
\label{eq:unit-normal-pairs}
\mathcal K
=
\left\{
(U,\nu)\in
\mathcal W^\star\times\R^{p\times p}:
\nu\in T_UO(p),\
\nu\perp T_U\mathcal W^\star,\
\|\nu\|_F=1
\right\}.
\end{equation}
The set $\mathcal K$ is compact.  Indeed, take any sequence
$(U_j,\nu_j)$ in $\mathcal K$.  Compactness of
$\mathcal W^\star$ and of the Euclidean unit sphere yields a
subsequence converging to some $(U,\nu)$.  The identities
$U_j^\top\nu_j+\nu_j^\top U_j=0$ pass to the limit, so
$\nu\in T_UO(p)$ by \eqref{eq:orthogonal-tangent-space}.  A smooth
local parametrization of $\mathcal W^\star$ around $U$ allows every
vector in $T_U\mathcal W^\star$ to be approximated by vectors in
$T_{U_j}\mathcal W^\star$.  Passing to the limit in the corresponding
inner products gives
$\nu\perp T_U\mathcal W^\star$.  Thus $(U,\nu)\in\mathcal K$, so
$\mathcal K$ is a closed subset of a compact set.

For $(U,\nu)\in\mathcal K$, define
\begin{equation}
\label{eq:normal-geodesic}
\gamma_{U,\nu}(r)
=
\operatorname{Exp}_U(r\nu),
\qquad
\zeta_{U,\nu}(r)
=
\frac{d}{dr}\gamma_{U,\nu}(r).
\end{equation}
\citet[Proposition~5.19(b)--(c)]{lee2018introduction} implies that
\eqref{eq:normal-geodesic} is a geodesic and depends smoothly on
$(U,\nu,r)$.  Therefore
\begin{align}
\Phi(U,\nu,r)
&=
\inner{\zeta_{U,\nu}(r)}{
\operatorname{Hess}_{WW}\mathcal J
(\gamma_{U,\nu}(r),u^\star)
[\zeta_{U,\nu}(r)]}
\end{align}
is continuous.  At $r=0$,
$\zeta_{U,\nu}(0)=\nu$, and
\eqref{eq:normal-curvature} gives
$\Phi(U,\nu,0)\le-\alpha_n$ on $\mathcal K$.
Compactness of \eqref{eq:unit-normal-pairs} therefore yields a
radius satisfying
\begin{equation}
\label{eq:curvature-radius-order}
0<\rho_{2,n}<\rho_{1,n}
\end{equation}
such that
\begin{equation}
\label{eq:uniform-nearby-curvature}
\Phi(U,\nu,r)
\le
-\frac{\alpha_n}{2},
\qquad
(U,\nu)\in\mathcal K,\quad
0\le r\le\rho_{2,n}.
\end{equation}

Fix $W$ with
$d_O(W,\mathcal W^\star)\le\rho_{2,n}$ and use
\eqref{eq:tubular-representation}.  If $\xi_W\ne0$, let
\begin{equation}
\nu_W
=
\frac{\xi_W}{\|\xi_W\|_F},
\qquad
\gamma_W(t)
=
\operatorname{Exp}_{U(W)}(t\xi_W),
\qquad
q_W(t)
=
\mathcal J(\gamma_W(t),u^\star).
\end{equation}
Then
\begin{equation}
\label{eq:normal-geodesic-velocity}
\dot\gamma_W(t)
=
\|\xi_W\|_F
\zeta_{U(W),\nu_W}
\!\left(t\|\xi_W\|_F\right),
\qquad
\|\dot\gamma_W(t)\|_F=\|\xi_W\|_F.
\end{equation}
Since $\gamma_W$ is a geodesic, \citet[Definition~5.5.1 and Proposition~5.5.2]{absil2008optimization} yields
\begin{align}
q_W''(t)
&=
\inner{\dot\gamma_W(t)}{
\operatorname{Hess}_{WW}\mathcal J
(\gamma_W(t),u^\star)[\dot\gamma_W(t)]}
\notag\\
&\le
-\frac{\alpha_n}{2}\|\xi_W\|_F^2,
\qquad
0\le t\le1,
\label{eq:geodesic-second-derivative}
\end{align}
where the inequality follows from
\eqref{eq:uniform-nearby-curvature} and
\eqref{eq:normal-geodesic-velocity}.  Equation
\eqref{eq:population-maximizer} gives
\begin{equation}
\label{eq:geodesic-first-derivative-zero}
q_W'(0)
=
\inner{
\operatorname{grad}_W
\mathcal J(U(W),u^\star)
}{\xi_W}
=
0.
\end{equation}
Integrating \eqref{eq:geodesic-second-derivative} once and using
\eqref{eq:normal-geodesic-velocity} yields
\begin{align}
\frac{\alpha_n}{2}\|\xi_W\|_F^2
\le
-q_W'(1)
\le
\left\|
\operatorname{grad}_W\mathcal J(W,u^\star)
\right\|_F
\|\xi_W\|_F.
\label{eq:geodesic-gradient-comparison}
\end{align}
Equations \eqref{eq:tubular-representation} and
\eqref{eq:geodesic-gradient-comparison} prove
\eqref{eq:local-gradient-growth}.  Taylor's expansion, together with
\eqref{eq:geodesic-first-derivative-zero} and
\eqref{eq:geodesic-second-derivative}, yields
\begin{align}
\mathcal J(W,u^\star)
=
J^\star
+
\int_0^1(1-t)q_W''(t)\,dt \leq
J^\star-\frac{\alpha_n}{4}\|\xi_W\|_F^2.
\label{eq:geodesic-value-comparison}
\end{align}
Equations \eqref{eq:tubular-representation} and
\eqref{eq:geodesic-value-comparison} prove
\eqref{eq:local-value-growth}.
\end{proof}

Lemma~\ref{lem:local-support} shows that the selected-coordinate derivatives of $\mathcal J(W,s)$ remain positive and uniformly exceed the unselected-coordinate derivatives throughout \eqref{eq:local-neighborhood-requirement} below.
\begin{lemma}
\label{lem:local-support}
Under Assumptions~\ref{ass:iid-symmetry}--\ref{ass:support-margin}, there is
$r_0\in(0,1)$, possibly depending on $n$ and $p$, such that every
$(W,s)\in O(p)\times\mathcal S_\kappa$ satisfying
\begin{equation}
\label{eq:local-neighborhood-requirement}
d_O(W,\mathcal W^\star)\le r_0,
\qquad
\|s-u^\star\|_2\le r_0
\end{equation}
also satisfies
\begin{align}
\min_{a\in V^\star}\{2D_0(W)s\}_a
-
\max\left\{
0,\max_{b\notin V^\star}\{2D_0(W)s\}_b
\right\}
&\ge
\frac{3m_n}{4}.
\label{eq:local-score-separation}
\end{align}
\end{lemma}

\begin{proof}
Define
\begin{align}
\Delta(W,s)
&=
\min_{a\in V^\star}\{2D_0(W)s\}_a
-
\max\left\{
0,\max_{b\notin V^\star}\{2D_0(W)s\}_b
\right\}.
\label{eq:score-gap-map}
\end{align}
The map in \eqref{eq:score-gap-map} is continuous.  By
\eqref{eq:support-margin} and compactness of
$\mathcal W^\star$, there exists $\rho_{3,n}>0$ such that
\begin{equation}
\label{eq:uniform-local-score-gap}
\Delta(W,s)\ge\frac{3m_n}{4}
\end{equation}
whenever
$d_O(W,\mathcal W^\star)\le\rho_{3,n}$ and
$\|s-u^\star\|_2\le\rho_{3,n}$.
Let
\begin{equation}
L_{0,n}
=
\sup_{W\in O(p)}
\|D_0(W)\|_{\mathrm{op}}.
\end{equation}
Equations \eqref{eq:Dn-D0} and \eqref{eq:M0} imply
$L_{0,n}\le pM_0<\infty$.  Choose $r_0$ so that
\begin{equation}
\label{eq:r0-choice}
0<r_0
\le
\min\left\{\rho_{2,n},\rho_{3,n},\frac{1}{2}\right\},
\qquad
r_0L_{0,n}\le\frac{m_n}{8}.
\end{equation}
The choice $r_0\le\rho_{3,n}$ and
\eqref{eq:uniform-local-score-gap} prove
\eqref{eq:local-score-separation}.
\end{proof}

Corollary~\ref{cor:local-population-gap} bounds $J^\star-\mathcal J(W,s)$ using $d_O(W,\mathcal W^\star)$ and $\|s-u^\star\|_1$.

\begin{corollary}
\label{cor:local-population-gap}
Under Assumptions~\ref{ass:iid-symmetry}--\ref{ass:support-margin}, the radius
$r_0$ satisfying \eqref{eq:r0-choice} can be chosen so that every
$(W,s)\in O(p)\times\mathcal S_\kappa$ satisfying
\eqref{eq:local-neighborhood-requirement} obeys
\begin{equation}
\label{eq:local-population-gap}
J^\star-\mathcal J(W,s)
\ge
\frac{\alpha_n}{4}d_O(W,\mathcal W^\star)^2
+
\frac{m_n}{4}\|s-u^\star\|_1.
\end{equation}
\end{corollary}

\begin{proof}
Let
\begin{equation}
h=s-u^\star,
\qquad
A_s
=
\sum_{a\in V^\star}(1-s_a),
\qquad
B_s
=
\sum_{b\notin V^\star}s_b.
\end{equation}
Feasibility in \eqref{eq:ossa-relaxed} gives
$0\le B_s\le A_s$ and
\begin{equation}
\label{eq:l1-mass-relation}
\|h\|_1=A_s+B_s\le2A_s.
\end{equation}
Equations \eqref{eq:local-neighborhood-requirement} and
\eqref{eq:r0-choice} give
$d_O(W,\mathcal W^\star)\le r_0\le\rho_{3,n}$.
Thus \eqref{eq:uniform-local-score-gap} applies with $s=u^\star$.
Using $0\le B_s\le A_s$ and
\eqref{eq:uniform-local-score-gap}, we obtain
\begin{align}
\inner{2D_0(W)u^\star}{u^\star-s}
&=\sum_{a\in V^\star}\{2D_0(W)u^\star\}_a(1-s_a)
 -\sum_{b\notin V^\star}\{2D_0(W)u^\star\}_b s_b
\notag\\
&\ge A_s\min_{a\in V^\star}\{2D_0(W)u^\star\}_a
 -B_s\max\left\{0,\max_{b\notin V^\star}\{2D_0(W)u^\star\}_b\right\}
\notag\\
&\ge A_s\left[
 \min_{a\in V^\star}\{2D_0(W)u^\star\}_a
 -\max\left\{0,\max_{b\notin V^\star}\{2D_0(W)u^\star\}_b\right\}
 \right]
\notag\\
&\ge\frac{3m_n}{4}A_s
 \ge\frac{3m_n}{8}\|s-u^\star\|_1.
\label{eq:population-support-linear}
\end{align}
The following identity
\begin{align}
\mathcal J(W,u^\star)-\mathcal J(W,s)
&=
\inner{2D_0(W)u^\star}{u^\star-s}
-
h^\top D_0(W)h
\end{align}
and \eqref{eq:population-support-linear} yield
\begin{align}
\mathcal J(W,u^\star)-\mathcal J(W,s)
&\ge
\frac{3m_n}{8}\|h\|_1
-
\|D_0(W)\|_{\mathrm{op}}\|h\|_2^2.
\label{eq:population-support-preliminary}
\end{align}
If $\|h\|_2\le r_0$, then
$\|h\|_2^2\le r_0\|h\|_1$.  Equations
\eqref{eq:population-support-preliminary} and
\eqref{eq:r0-choice} therefore give
\begin{equation}
\label{eq:population-support-growth}
\mathcal J(W,u^\star)-\mathcal J(W,s)
\ge
\frac{m_n}{4}\|s-u^\star\|_1.
\end{equation}
Equations \eqref{eq:local-neighborhood-requirement} and
\eqref{eq:r0-choice} give
$d_O(W,\mathcal W^\star)\le r_0\le\rho_{2,n}$, so
\eqref{eq:local-value-growth} applies.
Combining \eqref{eq:local-value-growth} and
\eqref{eq:population-support-growth} proves
\eqref{eq:local-population-gap}.
\end{proof}

For an empirical pair $(W,s)$, define
\begin{equation}
\label{eq:first-order-residuals}
r_W(W,s)
=
\|\operatorname{grad}_W\bar J_n(W,s)\|_F,
\qquad
r_s(W,s)
=
\sup_{v\in\mathcal S_\kappa}
\inner{\nabla_s\bar J_n(W,s)}{v-s}.
\end{equation}
The second residual is nonnegative because $v=s$ is feasible.  It
is the maximum of a linear function over the compact convex set in
\eqref{eq:ossa-relaxed}.  Both residuals are zero at a local maximizer.

For rank-$\kappa$ orthogonal projectors $P$ and $Q$,
$\Theta(P,Q)$ denotes the diagonal matrix of principal angles
between their ranges.

Let $(\widehat W,\widehat s)\in O(p)\times\mathcal S_\kappa$ and
$\varepsilon_W,\varepsilon_s\ge0$ satisfy
\begin{equation}
\label{eq:stationary-localization}
d_O(\widehat W,\mathcal W^\star)<r_0,
\qquad
\|\widehat s-u^\star\|_2<r_0,
\end{equation}
\begin{equation}
\label{eq:approximate-stationarity}
r_W(\widehat W,\widehat s)\le\varepsilon_W,
\qquad
r_s(\widehat W,\widehat s)\le\varepsilon_s.
\end{equation}
For the following residual bounds, suppose that
\begin{equation}
\label{eq:stationary-noise-condition}
\eta_n(x)\le\frac{m_n}{16\kappa}.
\end{equation}

Theorem~\ref{thm:local-residual} bounds $\|\widehat s-u^\star\|_1$ using $r_s(\widehat W,\widehat s)$.

\begin{theorem}
\label{thm:local-residual}
Under Assumptions~\ref{ass:iid-symmetry}--\ref{ass:support-margin}, on
$\mathcal G_x$, conditions
\eqref{eq:stationary-localization}--%
\eqref{eq:stationary-noise-condition} imply
\begin{align}
&\textnormal{(i)}\quad
\|\widehat s-u^\star\|_1
\le
\frac{4\varepsilon_s}{m_n},
\label{eq:stationary-support-error}\\
&\textnormal{(ii)}\quad
\min_{a\in V^\star}\widehat s_a
>
\max_{b\notin V^\star}\widehat s_b,
\quad \Top_\kappa(\widehat s)
=
V^\star
\label{eq:stationary-support-recovery}
\end{align}
In particular, $\varepsilon_s=0$ implies $\widehat s=u^\star$.
\end{theorem}

\begin{proof}
Equations \eqref{eq:local-score-separation} and
\eqref{eq:stochastic-support-gradient} imply
\begin{align}
\min_{a\in V^\star}
\{\nabla_s\bar J_n(\widehat W,\widehat s)\}_a
-
\max\left\{
0,\max_{b\notin V^\star}
\{\nabla_s\bar J_n(\widehat W,\widehat s)\}_b
\right\}
\geq
\frac{3m_n}{4}-4\kappa\eta_n(x)
\ge
\frac{m_n}{2},
\label{eq:empirical-score-separation}
\end{align}
where the last inequality uses \eqref{eq:stationary-noise-condition}.

Define
\begin{equation}
\label{eq:stationary-mass-definitions}
\widehat A
=
\sum_{a\in V^\star}(1-\widehat s_a),
\qquad
\widehat B
=
\sum_{b\notin V^\star}\widehat s_b.
\end{equation}
Feasibility gives $0\le\widehat B\le\widehat A$.  As in the
mass comparison leading to \eqref{eq:population-support-linear},
the empirical score gap in \eqref{eq:empirical-score-separation}
yields
\begin{align}
&\inner{\nabla_s\bar J_n(\widehat W,\widehat s)}{u^\star-\widehat s}
\notag\\
&\quad\ge\widehat A\left[
\min_{a\in V^\star}\{\nabla_s\bar J_n(\widehat W,\widehat s)\}_a
-\max\left\{0,\max_{b\notin V^\star}
\{\nabla_s\bar J_n(\widehat W,\widehat s)\}_b\right\}
\right]
\notag\\
&\quad\ge\frac{m_n}{2}\widehat A.
\label{eq:empirical-support-direction}
\end{align}
Since $u^\star\in\mathcal S_\kappa$,
\eqref{eq:first-order-residuals} and
\eqref{eq:approximate-stationarity} bound the left-hand side of
\eqref{eq:empirical-support-direction} by $\varepsilon_s$.  Thus
\begin{equation}
\widehat A
\le
\frac{2\varepsilon_s}{m_n},
\qquad
\|\widehat s-u^\star\|_1
=
\widehat A+\widehat B
\le
\frac{4\varepsilon_s}{m_n},
\end{equation}
which proves \eqref{eq:stationary-support-error}.

For every coordinate $j$, \eqref{eq:stationary-localization} and
\eqref{eq:r0-choice} give
\begin{equation}
|\widehat s_j-u_j^\star|
\le\|\widehat s-u^\star\|_2<r_0\le\frac12.
\end{equation}
Consequently, $\widehat s_a>1/2$ for $a\in V^\star$ and
$\widehat s_b<1/2$ for $b\notin V^\star$, which proves
\eqref{eq:stationary-support-recovery}.
The last assertion follows from \eqref{eq:stationary-support-error}
with $\varepsilon_s=0$.
\end{proof}
We denote
\begin{align}
b_n(x,\varepsilon_W,\varepsilon_s)
=
\frac{2}{\alpha_n}
\left\{
\varepsilon_W
+
2(\kappa-1)\sqrt{\kappa}\,\eta_n(x)
+
\frac{
8(\kappa+\sqrt{\kappa})
\{M_0+\eta_n(x)\}\varepsilon_s
}{m_n}
\right\}.
\label{eq:stationary-radius}
\end{align}

Corollary~\ref{cor:stationary-distance} bounds $d_O(\widehat W,\mathcal W^\star)$ in terms of $\varepsilon_W$, $\varepsilon_s$, and $\eta_n(x)$.

\begin{corollary}
\label{cor:stationary-distance}
Under Assumptions~\ref{ass:iid-symmetry}--\ref{ass:support-margin}, on
$\mathcal G_x$, conditions
\eqref{eq:stationary-localization}--%
\eqref{eq:stationary-noise-condition} imply
\begin{equation}
\label{eq:stationary-distance}
d_{V^\star}(\widehat W,\mathcal W^\star)
\le
d_O(\widehat W,\mathcal W^\star)
\le
b_n(x,\varepsilon_W,\varepsilon_s).
\end{equation}
\end{corollary}

\begin{proof}
For the alignment bound, the triangle inequality, followed by
\eqref{eq:empirical-W-lipschitz},
\eqref{eq:stochastic-W-gradient},
\eqref{eq:approximate-stationarity}, and
\eqref{eq:stationary-support-error}, yields
\begin{equation}\label{eq:population-gradient-upper}
\begin{split}
&
\left\|
\operatorname{grad}_W
\mathcal J(\widehat W,u^\star)
\right\|_F
\notag\\
&\quad\le
r_W(\widehat W,\widehat s)
+
\left\|
\operatorname{grad}_W\bar J_n(\widehat W,\widehat s)
-
\operatorname{grad}_W\bar J_n(\widehat W,u^\star)
\right\|_F
+
\left\|
\operatorname{grad}_W
(\bar J_n-\mathcal J)(\widehat W,u^\star)
\right\|_F
\notag\\
&\quad\le
\varepsilon_W
+
\frac{
8(\kappa+\sqrt{\kappa})
\{M_0+\eta_n(x)\}\varepsilon_s
}{m_n}
+
2(\kappa-1)\sqrt{\kappa}\,\eta_n(x).
\end{split}
\end{equation}
Combining \eqref{eq:local-gradient-growth} and
\eqref{eq:population-gradient-upper} gives the upper bound on
$d_O$ in \eqref{eq:stationary-distance}.

For every $U\in\mathcal W^\star$, the chordal distance is no larger
than the geodesic distance, so
\begin{equation}
\label{eq:selected-chordal-geodesic}
\|(\widehat W-U)P_{V^\star}\|_F
\le
\|\widehat W-U\|_F
\le
d_O(\widehat W,U).
\end{equation}
Taking the infimum in \eqref{eq:selected-chordal-geodesic} proves
$d_{V^\star}\le d_O$ in \eqref{eq:stationary-distance}.
\end{proof}

Corollary~\ref{cor:stationary-projector} transfers \eqref{eq:stationary-distance} to $\widehat W P_{V^\star}\widehat W^\top-\Pi^\star$.

\begin{corollary}
\label{cor:stationary-projector}
Under Assumptions~\ref{ass:iid-symmetry}--\ref{ass:support-margin}, on
$\mathcal G_x$, conditions
\eqref{eq:stationary-localization}--%
\eqref{eq:stationary-noise-condition} imply
\begin{align}
\textnormal{(i)}\quad
\left\|
\widehat WP_{V^\star}\widehat W^\top-\Pi^\star
\right\|_F
&\le
2b_n(x,\varepsilon_W,\varepsilon_s),
\label{eq:stationary-projector}\\
\textnormal{(ii)}\quad
\left\|
\sin\Theta\!\left(
\widehat WP_{V^\star}\widehat W^\top,\Pi^\star
\right)
\right\|_F^2
&\le
2b_n(x,\varepsilon_W,\varepsilon_s)^2.
\label{eq:stationary-angle}
\end{align}
\end{corollary}

\begin{proof}
For every $U\in\mathcal W^\star$,
\eqref{eq:target-projector} gives
\begin{align}
\widehat WP_{V^\star}\widehat W^\top-\Pi^\star
=
(\widehat W-U)P_{V^\star}\widehat W^\top
+
UP_{V^\star}(\widehat W-U)^\top.
\label{eq:projector-difference-expansion}
\end{align}
The Frobenius norm of each term on the right-hand side of
\eqref{eq:projector-difference-expansion} is at most
$\|(\widehat W-U)P_{V^\star}\|_F$.  Taking the infimum over
$U\in\mathcal W^\star$ yields
\begin{equation}
\left\|
\widehat WP_{V^\star}\widehat W^\top-\Pi^\star
\right\|_F
\le
2d_{V^\star}(\widehat W,\mathcal W^\star),
\end{equation}
which proves \eqref{eq:stationary-projector}.  Finally, rank-$\kappa$
orthogonal projectors $P$ and $Q$ satisfy
\begin{equation}
\label{eq:projector-principal-angle-identity}
\|P-Q\|_F^2
=
2\kappa-2\Tr(PQ)
=
2\|\sin\Theta(P,Q)\|_F^2.
\end{equation}
Equations \eqref{eq:stationary-projector} and
\eqref{eq:projector-principal-angle-identity} prove
\eqref{eq:stationary-angle}.
\end{proof}

\subsection{Existence of an empirical local maximizer}
\label{sec:empirical-local-maximizer}

All interiors and boundaries in this section are taken relative to
\begin{equation}
\mathcal X_\kappa
=
O(p)\times\mathcal S_\kappa.
\end{equation}
For $0<r<r_0$, define
\begin{equation}
\label{eq:closed-local-neighborhood}
\overline{\mathcal N}_r
=
\left\{
(W,s)\in\mathcal X_\kappa:
d_O(W,\mathcal W^\star)\le r,\
\|s-u^\star\|_2\le r
\right\},
\end{equation}
and
\begin{equation}
\label{eq:open-local-neighborhood}
\mathcal N_r
=
\left\{
(W,s)\in\mathcal X_\kappa:
d_O(W,\mathcal W^\star)<r,\
\|s-u^\star\|_2<r
\right\}.
\end{equation}
The following lemma characterizes the relative boundary of the set above.
\begin{lemma}\label{lemma:local relative bdry}
The relative boundary of
\eqref{eq:closed-local-neighborhood} is
\begin{align}
\partial_{\mathcal X_\kappa}\overline{\mathcal N}_r
=
\mathcal B_r
={}&
\left\{
(W,s)\in\mathcal X_\kappa:
d_O(W,\mathcal W^\star)=r,\
\|s-u^\star\|_2\le r
\right\}
\notag\\
&{}\qquad\cup
\left\{
(W,s)\in\mathcal X_\kappa:
d_O(W,\mathcal W^\star)\le r,\
\|s-u^\star\|_2=r
\right\}.
\label{eq:relative-local-boundary}
\end{align}
\end{lemma}
\begin{proof}
The continuity of the two distance maps shows that every point in
\eqref{eq:open-local-neighborhood} is in the relative interior of
$\overline{\mathcal N}_r$.  Conversely, suppose
$(W,s)\in\overline{\mathcal N}_r$ and
$d_O(W,\mathcal W^\star)=r$.  Equation
\eqref{eq:tubular-representation} gives
$U\in\mathcal W^\star$ and
$\xi\perp T_U\mathcal W^\star$ such that
\begin{equation}
\label{eq:boundary-W-normal-representation}
W=\operatorname{Exp}_U(\xi),
\qquad
\|\xi\|_F=r.
\end{equation}
For every sufficiently small $\delta>0$,
\begin{equation}
\label{eq:boundary-W-outward-extension}
W_\delta
=
\operatorname{Exp}_U\{(1+\delta)\xi\},
\qquad
d_O(W_\delta,\mathcal W^\star)
=(1+\delta)r>r,
\qquad
W_\delta\longrightarrow W.
\end{equation}
Equations \eqref{eq:r0-choice} and
\eqref{eq:curvature-radius-order} give
$r<r_0\le\rho_{2,n}<\rho_{1,n}$.  Hence $\delta$ can be chosen so
that
\eqref{eq:boundary-W-outward-extension} remains in the tubular
neighborhood; its asserted distance equality follows from
\eqref{eq:boundary-W-normal-representation} and
\eqref{eq:tubular-normal-distance}.  Thus $(W,s)$ is not in the
relative interior of $\overline{\mathcal N}_r$.

Now suppose $(W,s)\in\overline{\mathcal N}_r$ and
$\|s-u^\star\|_2=r$.  Set $h=s-u^\star$.  Feasibility gives
\begin{equation}
\label{eq:boundary-support-direction}
h_a\le0\quad(a\in V^\star),
\qquad
h_b\ge0\quad(b\notin V^\star),
\qquad
\one^\top h\le0.
\end{equation}
Since $r<r_0<1$, $\max_j|h_j|<1$.  Hence, for every sufficiently small
$\delta>0$,
\begin{equation}
\label{eq:boundary-support-outward-extension}
s_\delta
=
u^\star+(1+\delta)h
\in
\mathcal S_\kappa,
\qquad
\|s_\delta-u^\star\|_2
=(1+\delta)r>r,
\qquad
s_\delta\longrightarrow s.
\end{equation}
To verify the feasibility asserted in
\eqref{eq:boundary-support-outward-extension}, choose $\delta$
small enough that every coordinate remains in $[0,1]$; then
\begin{equation}
\one^\top s_\delta
=
\kappa+(1+\delta)\one^\top h
\le
\kappa
\end{equation}
by \eqref{eq:boundary-support-direction}.  Thus $(W,s)$ is not in
the relative interior of $\overline{\mathcal N}_r$.  Since
$\overline{\mathcal N}_r$ is relatively closed, these arguments
prove \eqref{eq:relative-local-boundary}.
\end{proof}

\begin{condition}\label{ass:local-separation}
For $0<r<r_0$,
\begin{subequations}
\label{eq:empirical-local-maximizer-conditions}
\begin{align}
\eta_n(x)&\le\frac{m_n}{16\kappa},
\label{eq:empirical-local-maximizer-noise}\\
2\kappa^2\eta_n(x)
&<
\frac{1}{4}\min\{\alpha_n r^2,m_n r\}.
\label{eq:empirical-local-maximizer-boundary}
\end{align}
\end{subequations}
\end{condition}
On $\mathcal G_x$, \eqref{eq:empirical-local-maximizer-noise} ensures that each $\{\nabla_s\bar J_n(W,s)\}_a$, $a\in V^\star$, exceeds both zero and every $\{\nabla_s\bar J_n(W,s)\}_b$, $b\notin V^\star$, by at least $m_n/2$ for $(W,s)$ satisfying \eqref{eq:local-neighborhood-requirement}, as in \eqref{eq:empirical-score-separation}.
Condition \eqref{eq:empirical-local-maximizer-boundary} makes the uniform bound in \eqref{eq:uniform-objective-deviation} sufficiently smaller than the lower bound for $J^\star-\mathcal J(W,s)$ on $\mathcal B_r$ as in \eqref{eq:population-gap-on-local-boundary} below.

Theorem~\ref{thm:empirical-local-maximizer} establishes that $\bar J_n$ has local maximizers in $\mathcal N_r$.
\begin{theorem}
\label{thm:empirical-local-maximizer}
Under Assumptions~\ref{ass:iid-symmetry}--\ref{ass:local-separation}, on $\mathcal G_x$.  Every
maximizer $(\widehat W,\widehat s)$ of $\bar J_n$ over
$\overline{\mathcal N}_r$ is a local maximizer on
$\mathcal X_\kappa$ and satisfies
\begin{equation}
\label{eq:empirical-local-maximizer-interior}
d_O(\widehat W,\mathcal W^\star)<r,
\qquad
\|\widehat s-u^\star\|_2<r.
\end{equation}
At least one such maximizer exists.
\end{theorem}

\begin{proof}
The set $\mathcal X_\kappa$ is compact.  Equation
\eqref{eq:set-distance-lipschitz} gives continuity of the
set-distance, so $\overline{\mathcal N}_r$ in
\eqref{eq:closed-local-neighborhood} is a nonempty compact subset
of $\mathcal X_\kappa$.  Since $\bar J_n$ is continuous, it attains
its maximum over $\overline{\mathcal N}_r$.

We adapt the argument used in the proof of Theorem~3 in \citet{ning2013high}.  
The comparison is over the relative boundary in \eqref{eq:relative-local-boundary}.  
For every $(W,s)\in\mathcal B_r$, \eqref{eq:local-population-gap} and $\|s-u^\star\|_1\ge\|s-u^\star\|_2$ give
\begin{equation}
\label{eq:population-gap-on-local-boundary}
J^\star-\mathcal J(W,s)
\ge
\frac{1}{4}\min\{\alpha_n r^2,m_n r\}.
\end{equation}
The curvature term in \eqref{eq:local-population-gap} gives the
$\alpha_n r^2/4$ bound on the first face of
\eqref{eq:relative-local-boundary}, whereas its support term gives
the $m_n r/4$ bound on the second face.  For any
$U\in\mathcal W^\star$, equations
\eqref{eq:uniform-objective-deviation} and
\eqref{eq:population-gap-on-local-boundary} yield
\begin{equation}\label{eq:empirical-comparison-on-local-boundary}
\begin{split}
\bar J_n(U,u^\star)-\bar J_n(W,s)
&=
J^\star-\mathcal J(W,s)
+
(\bar J_n-\mathcal J)(U,u^\star)
-
(\bar J_n-\mathcal J)(W,s)\\
&\ge
\frac{1}{4}\min\{\alpha_n r^2,m_n r\}
-
2\kappa^2\eta_n(x)
>
0,
\end{split}
\end{equation}
where the strict inequality is due to \eqref{eq:empirical-local-maximizer-boundary}.  
The point $(U,u^\star)$ belongs to $\overline{\mathcal N}_r$.  Hence \eqref{eq:empirical-comparison-on-local-boundary} excludes
$\mathcal B_r$ from the set of restricted maximizers.  
By \eqref{eq:relative-local-boundary}, every restricted maximizer
belongs to $\mathcal N_r$, proving \eqref{eq:empirical-local-maximizer-interior}.

Since $\mathcal N_r$ is open relative to
$\mathcal X_\kappa$, every restricted maximizer has a relative
neighborhood in $\mathcal X_\kappa$ contained in
$\overline{\mathcal N}_r$.  
It is therefore a local maximizer of $\bar J_n$ on the original feasible set $\mathcal X_\kappa$, including when its support vector lies on a coordinate or respective face of $\mathcal S_\kappa$.
\end{proof}

Corollary~\ref{cor:empirical-local-maximizer-residuals} evaluates $r_W$ and $r_s$ at the maximizers in Theorem~\ref{thm:empirical-local-maximizer}.

\begin{corollary}
\label{cor:empirical-local-maximizer-residuals}
Under Assumptions~\ref{ass:iid-symmetry}--\ref{ass:local-separation}, on $\mathcal G_x$.
Every maximizer in Theorem~\ref{thm:empirical-local-maximizer} satisfies
\begin{equation}
\label{eq:local-maximizer-zero-residuals}
\textnormal{(i)}\quad
r_W(\widehat W,\widehat s)=0,
\qquad
\textnormal{(ii)}\quad
r_s(\widehat W,\widehat s)=0.
\end{equation}
\end{corollary}

\begin{proof}
Fix any such maximizer.  Holding its support block fixed and varying
the orthogonal block along smooth curves gives
\begin{equation}
\label{eq:local-maximizer-W-residual}
\operatorname{grad}_W
\bar J_n(\widehat W,\widehat s)
=
0.
\end{equation}
For every $v\in\mathcal S_\kappa$, the segment
\begin{equation}
t
\longmapsto
\widehat s+t(v-\widehat s),
\qquad
0\le t\le1,
\end{equation}
is feasible.  The one-sided derivative at $t=0$ is nonpositive by
local maximality, and hence
\begin{equation}
\label{eq:local-maximizer-support-direction}
\inner{
\nabla_s\bar J_n(\widehat W,\widehat s)
}{v-\widehat s}
\le
0,
\qquad
v\in\mathcal S_\kappa.
\end{equation}
Taking the supremum in
\eqref{eq:local-maximizer-support-direction}, with
$v=\widehat s$ attaining zero, gives
$r_s(\widehat W,\widehat s)=0$ by
\eqref{eq:first-order-residuals}.  Equation
\eqref{eq:local-maximizer-W-residual} gives
$r_W(\widehat W,\widehat s)=0$, proving
\eqref{eq:local-maximizer-zero-residuals}.
\end{proof}

Corollary~\ref{cor:empirical-local-maximizer-support} combines \eqref{eq:local-maximizer-zero-residuals} and \eqref{eq:stationary-support-error} to obtain $\widehat s=u^\star$.

\begin{corollary}
\label{cor:empirical-local-maximizer-support}
Under Assumptions~\ref{ass:iid-symmetry}--\ref{ass:local-separation}, on $\mathcal G_x$.
Every
maximizer in Theorem~\ref{thm:empirical-local-maximizer} satisfies
\begin{equation}
\label{eq:empirical-local-maximizer-support}
\begin{aligned}
&\textnormal{(i)}\quad
\widehat s
=
u^\star,\\
&\textnormal{(ii)}\quad
\operatorname{supp}(\widehat s)
=
\Top_\kappa(\widehat s)
=
V^\star.
\end{aligned}
\end{equation}
\end{corollary}

\begin{proof}
Equation \eqref{eq:empirical-local-maximizer-interior},
\eqref{eq:local-maximizer-zero-residuals}, and
\eqref{eq:empirical-local-maximizer-noise} verify
\eqref{eq:stationary-localization},
\eqref{eq:approximate-stationarity}, and
\eqref{eq:stationary-noise-condition}, respectively, with
$\varepsilon_W=\varepsilon_s=0$.  Therefore
Theorem~\ref{thm:local-residual} applies, and
\eqref{eq:stationary-support-error} gives
$\widehat s=u^\star$.  Since
$u^\star=\one_{V^\star}$ by definition, its support and its
$\kappa$ largest coordinates both equal $V^\star$, proving
\eqref{eq:empirical-local-maximizer-support}.
\end{proof}

Corollary~\ref{cor:empirical-local-maximizer-distance} specializes \eqref{eq:stationary-distance} to $\varepsilon_W=\varepsilon_s=0$.

\begin{corollary}
\label{cor:empirical-local-maximizer-distance}
Under Assumptions~\ref{ass:iid-symmetry}--\ref{ass:local-separation}, on $\mathcal G_x$.
Every maximizer in Theorem~\ref{thm:empirical-local-maximizer} satisfies
\begin{equation}
\label{eq:empirical-local-maximizer-distance}
d_{V^\star}(\widehat W,\mathcal W^\star)
\le
d_O(\widehat W,\mathcal W^\star)
\le
\frac{
4(\kappa-1)\sqrt{\kappa}\,\eta_n(x)
}{\alpha_n}.
\end{equation}
\end{corollary}

\begin{proof}
Equations \eqref{eq:empirical-local-maximizer-interior} and
\eqref{eq:local-maximizer-zero-residuals}, together with
\eqref{eq:empirical-local-maximizer-noise}, verify
\eqref{eq:stationary-localization}--%
\eqref{eq:stationary-noise-condition} with
$\varepsilon_W=\varepsilon_s=0$.  Hence
Corollary~\ref{cor:stationary-distance} applies.  Equation
\eqref{eq:stationary-radius} gives
\begin{equation}
\label{eq:local-maximizer-specialized-radius}
b_n(x,0,0)
=
\frac{
4(\kappa-1)\sqrt{\kappa}\,\eta_n(x)
}{\alpha_n}.
\end{equation}
Substitution of
\eqref{eq:local-maximizer-specialized-radius} into
\eqref{eq:stationary-distance} yields
\eqref{eq:empirical-local-maximizer-distance}.
\end{proof}

Corollary~\ref{cor:empirical-local-maximizer-projector} specializes \eqref{eq:stationary-projector} and \eqref{eq:stationary-angle} to $\varepsilon_W=\varepsilon_s=0$.

\begin{corollary}
\label{cor:empirical-local-maximizer-projector}
Under Assumptions~\ref{ass:iid-symmetry}--\ref{ass:local-separation}, on $\mathcal G_x$.
Every maximizer in Theorem~\ref{thm:empirical-local-maximizer} satisfies
\begin{align}
\textnormal{(i)}\quad
\left\|
\widehat WP_{V^\star}\widehat W^\top-\Pi^\star
\right\|_F
&\le
\frac{
8(\kappa-1)\sqrt{\kappa}\,\eta_n(x)
}{\alpha_n},
\label{eq:empirical-local-maximizer-projector}\\
\textnormal{(ii)}\quad
\left\|
\sin\Theta\!\left(
\widehat WP_{V^\star}\widehat W^\top,\Pi^\star
\right)
\right\|_F^2
&\le
\frac{
32\kappa(\kappa-1)^2\eta_n(x)^2
}{\alpha_n^2}.
\label{eq:empirical-local-maximizer-angle}
\end{align}
\end{corollary}

\begin{proof}
The same substitutions into
\eqref{eq:stationary-localization}--%
\eqref{eq:stationary-noise-condition} allow
Corollary~\ref{cor:stationary-projector} to be applied with
$\varepsilon_W=\varepsilon_s=0$.
Substitution of
\eqref{eq:local-maximizer-specialized-radius} into
\eqref{eq:stationary-projector} and
\eqref{eq:stationary-angle} yields
\eqref{eq:empirical-local-maximizer-projector} and
\eqref{eq:empirical-local-maximizer-angle}, respectively.
\end{proof}

If the deterministic inequalities in
\eqref{eq:empirical-local-maximizer-conditions} hold, then
\eqref{eq:good-event-probability} and
Theorem~\ref{thm:empirical-local-maximizer} together with
Corollaries~\ref{cor:empirical-local-maximizer-residuals}--%
\ref{cor:empirical-local-maximizer-projector} imply that
\eqref{eq:empirical-local-maximizer-interior},
\eqref{eq:empirical-local-maximizer-support},
\eqref{eq:empirical-local-maximizer-distance},
\eqref{eq:empirical-local-maximizer-projector}, and
\eqref{eq:empirical-local-maximizer-angle}
hold simultaneously with probability at least $1-2e^{-x}$.

Theorem~\ref{thm:canonical-ossa} follows from Theorem~\ref{thm:empirical-local-maximizer}, Theorem~\ref{thm:local-residual}, and Corollary~\ref{cor:stationary-projector}.

\begin{proof}[Proof of Theorem~\ref{thm:canonical-ossa}]
By Lemma~\ref{lem:operator-concentration},
$\mathbb P(\mathcal G_x)\ge1-2e^{-x}$.
On $\mathcal G_x$, Theorem~\ref{thm:empirical-local-maximizer}
and \eqref{eq:empirical-local-maximizer-conditions} give at least
one local maximizer of $\bar J_n$ satisfying
\begin{equation}\label{eq:main-localization}
d_O(\widehat W,\mathcal W^\star)<r,\quad 
\|\widehat s-u^\star\|_2<r.
\end{equation}
Since $J=n\bar J_n$ with $n>0$,
the local maximizers of $J$ and $\bar J_n$ coincide.

Now let $(\widehat W,\widehat s)$ be any local maximizer satisfying
\eqref{eq:main-localization}.  Varying $W$ along smooth curves in
$O(p)$ and $s$ along the feasible segments
$\widehat s+t(v-\widehat s)$ yields
\begin{equation}
\operatorname{grad}_W\bar J_n(\widehat W,\widehat s)=0,
\qquad
\inner{\nabla_s\bar J_n(\widehat W,\widehat s)}{v-\widehat s}
\le0,
\quad v\in\mathcal S_\kappa.
\end{equation}
Consequently, both residuals in \eqref{eq:first-order-residuals}
are zero.  
Since $r<r_0$, \eqref{eq:main-localization} implies \eqref{eq:stationary-localization}.
Thus Theorem~\ref{thm:local-residual}, with
$\varepsilon_W=\varepsilon_s=0$, yields
$\widehat s=u^\star$ and
$\widehat{\mathcal V}=\Top_\kappa(\widehat s)=V^\star$.
Corollary~\ref{cor:stationary-projector} and
\eqref{eq:local-maximizer-specialized-radius} give the projector
bound in \eqref{eq:main-local-conclusions}.
All these conclusions hold on the event $\mathcal G_x$, which proves the theorem.
\end{proof}

\subsection{Sufficient conditions}
\label{sec:shared-signal}

We verify the assumptions for a specified shared-signal family of the
observed matrices. For the fixed $V^\star$ introduced above, suppose
\begin{equation}
\label{eq:shared-dimensions}
2\le |V^\star|=\kappa<p.
\end{equation}
For $\ell=1,\ldots,L$, let the fixed matrices
$C^{(\ell)}\in\R^{p\times p}$ satisfy
\begin{equation}
\label{eq:shared-basis}
C^{(\ell)}=C^{(\ell)\top},
\qquad
\diag\!\left(C^{(\ell)}\right)=0,
\qquad
C^{(\ell)}=P_{V^\star}C^{(\ell)}P_{V^\star}.
\end{equation}
Suppose that
\begin{equation}
\label{eq:shared-latent-signal}
C_i=\sum_{\ell=1}^Lz_{i\ell}C^{(\ell)},
\qquad
z_i=(z_{i1},\ldots,z_{iL})^\top,
\qquad
\Sigma_z=\mathbb E(z_i z_i^\top)\succ0,
\end{equation}
where $\Sigma_z$ is a finite second-moment matrix, and that, for a
fixed $W^\star\in O(p)$ and $\sigma_A,\sigma_B\ge0$,
\begin{equation}
\label{eq:shared-observed-model}
A_i=W^\star C_iW^{\star\top}+\sigma_AE_i^A,
\qquad
B_i=C_i+\sigma_BE_i^B.
\end{equation}
The noise matrices satisfy
\begin{equation}
\label{eq:shared-noise-structure}
\mathbb E(E_i^A)=\mathbb E(E_i^B)=0_{p\times p},
\qquad
E_i^A=E_i^{A\top},\quad E_i^B=E_i^{B\top},\quad
\diag(E_i^B)=0
\quad\text{almost surely}.
\end{equation}
Assume that, for every $a,b$,
\begin{equation}
\label{eq:shared-cross-moments}
\begin{aligned}
\mathbb E(C_{i,ab}E_i^A)&=0_{p\times p},\\
\mathbb E\!\left(E_{i,ab}^B W^\star C_iW^{\star\top}\right)
&=0_{p\times p},\\
\mathbb E(E_{i,ab}^B E_i^A)&=0_{p\times p},
\end{aligned}
\end{equation}
and that
\begin{equation}
\label{eq:shared-common-null-space}
\bigcap_{\ell=1}^L\ker\!\left(C^{(\ell)}\right)
=\operatorname{span}\{e_j:j\notin V^\star\}
=\ker(P_{V^\star}).
\end{equation}
Suppose further that
\begin{equation}
\label{eq:shared-iid-tuples}
\{(z_i,E_i^A,E_i^B)\}_{i=1}^n
\quad\text{are independent and identically distributed},
\end{equation}
and that finite, possibly dimension-dependent constants $K_A,K_B$
satisfy
\begin{equation}
\label{eq:shared-sub-gaussian-tails}
\max_{a,b}\|B_{i,ab}\|_{\psi_2}\le K_B,
\qquad
\sup_{\|x\|_2=\|y\|_2=1}\|x^\top A_i y\|_{\psi_2}\le K_A.
\end{equation}
For this model, the quantities in Assumption~\ref{ass:population-target}
are given by
\begin{equation}
\label{eq:shared-population-target}
\begin{aligned}
\mathcal W^\star
&=\{W^\star Q:Q\in O(p),\ Q^\top C^{(\ell)}Q=C^{(\ell)}
       \text{ for }\ell=1,\ldots,L\},\\
\Pi^\star&=W^\star P_{V^\star}W^{\star\top},
\qquad
J^\star=\mathbb E\|C_i\|_F^2.
\end{aligned}
\end{equation}
Set
\begin{equation}
\label{eq:shared-explicit-margin}
m_n=2\min_{a\in V^\star}
\sum_{b\in V^\star\setminus\{a\}}\mathbb E(C_{i,ab}^2).
\end{equation}

Proposition~\ref{prop:shared-signal} verifies
Assumptions~\ref{ass:iid-symmetry}--\ref{ass:support-margin} for the
model in \eqref{eq:shared-observed-model}.
\begin{proposition}
\label{prop:shared-signal}
Under \eqref{eq:shared-dimensions}--\eqref{eq:shared-sub-gaussian-tails},
Assumptions~\ref{ass:iid-symmetry}--\ref{ass:support-margin} hold with
\eqref{eq:shared-population-target}, the strictly positive margin in
\eqref{eq:shared-explicit-margin},
$\nu_n=C K_AK_B$ for a universal constant $C>0$, and the strictly
positive curvature constant $\alpha_n=\alpha_C$ defined in
\eqref{eq:shared-curvature-constant} below.
\end{proposition}

\begin{proof}
Every $W\in O(p)$ has the unique representation $W=W^\star Q$, where
$Q=W^{\star\top}W\in O(p)$. Expanding the population criterion yields
\begin{equation}
\label{eq:shared-population-objective}
\mathcal J(W^\star Q,u^\star)
=\mathbb E\Tr(Q^\top C_iQ C_i).
\end{equation}
Indeed, for every $a,b\in V^\star$, the three remaining terms in the
entrywise expansion have coefficients $\sigma_A$, $\sigma_B$, and
$\sigma_A\sigma_B$, respectively, and their unscaled expectations
vanish by \eqref{eq:shared-cross-moments}:
\begin{align}
\mathbb E\!\left[C_{i,ab}
 e_a^\top Q^\top W^{\star\top}E_i^A W^\star Qe_b\right]
&=e_a^\top Q^\top W^{\star\top}
  \mathbb E(C_{i,ab}E_i^A)W^\star Qe_b=0,
\label{eq:shared-cross-term-A}\\
\mathbb E\!\left[E_{i,ab}^B e_a^\top Q^\top C_iQe_b\right]
&=e_a^\top Q^\top W^{\star\top}
  \mathbb E\!\left(E_{i,ab}^B W^\star C_iW^{\star\top}\right)
  W^\star Qe_b=0,\notag\\
\mathbb E\!\left[E_{i,ab}^B
 e_a^\top Q^\top W^{\star\top}E_i^A W^\star Qe_b\right]
&=e_a^\top Q^\top W^{\star\top}
  \mathbb E(E_{i,ab}^B E_i^A)W^\star Qe_b=0.
\label{eq:shared-cross-term-AB}
\end{align}
Symmetry identifies the remaining sum with the trace in
\eqref{eq:shared-population-objective}. Orthogonal invariance of the
Frobenius norm therefore yields
\begin{equation}
\label{eq:shared-global-gap}
J^\star-\mathcal J(W^\star Q,u^\star)
=\frac12\mathbb E\|Q^\top C_iQ-C_i\|_F^2.
\end{equation}
Moreover, expansion in the fixed matrices $C^{(\ell)}$ gives the
second-moment identity
\begin{align}
\mathbb E\|Q^\top C_iQ-C_i\|_F^2
&=\sum_{\ell,m=1}^L(\Sigma_z)_{\ell m}
 \inner{Q^\top C^{(\ell)}Q-C^{(\ell)}}
       {Q^\top C^{(m)}Q-C^{(m)}}\notag\\
&\ge\lambda_{\min}(\Sigma_z)
 \sum_{\ell=1}^L\|Q^\top C^{(\ell)}Q-C^{(\ell)}\|_F^2.
\label{eq:shared-gap-covariance}
\end{align}
The inequality follows by applying the smallest-eigenvalue bound for
$\Sigma_z$ to the $L$ coefficients at each matrix entry and summing
over the entries. Since $\Sigma_z\succ0$, the expression in
\eqref{eq:shared-gap-covariance} is zero exactly when every
$C^{(\ell)}$ is preserved by $Q$. Thus
\eqref{eq:shared-population-target} gives the full population
maximizer set at $u^\star$.

Define
\begin{equation}
\label{eq:shared-centralizer}
\mathcal H_C=\{Q\in O(p):Q^\top C^{(\ell)}Q=C^{(\ell)}
            \text{ for }\ell=1,\ldots,L\}.
\end{equation}
This is a closed subgroup of the compact group $O(p)$.
Cartan's closed subgroup theorem implies that it is an embedded Lie
subgroup \citep[Theorem~20.12]{lee2013smooth}. Consequently,
$\mathcal W^\star=W^\star\mathcal H_C$ is a nonempty compact smooth
embedded submanifold of $O(p)$. Each of its points maximizes
$\mathcal J(\cdot,u^\star)$ by \eqref{eq:shared-global-gap}; hence its
Riemannian gradient is zero.

For $Q\in\mathcal H_C$, the defining equalities are equivalent to
\begin{equation}
\label{eq:shared-commutation}
C^{(\ell)}Q=QC^{(\ell)},\qquad \ell=1,\ldots,L.
\end{equation}
It follows that $Q$ preserves the common kernel in
\eqref{eq:shared-common-null-space}. Orthogonality then implies
preservation of its orthogonal complement,
$\operatorname{range}(P_{V^\star})$, and therefore
\begin{equation}
\label{eq:shared-projector-invariance}
QP_{V^\star}Q^\top=P_{V^\star},\qquad
(W^\star Q)P_{V^\star}(W^\star Q)^\top=\Pi^\star.
\end{equation}
If $R\in O(p)$ satisfies $RP_{V^\star}=P_{V^\star}$, orthogonality
gives
\begin{equation}
\label{eq:shared-completion-projector-identities}
R^\top P_{V^\star}=P_{V^\star}R=P_{V^\star}.
\end{equation}
The support condition in \eqref{eq:shared-basis} now implies
\begin{equation}
\label{eq:shared-completion-invariance}
R^\top C^{(\ell)}R
=R^\top P_{V^\star}C^{(\ell)}P_{V^\star}R
=C^{(\ell)}.
\end{equation}
Thus $R\in\mathcal H_C$, and the group property gives
$UR\in\mathcal W^\star$ whenever $U\in\mathcal W^\star$.
These identities verify every part of
Assumption~\ref{ass:population-target}.

The Lie algebra of $\mathcal H_C$ is
\begin{equation}
\mathfrak h_C
=\{\Omega\in\R^{p\times p}:\Omega^\top=-\Omega,\ 
 C^{(\ell)}\Omega=\Omega C^{(\ell)}\text{ for }\ell=1,\ldots,L\}.
\end{equation}
Indeed, differentiating the defining equalities of $\mathcal H_C$
at the identity gives the displayed commutation relations;
conversely, these relations imply that $\exp(t\Omega)$ commutes
with every $C^{(\ell)}$ for all $t$.
For a skew-symmetric $\Omega$, differentiating
\eqref{eq:shared-global-gap} along $Q(t)=\exp(t\Omega)$ yields
\begin{equation}
\label{eq:shared-curvature-identity}
-\left.\frac{d^2}{dt^2}
 \mathcal J(W^\star\exp(t\Omega),u^\star)\right|_{t=0}
=\mathbb E\|C_i\Omega-\Omega C_i\|_F^2.
\end{equation}
To see the coefficient explicitly, the matrix
$F_i(t)=\exp(-t\Omega)C_i\exp(t\Omega)-C_i$ satisfies
$F_i(0)=0$ and $F_i'(0)=C_i\Omega-\Omega C_i$; hence the second
derivative of $\|F_i(t)\|_F^2/2$ at zero is
$\|F_i'(0)\|_F^2$. The finite second moment in
\eqref{eq:shared-latent-signal} justifies differentiation after
expectation, or equivalently differentiation of the finite
second-moment expansion.

The same expansion as in \eqref{eq:shared-gap-covariance} gives
\begin{align}
\mathbb E\|C_i\Omega-\Omega C_i\|_F^2
&=\sum_{\ell,m=1}^L(\Sigma_z)_{\ell m}
 \inner{C^{(\ell)}\Omega-\Omega C^{(\ell)}}
       {C^{(m)}\Omega-\Omega C^{(m)}}\notag\\
&\ge\lambda_{\min}(\Sigma_z)
 \sum_{\ell=1}^L\|C^{(\ell)}\Omega-\Omega C^{(\ell)}\|_F^2.
\label{eq:shared-curvature-second-moment}
\end{align}
Consequently,
\begin{equation}
\label{eq:shared-curvature-kernel}
\mathbb E\|C_i\Omega-\Omega C_i\|_F^2=0
\quad\Longleftrightarrow\quad \Omega\in\mathfrak h_C.
\end{equation}
The orthogonal complement of $\mathfrak h_C$ within the
skew-symmetric matrices has positive dimension. To verify this,
choose unit vectors $v\in\operatorname{range}(P_{V^\star})$ and
$w\in\ker(P_{V^\star})$. By
\eqref{eq:shared-common-null-space}, there is an $\ell$ such that
$C^{(\ell)}v\ne0$. The skew-symmetric matrix
$\Omega_{v,w}=vw^\top-wv^\top$ satisfies
\begin{equation}
\{C^{(\ell)}\Omega_{v,w}-\Omega_{v,w}C^{(\ell)}\}w
=C^{(\ell)}v\ne0,
\end{equation}
so $\mathfrak h_C$ is a proper subspace. Define
\begin{equation}
\label{eq:shared-curvature-constant}
\alpha_C
=\min_{\substack{\Omega^\top=-\Omega,\ \Omega\perp\mathfrak h_C\\
                 \|\Omega\|_F=1}}
 \mathbb E\|C_i\Omega-\Omega C_i\|_F^2.
\end{equation}
The constraint set is nonempty and compact, and the objective is
continuous and strictly positive there by
\eqref{eq:shared-curvature-kernel}. Hence $\alpha_C>0$.

For $Q_0\in\mathcal H_C$ and $U=W^\star Q_0$, left multiplication
by $U$ is an isometry for the Frobenius metric, and
\begin{equation}
\label{eq:shared-tangent-normal}
\begin{aligned}
T_U\mathcal W^\star
&=\{U\Omega:\Omega\in\mathfrak h_C\},\\
\{\xi\in T_UO(p):\xi\perp T_U\mathcal W^\star\}
&=\{U\Omega:\Omega^\top=-\Omega,\ \Omega\perp\mathfrak h_C\}.
\end{aligned}
\end{equation}
For such $\xi=U\Omega$, the curve $t\mapsto U\exp(t\Omega)$ is
the geodesic with initial velocity $\xi$
\citep[Section~2.2.2, Eq.~(2.14)]{edelman1998geometry}.
Since $Q_0$ preserves every $C_i$, one has
\begin{equation}
\label{eq:shared-left-invariance}
\mathcal J(W^\star Q_0\exp(t\Omega),u^\star)
=\mathcal J(W^\star\exp(t\Omega),u^\star).
\end{equation}
The geodesic second-derivative identity
\citep[Proposition~5.5.4]{absil2008optimization},
\eqref{eq:shared-curvature-identity}, and
\eqref{eq:shared-curvature-constant} therefore imply
\begin{equation}
-\inner{\xi}{\operatorname{Hess}_{WW}\mathcal J(U,u^\star)[\xi]}
\ge\alpha_C\|\Omega\|_F^2
=\alpha_C\|\xi\|_F^2.
\end{equation}
This proves Assumption~\ref{ass:normal-curvature} with
$\alpha_n=\alpha_C$.

For $U=W^\star Q\in\mathcal W^\star$, expansion of
\eqref{eq:Dn-D0}, followed by
\eqref{eq:shared-commutation} and \eqref{eq:shared-cross-moments}, yields
\begin{equation}
\label{eq:shared-D0}
\{D_0(U)\}_{ab}
=\begin{cases}
\mathbb E(C_{i,ab}^2),&a,b\in V^\star,\\
0,&\{a,b\}\not\subseteq V^\star.
\end{cases}
\end{equation}
As $\diag(C_i)=0$, this implies
\begin{equation}
\label{eq:shared-score-coordinates}
\{2D_0(U)u^\star\}_a
=\begin{cases}
2\displaystyle\sum_{b\in V^\star\setminus\{a\}}
 \mathbb E(C_{i,ab}^2),&a\in V^\star,\\
0,&a\notin V^\star.
\end{cases}
\end{equation}
The selected sums are positive. Indeed, if such a sum were zero
for some $a\in V^\star$, then $\mathbb E(C_{i,ab}^2)=0$ for every
$b$. By \eqref{eq:shared-latent-signal} and $\Sigma_z\succ0$, this
would give $C_{ab}^{(\ell)}=0$ for all $b,\ell$. 
Its symmetry would then imply $C^{(\ell)}e_a=0$ for every $\ell$, contradicting
\eqref{eq:shared-common-null-space}. Thus
\eqref{eq:shared-explicit-margin} is strictly positive, and
\eqref{eq:shared-score-coordinates} proves
Assumption~\ref{ass:support-margin}.

The structural conditions and \eqref{eq:shared-iid-tuples} give
Assumption~\ref{ass:iid-symmetry}. For unit vectors $x,y$, Lemma~2.7.7 of \citep{vershynin2018highdimensional} implies
\begin{equation}
\label{eq:shared-product-tail}
\|B_{i,ab}x^\top A_i y\|_{\psi_1}
\le C\|B_{i,ab}\|_{\psi_2}\|x^\top A_i y\|_{\psi_2}
\le C K_AK_B.
\end{equation}
This proves Assumption~\ref{ass:sub-exponential-tail} and completes
the verification.
\end{proof}

For a concrete instance, take $L=1$, $C_i=z_iC$, and
\begin{equation}
\label{eq:shared-complete-block}
C_{ab}
=\begin{cases}
c,&a,b\in V^\star,\ a\ne b,\\
0,&\text{otherwise},
\end{cases}
\qquad c\ne0.
\end{equation}
Let $z_i\sim N(0,1)$, and let $G_i^A,G_i^B\in\R^{p\times p}$
have independent $N(0,1)$ entries, with all these variables and
matrices mutually independent across modalities and subjects.
Set
\begin{equation}
\label{eq:shared-gaussian-noise}
E_i^A=\frac12(G_i^A+G_i^{A\top}),
\qquad
E_{i,ab}^B=
\begin{cases}
\dfrac12(G_{i,ab}^B+G_{i,ba}^B),&a\ne b,\\
0,&a=b.
\end{cases}
\end{equation}
Define $A_i,B_i$ by \eqref{eq:shared-observed-model} for any fixed
$W^\star\in O(p)$. Symmetry, centering, and independence verify
\eqref{eq:shared-noise-structure}, \eqref{eq:shared-cross-moments},
and \eqref{eq:shared-iid-tuples}.

Note that three distinct eigenvalues of $C$ are $c(\kappa-1)$, $-c$, 0, with multiplicities $1$, $\kappa-1$, and $p-\kappa$, respectively.
Thus its kernel is exactly $\ker(P_{V^\star})$.
Let $\lambda_1,\ldots,\lambda_p$ denote these eigenvalues in an
orthonormal eigenbasis. In this basis, the identity
$(C\Omega-\Omega C)_{ab}=(\lambda_a-\lambda_b)\Omega_{ab}$
shows that $\mathfrak h_C$ consists of the skew-symmetric blocks
within the eigenspaces and that its orthogonal complement consists
of the blocks between distinct eigenspaces. The smallest absolute
difference between distinct eigenvalues within the spectrum is $|c|$. Hence
\begin{equation}
\|C\Omega-\Omega C\|_F^2\ge c^2\|\Omega\|_F^2,
\qquad \Omega^\top=-\Omega,\quad\Omega\perp\mathfrak h_C.
\end{equation}
Since $\mathbb E(z_i^2)=1$, the constants are therefore $\alpha_n=c^2$, $m_n=2(\kappa-1)c^2$, and $J^\star=\kappa(\kappa-1)c^2$.
For every pair of unit vectors $x,y$, independence and the
symmetrized Gaussian construction yield
\begin{align}
\operatorname{Var}(x^\top A_i y)
&=(x^\top W^\star C W^{\star\top}y)^2
  +\frac{\sigma_A^2}{2}\{1+(x^\top y)^2\}
\le c^2(\kappa-1)^2+\sigma_A^2,
\label{eq:shared-gaussian-A-variance}\\
\max_{a,b}\operatorname{Var}(B_{i,ab})
&=c^2+\frac{\sigma_B^2}{2}.
\label{eq:shared-gaussian-B-variance}
\end{align}
Therefore, for a
universal constant $C_\psi>0$ determined by the convention for the
$\psi_2$ norm, \eqref{eq:shared-sub-gaussian-tails} holds with $K_A=C_\psi\sqrt{c^2(\kappa-1)^2+\sigma_A^2}$, and $K_B=C_\psi\sqrt{c^2+\sigma_B^2/2}$.
The population coefficient bound \eqref{eq:M0} holds with $\mathbb E(B_{i,ab}A_i)=C_{ab}W^\star C W^{\star\top}$, and $M_0=c^2(\kappa-1)$.

\section{Full \textsc{Ossa} algorithm pseudocode}
\label{app:multi-algo}
 
Algorithm~\ref{alg:single-ossa} gives the canonical \textsc{Ossa} single-support \textsc{Ossa} solver.
Algorithm~\ref{alg:multi-ossa} gives the overlap-regularized \(K\)-support
extension. 

When $K=1$ and $\lambda=0$, the inner $\ell$-loop collapses to a single $s$-update, the aggregate $M_i$ reduces to $SB_iS$, and the algorithm reduces exactly to Algorithm~\ref{alg:single-ossa}. For $K\ge2$, the Gauss-Seidel sweep updates each support against the current $W$ and the current values of the other supports, with the overlap penalty entering linearly through $-\lambda\sum_{r\ne\ell}s^{(r)}$. As in the canonical \textsc{Ossa} case, backtracking guarantees a monotone objective sequence, and multiple random restarts are used to mitigate local optima.
 
\begin{algorithm}[H]
\caption{Overlap-regularized K-clique \textsc{Ossa}}
\label{alg:multi-ossa}
\begin{algorithmic}[1]
\Require Symmetric matrices $\{A_i,B_i\}_{i=1}^n$ in one community; number of supports $K$; per-support budget $\kappa$; overlap penalty $\lambda\ge0$; initial steps $\eta_W,\eta_s$; backtracking factor $\beta\in(0,1)$
\Ensure Alignment $W$, relaxed supports $\{s^{(\ell)}\}_{\ell=1}^K$, and hard supports $\{\widehat{\mathcal V}^{(\ell)}\}_{\ell=1}^K$
\State Initialize $W\leftarrow I_p$ or a random orthogonal matrix; initialize $s^{(\ell)}\in\mathcal S_\kappa$ for $\ell=1,\ldots,K$
\Repeat
    \State $S^{(\ell)}\leftarrow\diag(s^{(\ell)})$ for $\ell=1,\ldots,K$
    \State $M_i\leftarrow\sum_{\ell=1}^K S^{(\ell)} B_i S^{(\ell)}$ for $i=1,\ldots,n$
    \State $G_W\leftarrow 2\sum_{i=1}^n A_i W M_i$ and $W^+(\eta)\leftarrow\Pi_{O(p)}(W+\eta G_W)$
    \State Choose the largest $\eta\in\{\eta_W\beta^m:m\ge0\}$ satisfying
    \Statex \hskip\algorithmicindent $J_K(W^+(\eta),\{s^{(\ell)}\})\ge J_K(W,\{s^{(\ell)}\})$, and set $W\leftarrow W^+(\eta)$
    \State $D\leftarrow\sym\{\sum_{i=1}^n (W^\top A_i W)\circ B_i\}$
    \For{$\ell=1,\ldots,K$} \Comment{Gauss-Seidel sweep over supports}
        \State $g^{(\ell)}\leftarrow 2D s^{(\ell)}-\lambda\sum_{r\ne\ell} s^{(r)}$
        \State $s^{(\ell,+)}(\eta)\leftarrow\Pi_{\mathcal S_\kappa}(s^{(\ell)}+\eta g^{(\ell)})$
        \State Choose the largest $\eta\in\{\eta_s\beta^m:m\ge0\}$ satisfying
        \Statex \hskip\algorithmicindent $J_K(W,\ldots,s^{(\ell,+)}(\eta),\ldots)\ge J_K(W,\ldots,s^{(\ell)},\ldots)$,
        \Statex \hskip\algorithmicindent and set $s^{(\ell)}\leftarrow s^{(\ell,+)}(\eta)$
    \EndFor
\Until{the objective and iterate changes satisfy the stopping tolerances}
\State $\widehat{\mathcal V}^{(\ell)}\leftarrow\Top_\kappa(s^{(\ell)})$ for $\ell=1,\ldots,K$
\State \Return $W$, $\{s^{(\ell)}\}_{\ell=1}^K$, $\{\widehat{\mathcal V}^{(\ell)}\}_{\ell=1}^K$
\end{algorithmic}
\end{algorithm}

\paragraph{Implementation constants.}
Unless otherwise stated, we use initial step sizes $\eta_W = 5\times 10^{-2}$, $\eta_s = 5\times 10^{-2}$, backtracking factor $\beta = 0.5$ with a floor $\eta_{\min}=10^{-12}$, relative objective tolerance $\mathrm{tol}_{\text{outer}}=10^{-5}$, per-block objective-improvement threshold $\mathrm{tol}_{\text{inner}}=10^{-7}$, and a maximum of $20$ outer block-coordinate iterations (with up to $30$ projected-gradient iterations per $W$-block and per support-vector update, each using backtracking line search). Random orthogonal initializations are generated by QR decomposition of an i.i.d.\ $N(0,1)$ Gaussian matrix (equivalently, samples from the Haar measure on $O(p)$). Support initializations are generated by drawing each entry i.i.d.\ uniform on $[0,1]$ and then projecting onto $\mathcal S_\kappa$ via the clip-and-soft-threshold simplex projection, we verified that uniform $s = (\kappa/p)\one_p$ and binary $\kappa$-hot random initializations produce indistinguishable selections within restart-to-restart variability. For the ABCD experiments, we use $16$ random restarts for canonical \textsc{Ossa} evaluation (Algorithm~\ref{alg:single-ossa}), $4$ for per-$\kappa$ selection, and $4$ for multi-clique fits (Algorithm~\ref{alg:multi-ossa}). The canonical single-clique simulation sweeps use $8$ random restarts; settings for the remaining simulation diagnostics are reported in Appendices~\ref{app:budget}--\ref{app:lambda-traj}. The restart with the largest training objective is retained. For the ABCD bootstrap stability analysis we use $2$ restarts with a lighter $(\mathrm{max\_outer},\mathrm{max\_iter}_{W,s})= (10,15)$ schedule and $\mathrm{tol}_{\text{outer}}=10^{-4}$, $\mathrm{tol}_{\text{inner}}=10^{-6}$ to keep the $100$-bootstrap budget tractable, retaining the same step sizes and backtracking factor. Detailed optimization settings are listed in Table~\ref{tab:ossa-budgets}.

\begin{table}[h]
\centering
\caption{\textsc{Ossa} optimization budgets used at each stage of the pipeline. Joint support-and-alignment fits use initial step sizes $\eta_W=\eta_s=5\!\times\!10^{-2}$, backtracking factor $\beta=0.5$ with floor $\eta_{\min}=10^{-12}$, and projected-uniform $s$ initialization. Fixed-support baseline $W$-refits hold $s$ fixed and use three $W$ initializations, including the identity (Appendix~\ref{app:abcd-baseline-W}). ``Outer'' is the maximum number of block-coordinate sweeps for joint fits; for fixed-support $W$-refits, it gives the maximum number of $W$-update iterations per initialization. ``Inner'' is the maximum number of projected-gradient iterations per $W$-block and per support-vector update; each iteration uses backtracking line search. ``Restarts'' is the number of optimization runs from the specified initializations; the run with the largest training objective is retained.}

\label{tab:ossa-budgets}
\begin{tabular}{lccccc}
\toprule
Stage & Restarts & Outer & Inner & $\eta_W=\eta_s$ & $(\mathrm{tol}_{\text{outer}},\mathrm{tol}_{\text{inner}})$ \\
\midrule
Canonical \textsc{Ossa} evaluation (Table~\ref{tab:abcd-main})         & $16$ & $20$  & $30$ & $5\!\times\!10^{-2}$ & $(10^{-5},10^{-7})$ \\
Per-$\kappa$ selection                            & $4$  & $15$  & $20$ & $5\!\times\!10^{-2}$ & $(10^{-5},10^{-7})$ \\
Multi-clique fits ($K\!\ge\!2$)                              & $4$  & $15$  & $20$ & $5\!\times\!10^{-2}$ & $(10^{-5},10^{-7})$ \\
ABCD bootstrap stability                                     & $2$  & $10$  & $15$ & $5\!\times\!10^{-2}$ & $(10^{-4},10^{-6})$ \\
Canonical single-clique sweeps                              & $8$  & $20$  & $40$ & $5\!\times\!10^{-2}$ & $(10^{-5},10^{-7})$ \\
Per-baseline $W$-refit (fixed $s$)                           & $3$ & $200$ & -   & $5\!\times\!10^{-2}$ & $\mathrm{tol}_{\text{improve}}=10^{-9}$ \\
\bottomrule
\end{tabular}\\
\end{table}

\paragraph{Computational cost.}
In a community of size \(p\), one dense outer iteration costs \(O(np^3)\) for the \(W\)-block matrix products and \(O(np^2)\) for forming \(D(W)\), plus \(O(Kp^2)\) for support gradients and \(O(Kp\log p)\) for capped-simplex projections. The polar projection costs \(O(p^3)\). Memory is \(O(np^2)\) for dense subject matrices. In the ABCD experiments, the largest community has about 30 nodes, so runtime is dominated by the number of subject-level matrix products and restarts rather than by the cubic polar step.
 
\section{Additional experimental results and diagnostics}
\label{app:robustness}
This appendix collects simulation sweep curves, robustness analyses, and ABCD training-side diagnostics referenced from Sections~\ref{sec:experiments} and~\ref{subsec:real}. Unless otherwise stated, the simulation experiments use the canonical \textsc{Ossa} generative model of Section~\ref{subsec:single-sim} or the multi-clique extension of Section~\ref{subsec:multi-sim}, with the same nuisance structure.

\subsection{Complete simulation generator}
\label{app:simu-setup}
In Section~\ref{subsec:single-sim}, the main text writes $\zeta_i^A=\xi_i^A+\tau_i^A$ and $\zeta_i^B=\xi_i^B+\tau_i^B$ to keep the simulation model concise. For each replicate we draw the data $\{(A_i,B_i)\}_{i=1}^n$ from the following procedure (single-clique unless noted; multi-clique substitutions in brackets). Default ambient size is $p=40$ for the canonical \textsc{Ossa} experiments and $p=50$ for the multi-clique experiments, with $n=30$ subjects per replicate. ``$\mathrm{LR}_r(V,c)$'' denotes a symmetric rank-$r$ random block of magnitude $c$, supported on $V\subseteq[p]$, generated as $UU^\top\!\circ\!\Sigma$ with $U\in\mathbb R^{|V|\times r}$ having i.i.d.\ $N(0,c^2/r)$ entries and $\Sigma$ a symmetrized random $\pm1$ sign matrix (entries $\Sigma_{ab}=\tfrac12(\sigma_{ab}+\sigma_{ba})$ with $\sigma_{ab}\stackrel{\text{i.i.d.}}{\sim}\mathrm{Unif}\{-1,+1\}$), embedded as zero outside $V\!\times\!V$.

\begin{itemize}
\item \textbf{Draw planted support(s).}
Sample $\mathcal V^\star\subseteq[p]$ uniformly without replacement of size $\kappa=10$. For the multi-clique generator, sample $K\in\{2,3\}$ supports $\mathcal V^{\star,(\ell)}\subseteq[p]$ via a random permutation of $[p]$ and consecutive blocks of size $\kappa$, with adjacent supports sharing exactly $r\in\{0,4\}$ nodes (so they are pairwise disjoint when $r=0$). Defaults: $K{=}2,\kappa{=}10,r{=}0$ (\emph{disjoint}); $K{=}2,\kappa{=}10,r{=}4$ (\emph{overlap-4}); $K{=}3,\kappa{=}8,r{=}0$ (\emph{$K{=}3$ disjoint}).

\item \textbf{Draw low-rank block $U_i$ and sign matrix $\Sigma_i$.} Sample $U_i\in\mathbb R^{\kappa\times r_{\text{sig}}}$ with i.i.d.\ $N(0,c_{\text{sig}}^2/r_{\text{sig}})$ entries and the symmetric sign mask $\Sigma_i$ as above. Defaults $r_{\text{sig}}=3$, $c_{\text{sig}}=1.5$.

\item \textbf{Form $C_i$ (Canonical \textsc{Ossa}) or $C_i^{\mathrm{tot}}$ (multi) and embed.} Canonical \textsc{Ossa}: $C_i=\mathrm{LR}_{r_{\text{sig}}}(\mathcal V^\star,c_{\text{sig}})$. Multi: $C_i^{\mathrm{tot}}=\sum_{\ell=1}^K \mathrm{LR}_{r_{\text{sig}}}(\mathcal V^{\star,(\ell)},c_{\text{sig}})$ with an independent $(U_i^{(\ell)},\Sigma_i^{(\ell)})$ for each support. Both are zero outside their planted supports.

\item \textbf{Draw background matrices $\xi_i^A,\xi_i^B$.} Each is a fresh i.i.d.\ rank-$r_{\text{bg}}$ symmetric block with the same $\mathrm{LR}$ recipe but supported on all $p$ nodes, with amplitudes $c_{\text{bg}}^A=c_{\text{bg}}^B=0.6$ (Canonical \textsc{Ossa}) and $c_{\text{bg}}^A=c_{\text{bg}}^B=0.4$ (multi-clique), $r_{\text{bg}}=4$. Backgrounds are independent across modalities.

\item \textbf{Draw modality-specific nuisance blocks $\tau_i^A,\tau_i^B$ (Canonical \textsc{Ossa} only).}
Two disjoint modality-specific nuisance variation node sets $\mathcal D^A,\mathcal D^B\subseteq [p]\setminus\mathcal V^\star$, each of size $|\mathcal D|=12$, are sampled once per replicate (not per subject) by random permutation of $[p]\setminus\mathcal V^\star$ and consecutive slicing. For each subject $i$ we then draw $\tau_i^B=\mathrm{LR}_{r_{\text{dec}}}(\mathcal D^B,c_{\text{dec}}^B)$ and $\tau_i^A=\mathrm{LR}_{r_{\text{dec}}}(\mathcal D^A,c_{\text{dec}}^A)$ independently, with $r_{\text{dec}}=3$ and $c_{\text{dec}}^A=c_{\text{dec}}^B=1.5$. The multi-clique generator omits this step.

\item \textbf{Draw noise matrices $E_i^A,E_i^B$.}
Symmetric Gaussian: $E_i^A=\tfrac12(G_i^A+G_i^{A\top})$ with $G_i^A\!\sim\!N(0,1)^{p\times p}$ (and analogously for $E_i^B$), then scaled by $\sigma_A,\sigma_B$. Defaults $\sigma_A=\sigma_B=0.3$ (Canonical \textsc{Ossa}) and $\sigma_A=\sigma_B=0.4$ (multi-clique). Sweep ranges in the appendix vary $\sigma\in[0.1,0.5]$.

\item \textbf{Symmetrize and zero the diagonal.}
Set \[ B_i \;=\; \tfrac12(\widetilde B_i+\widetilde B_i^\top), \qquad A_i \;=\; \tfrac12(\widetilde A_i+\widetilde A_i^\top), \] where $\widetilde B_i = C_i + \xi_i^B + \tau_i^B + E_i^B$ and $\widetilde A_i = W^\star C_i (W^\star)^\top + \xi_i^A + \tau_i^A + E_i^A$ (replace $C_i$ by $C_i^{\mathrm{tot}}$ and drop the $\tau$ terms in the multi-clique case), then enforce $\mathrm{diag}(A_i)=\mathrm{diag}(B_i)=0$.

\item \textbf{Draw $W^\star$.}
Sample $G\sim N(0,1)^{p\times p}$, take its (sign-corrected) QR factor $Q$ so that $Q\sim\text{Haar}(O(p))$, and set \[ W^\star \;=\; \mathrm{polar}\!\bigl((1-\alpha)\,I_p + \alpha\,Q\bigr)        \;=\; \tilde U\tilde V^\top,\qquad \tilde U\,\Sigma\,\tilde V^\top        \;=\;\mathrm{SVD}\!\bigl((1-\alpha)\,I_p + \alpha\,Q\bigr), \] which equals $I_p$ at $\alpha{=}0$ and a Haar-random orthogonal matrix at $\alpha{=}1$, and is itself orthogonal for every $\alpha\in[0,1]$. The rotation strength is swept over $\alpha\in\{0,0.25,0.5,0.75,1\}$ for \textsc{Ossa}, \textsc{Ossa}-\textsc{noW}, and the support-selection baselines; the \textsc{Ossa}-\textsc{oracle} reference uses $\alpha\in\{0,0.2,0.4,0.6,0.8,1\}$. The headline canonical \textsc{Ossa} configuration uses $\alpha{=}1$ and the multi-clique configuration uses $\alpha{=}0.6$.
\end{itemize}

\subsection{Full canonical \textsc{Ossa} sweep curves}
\label{app:single-sweeps}

\begin{figure}[h]
\centering
\includegraphics[width=0.7\linewidth]{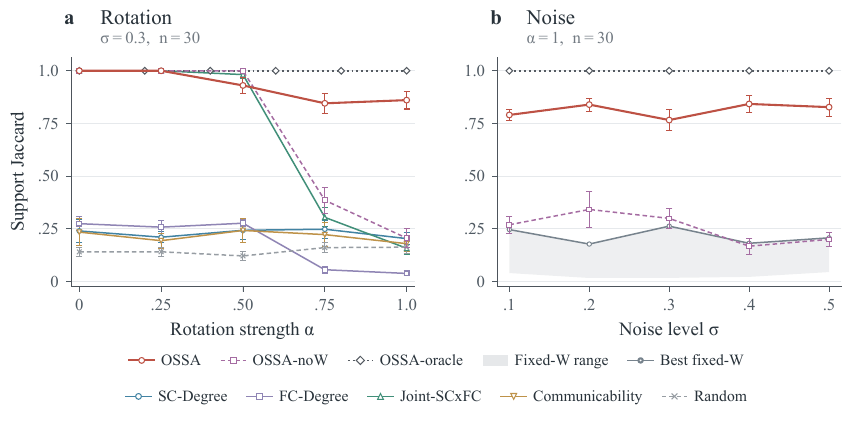}
\caption{Canonical \textsc{Ossa} support recovery (Jaccard with $\mathcal V^\star$, mean $\pm$ s.e., $10$ replicates). 
(a) Rotation sweep $\alpha \in [0, 1]$ at $\sigma = 0.3$, $n = 30$. 
(b) Noise sweep $\sigma \in [0.1, 0.5]$ at $\alpha = 1$, $n = 30$. In panel (b), the grey band spans the minimum and maximum mean Jaccard values of SC-Degree, FC-Degree, Joint-SCxFC, Spectral-SC, and Spectral-FC at each noise level; the ``Best fixed-$W$'' line follows the upper boundary.}
\label{fig:single-sweeps}
\end{figure}

Figure~\ref{fig:single-sweeps} reports the full rotation and noise trajectories summarized by Table~\ref{tab:single-summary}. The rotation sweep isolates the coordinate-mismatch setting, while the noise sweep shows that \textsc{Ossa} remains above the fixed-$W$ baselines even when the planted support is observed through stronger subject-level noise.

\subsection{ABCD community-detection diagnostics}
\label{app:abcd-community}

\begin{figure}[h]
  \centering
  \includegraphics[width=\linewidth]{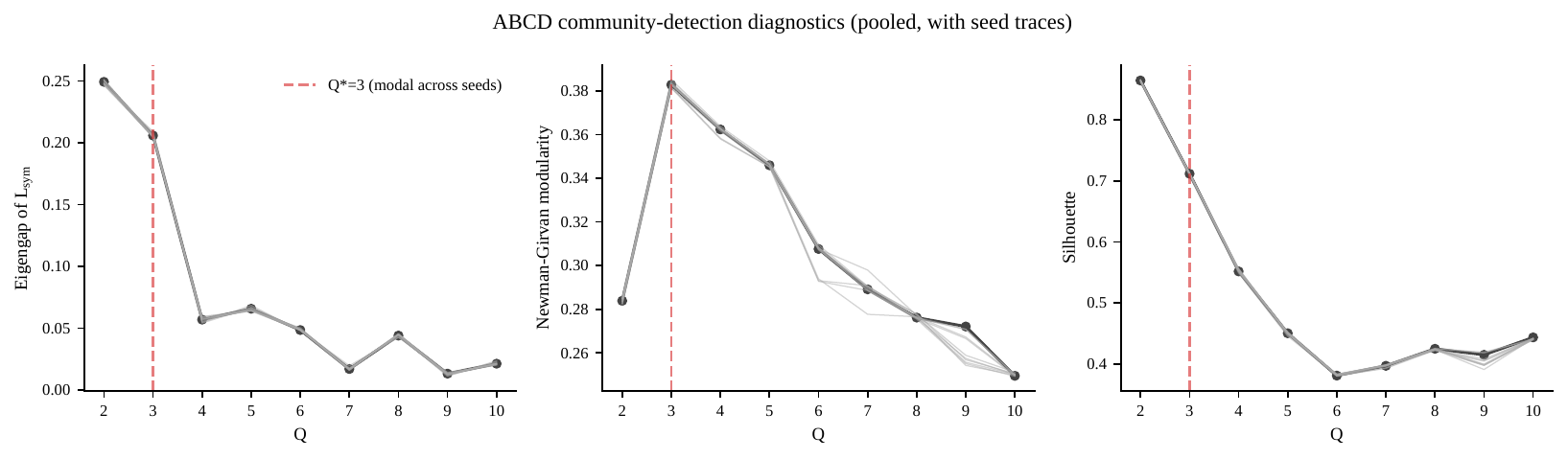}
  \caption{ABCD community-detection diagnostics on the training absolute-FC mean. Light gray traces are per-seed; dark line is the pooled training average. The diagnostics support a stable \(Q=3\) working partition: modularity peaks at \(Q=3\), and eigengap/silhouette show a coarse \(Q=2\) split with a clear elbow at \(Q=3\). No split seed selects a different working partition under this rule.}
  \label{fig:abcd-community}
\end{figure}

Figure~\ref{fig:abcd-community} reports the training-side diagnostics used to fix the number of FC communities in the baseline ABCD cohort. The diagnostics are computed only from the training absolute-FC mean within each split and therefore do not use validation or test data. Modularity peaks at \(Q=3\), while eigengap and silhouette show a strong coarser \(Q=2\) split followed by a stable elbow at \(Q=3\). We use \(Q=3\) as the smallest stable nontrivial partition beyond the coarse two-community split, and the choice is stable across split seeds. In the experiments, communities are estimated by spectral clustering on the training mean absolute-FC matrix. The multilayer DCSBM result motivates the shared-community assumption but is not used as a literal generative claim for ABCD.

\subsection{Sparsity-budget misspecification}
\label{app:budget}
In practice the budget $\kappa$ is rarely known exactly. We sweep the supplied $\kappa \in \{4, 6, 8, 10, 12, 14, 16\}$ at planted $\kappa^\star = 10$, holding all other settings at the moderate-rotation, moderate-noise slice $p = 40$, $n = 30$, $\alpha = 0.6$, $\sigma = 0.4$ ($8$ replicates per supplied $\kappa$, $6$ random restarts), and report Jaccard, precision, and recall against $\mathcal V^\star$.

\begin{figure}[h]
\centering
\includegraphics[width=\linewidth]{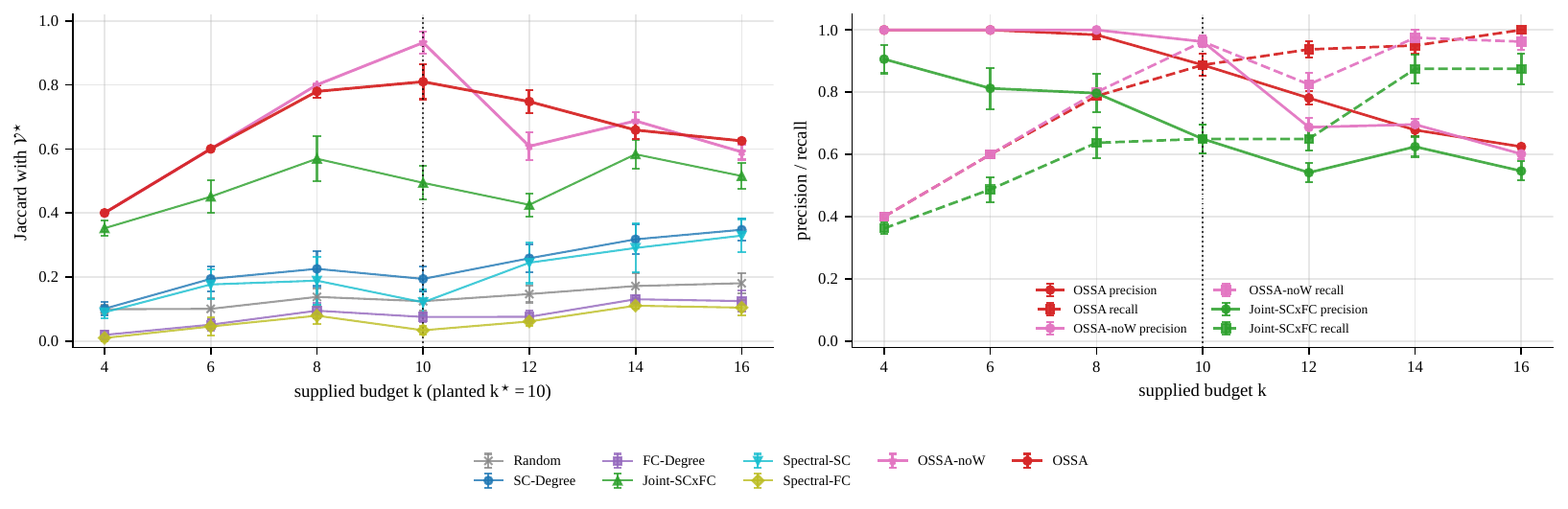}
\caption{Robustness to sparsity-budget misspecification at $\kappa^\star = 10$, $\alpha = 0.6$, $\sigma = 0.4$ (mean $\pm$ s.e., $8$ replicates). (a) Jaccard with $\mathcal V^\star$ vs.\ supplied $\kappa$. (b) Precision and recall vs.\ supplied $\kappa$ for \textsc{Ossa}, \textsc{Ossa}-\textsc{noW}, and the strongest non-\textsc{Ossa} baseline (\textsc{Joint-SCxFC}); solid = precision, dashed = recall.}
\label{fig:budget}
\end{figure}

\textsc{Ossa} degrades gracefully on both sides of the truth (Figure~\ref{fig:budget}). Under-specification ($\kappa < \kappa^\star$) yields high precision but limited recall (e.g., $\kappa = 4$: precision $1.00$, recall $0.40$, Jaccard $0.40$); over-specification ($\kappa > \kappa^\star$) yields perfect recall and reduced precision (e.g., $\kappa = 16$: precision $0.62$, recall $1.00$, Jaccard $0.62$). At the truth $\kappa = 10$, \textsc{Ossa} attains Jaccard $0.81 \pm 0.06$ and precision $0.89$. The strongest non-\textsc{Ossa} baseline at $\kappa = 10$ is \textsc{Joint-SCxFC} at $0.49$, which tracks below \textsc{Ossa} across the entire grid. \textsc{Ossa}-\textsc{noW} matches or slightly exceeds \textsc{Ossa} at $\alpha = 0.6$ for $\kappa \leq 10$ (where $W = I$ is a useful approximation at moderate rotation) but degrades faster on the over-specified side. For this budget-misspecification sweep, we also include \textsc{Spectral-SC} and \textsc{Spectral-FC}, which rank nodes by the absolute loading of the leading eigenvector of the training mean SC and FC matrices, respectively. Missing or extra budget translates into the corresponding precision/recall trade-off at the planted nodes rather than catastrophic Jaccard loss, indicating that \textsc{Ossa} is a usable estimator when $\kappa$ is supplied heuristically.

\subsection{Selection stability}
\label{app:stability}
\begin{figure}[h]
\centering
\includegraphics[width=\linewidth]{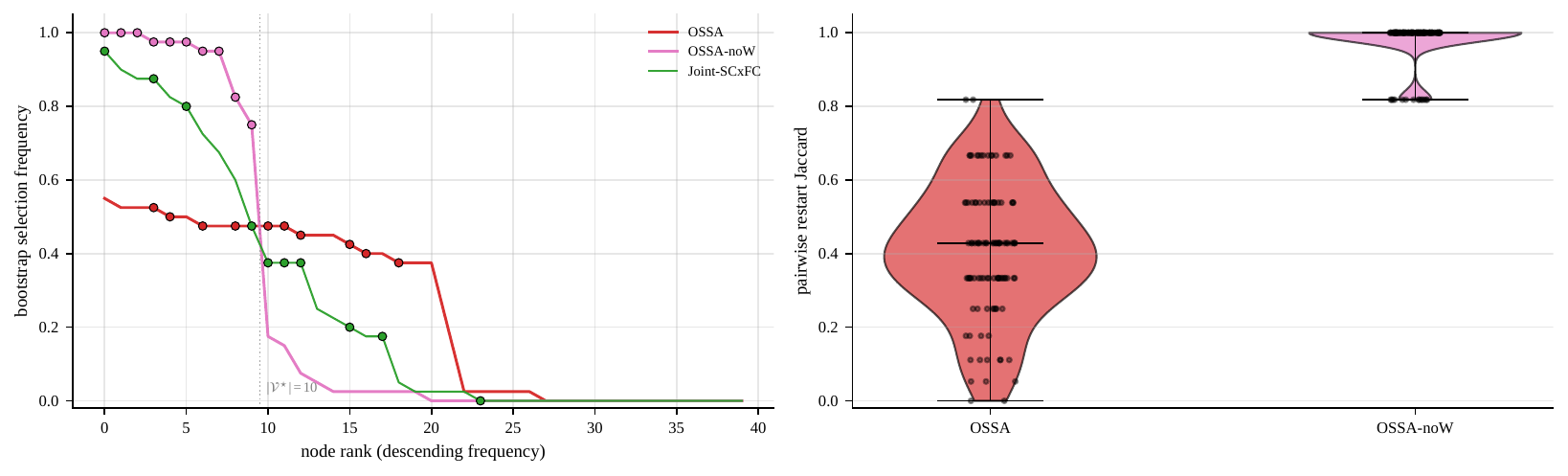}
\caption{Selection stability of \textsc{Ossa} at $\alpha = 0.6$, $\sigma = 0.4$, $n = 30$. (a) Per-node bootstrap selection frequency, sorted descending; circled markers indicate planted nodes ($|\mathcal V^\star| = 10$). (b) Violin distribution of pairwise Jaccard between $15$ random restarts on the same data, for \textsc{Ossa} and \textsc{Ossa}-\textsc{noW}.}
\label{fig:stability}
\end{figure}

We assess stability of the selected support along two complementary axes on a single base dataset at $\alpha = 0.6$, $\sigma = 0.4$, $n = 30$, $\kappa^\star = 10$: per-node selection frequency across $40$ bootstrap resamples (with replacement) of the subjects. Each bootstrap fit uses one random initialization, so the bootstrap curve captures both subject-resampling variability and single-start optimization variability. We also examine the distribution of pairwise Jaccard between $15$ random initializations on the same (non-resampled) data for \textsc{Ossa} and \textsc{Ossa}-\textsc{noW} (Figure~\ref{fig:stability}b).

Bootstrap selection frequency on the $10$ planted nodes is substantially higher than on non-planted nodes for \textsc{Ossa} ($0.46$ vs.\ $0.18$; Figure~\ref{fig:stability}a), consistent with concentration on $\mathcal V^\star$ rather than a restart-dependent local optimum. 
The moderate pairwise Jaccard similarity across restarts indicates sensitivity of the fitted support to initialization, consistent with the nonconvexity of the joint optimization problem. For the main experiments, we use multiple restarts and keep the run with the largest training objective. 
However, we note that this selection rule does not guarantee a global maximizer or eliminate variation across independently generated sets of restarts. 
\textsc{Ossa}-\textsc{noW} achieves higher restart concentration at this moderate rotation ($\alpha = 0.6$, where $W = I$ is a useful approximation to $W^\star$ and the smaller-dimensional optimization is easier to stabilize) but pays the cost in recovery quality at $\alpha = 1$ (Section~\ref{subsec:single-sim}, Table~\ref{tab:single-summary}).

\subsection{Multi-clique simulation details}
\label{app:multi-sim-details}

This subsection gives the full design behind Section~\ref{subsec:multi-sim}. We use one community with $p=50$ nodes and $n=30$ subjects. Each replicate contains $K$ planted supports $\{\mathcal V_\ell^\star\}_{\ell=1}^K$, each carrying an independently sampled mixed-sign low-rank signal block. The subject-level signal is
\[
C_i^{\mathrm{tot}}=\sum_{\ell=1}^K C_i^{(\ell)},
\]
and the observed matrices are
\[
B_i = C_i^{\mathrm{tot}} + \xi_i^B + \sigma_B E_i^B,
\qquad
A_i = W^\star C_i^{\mathrm{tot}} {W^\star}^{\top} + \xi_i^A + \sigma_A E_i^A .
\]
We fix rotation strength $\alpha=0.6$ and noise level $\sigma=0.4$. We evaluate three planted regimes: two disjoint settings, $(K,\kappa)=(2,10)$ and $(3,8)$, and one overlapping setting with $K=2$, $\kappa=10$, and exactly four shared nodes. The overlapping nodes receive additive signal from both planted supports. No extra modality-specific nuisance variation blocks are added, because the other planted supports already act as structured distractors for each target support.

We compare \textsc{Ossa}-$K$ against four size-matched competitors. \textsc{Spectral-SC}-\(K\): compute the top \(K\) eigenvectors of the training mean SC matrix \(\bar B\). For component \(\ell\), rank nodes by \(|v_\ell(j)|\) and select the top \(\kappa\) nodes. If a disjoint version is used, previously selected nodes are removed before the next component. \textsc{Communicability-$K$} clusters nodes using communicability distance. \textsc{Sequential-Ossa} repeatedly fits canonical \textsc{Ossa}, masks the selected nodes, and refits until $K$ supports are obtained. \textsc{Random-$K$} samples $K$ supports uniformly. Unlike \textsc{Ossa}-$K$, \textsc{Sequential-Ossa} estimates a fresh rotation at each step and enforces disjoint supports by construction.

For this diagnostic simulation, $\lambda$ for \textsc{Ossa}-$K$ is selected on the grid reported in Appendix~\ref{app:lambda-traj}, choosing the smallest value within one standard error of the peak best-match Jaccard. Because planted labels are available here, this $\lambda_{\mathrm{best}}$ rule is used only to summarize the recoverability envelope of the method; it is not used as a real-data tuning rule. We report best-match Jaccard after Hungarian assignment to the planted supports, together with the mean pairwise overlap of the fitted supports. All methods use six random restarts, with eight replicates for $K=2$ and six for $K=3$.

\begin{table}[h]
\caption{Full multi-clique support recovery (mean $\pm$ s.e.). \textsc{Jacc}: best-match Jaccard. \textsc{Ovl}: mean pairwise overlap of fitted supports. In the two disjoint regimes all methods produce fitted overlap $0.00\pm0.00$; overlap is therefore reported only for the $r=4$ regime.}
\label{tab:multi-summary-full}
\centering
\tablesetup
\begin{tabular}{lcccc}
\toprule
\textbf{Method} & \textbf{Disjoint } $K{=}2$ & \multicolumn{2}{c}{\textbf{Overlap } $r{=}4, K{=}2$} & \textbf{Disjoint } $K{=}3$ \\
\cmidrule(lr){3-4}
& \textsc{Jacc} & \textsc{Jacc} & \textsc{Ovl} & \textsc{Jacc} \\
\midrule
\textsc{Random}-$K$            & $0.150 \pm 0.009$ & $0.123 \pm 0.016$ & $0.00 \pm 0.00$ & $0.133 \pm 0.012$ \\
\textsc{Spectral-SC}-$K$       & $0.522 \pm 0.048$ & $0.368 \pm 0.026$ & $0.00 \pm 0.00$ & $0.401 \pm 0.039$ \\
\textsc{Communicability}-$K$   & $0.652 \pm 0.056$ & $0.284 \pm 0.023$ & $0.00 \pm 0.00$ & $0.552 \pm 0.061$ \\
\textsc{Sequential-Ossa}       & $0.710 \pm 0.040$ & $0.448 \pm 0.026$ & $0.00 \pm 0.00$ & $0.744 \pm 0.063$ \\
\textsc{Ossa}-$K$ ($\lambda_{\mathrm{best}}$) & $\mathbf{0.945 \pm 0.023}$ & $\mathbf{0.910 \pm 0.066}$ & $\mathbf{3.75 \pm 0.56}$ & $\mathbf{0.975 \pm 0.016}$ \\
\bottomrule
\end{tabular}
\end{table}

\subsection{Continuous $\lambda$-trajectory}
\label{app:lambda-traj}
\begin{figure}[h]
\centering
\includegraphics[width=\linewidth]{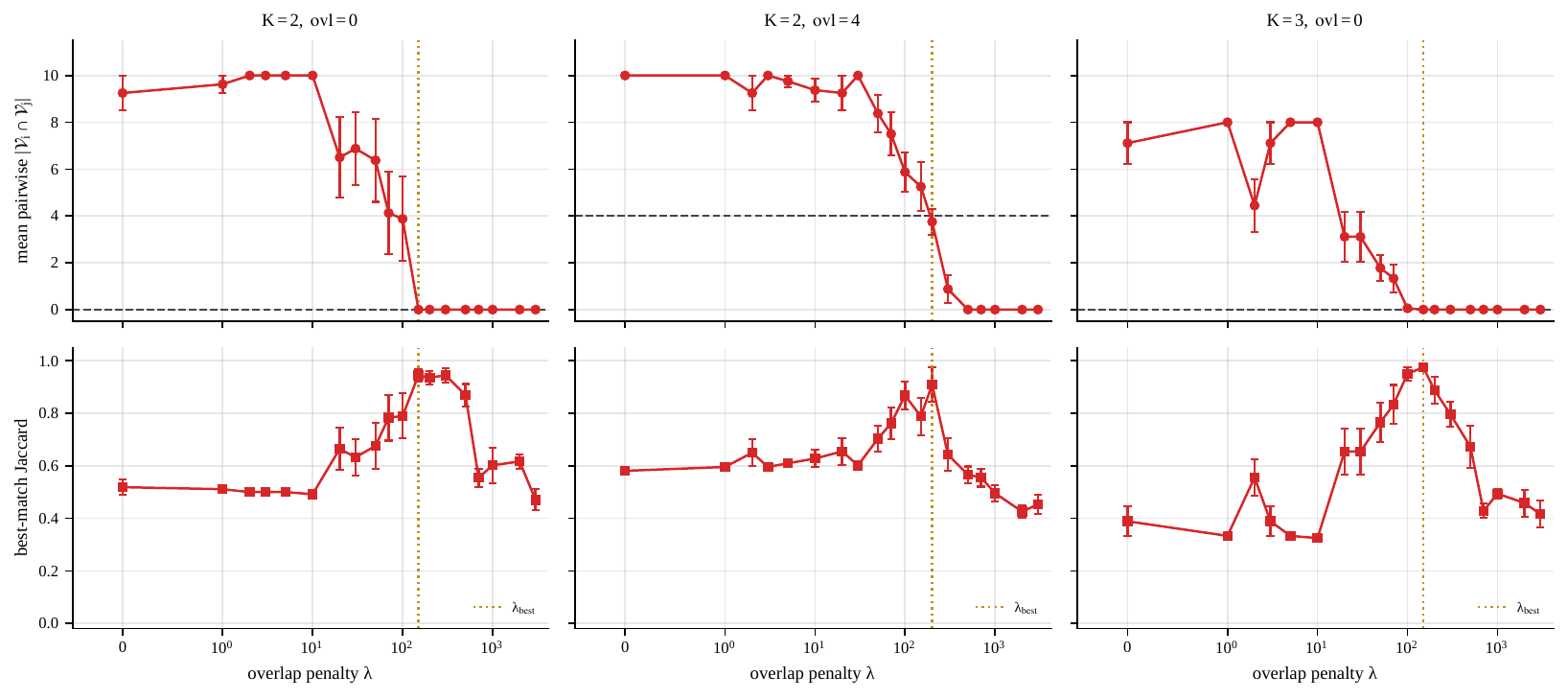}
\caption{Continuous $\lambda$-trajectory of \textsc{Ossa}-$K$ across three planted regimes (mean $\pm$ s.e.; $8$ replicates for $K = 2$, $6$ for $K = 3$). Top row: mean pairwise overlap of fitted supports; dashed line marks the planted overlap. Bottom row: best-match Jaccard with the planted supports. Columns: $K = 2$ disjoint, $K = 2$ with $r = 4$ shared nodes, $K = 3$ disjoint. Vertical dotted line marks $\lambda_{\text{best}}$ summarized in Table~\ref{tab:multi-summary}. Full baseline values are reported in Appendix Table~\ref{tab:multi-summary-full}.}
\label{fig:lambda-traj}
\end{figure}

The $\lambda_{\text{best}}$ values summarized in Section~\ref{subsec:multi-sim} and Appendix~\ref{app:multi-sim-details} are read off a dense log-spaced grid spanning four decades. We fit \textsc{Ossa}-$K$ at $\lambda \in \{0, 1, 2, 3, 5, 10, 20, 30, 50, 70, 100, 150, 200,$ $300, 500, 700, 1000, 2000, 3000\}$ on each of the three regimes from Section~\ref{subsec:multi-sim} ($K = 2$ disjoint, $K = 2$ with $r = 4$ shared nodes, $K = 3$ disjoint), using the same seeds and replicate counts as the main experiment ($8$ replicates for $K = 2$, $6$ for $K = 3$, $6$ random restarts per fit). The vertical line in each panel of Figure~\ref{fig:lambda-traj} marks the selected $\lambda_{\text{best}}$ for that regime. This oracle selection is used only to summarize whether the \(\lambda\)-path contains a recoverability region. It is not a deployable tuning rule and is not used in ABCD.

The trajectories share a common shape: at small $\lambda$ the supports collapse (high pairwise overlap, low Jaccard against the planted partition); raising $\lambda$ separates the supports until fitted overlap matches the planted value (with $\lambda_{\text{best}}$ near this crossing); further increases drive the supports into full disjointness. In the disjoint regimes, fitted overlap reaches zero between $\lambda = 100$ and $\lambda = 200$. For $K = 2$, Jaccard remains near $0.95$ across $\lambda \in [150, 300]$, whereas for $K = 3$ it peaks at $\lambda = 150$ and declines over the same interval. In the $r = 4$ regime, fitted overlap shows an overall decrease from $\kappa = 10$ at $\lambda = 0$ to approximately $r = 4$ near $\lambda_{\text{best}} = 200$ ($3.75 \pm 0.56$), with Jaccard peaking near this crossing and declining as larger penalties force the supports toward full disjointness.

\subsection{ABCD $\kappa$-robustness}
\label{app:abcd-kappa-robust}
We verify that \textsc{Ossa}'s lead over the nine comparison methods in Table~\ref{tab:abcd-kappa-robust} is preserved at $\kappa \in \{3, 5, 8, 10\}$, not only at the design choice $\kappa = 3$. Each row summarizes the same $30$ (seed, community) units. The SC-degree-quartile-matched null uses $1000$ draws per unit at $\kappa=3$ and $30$ draws per unit at $\kappa\in\{5,8,10\}$ (Table~\ref{tab:abcd-kappa-robust}). 

The headline pattern is that \textsc{Ossa} achieves the largest mean null-corrected gain $M_{3,\text{aln}}$ at every tested budget. At $\kappa=3$, \textsc{SC-Degree}, \textsc{Joint-SCxFC}, and \textsc{Communicability} have positive mean null-corrected gains ($+0.59$ to $+0.69$), showing that these selectors achieve higher aligned correlations than the degree-matched null average. At the tested budgets $\kappa\geq5$, \textsc{SC-Degree}, \textsc{Joint-SCxFC}, \textsc{Communicability}, \textsc{FC-Degree}, and \textsc{SDI} have near-zero or negative mean null-corrected gains $M_{3,\text{aln}}$. \textsc{Ossa}, in contrast, retains $M_{3,\text{aln}}$ gain of $+0.37$ to $+0.41$ at $\kappa \in \{5, 8, 10\}$, exceeding all compared baselines at each of these budgets.

\begin{table}[h]
\caption{ABCD \(\kappa\)-robustness: \textsc{Ossa} vs. nine comparison methods at \(\kappa\in\{3,5,8,10\}\), mean \(\pm\) s.e. across 30 (seed, community) units. The table reports \(M_{3,\mathrm{aln}}\) gain, defined as \(M_{2,\mathrm{aln}}\) minus the mean SC-degree-quartile-matched
null.}
\label{tab:abcd-kappa-robust}
\centering
\begin{tabular}{lcccc}
\toprule
& \multicolumn{4}{c}{\textbf{$M_{3,\text{aln}}$ gain}} \\
\cmidrule(l){2-5}
Method & $\kappa = 3$ & $\kappa = 5$ & $\kappa = 8$ & $\kappa = 10$ \\
\midrule
\textsc{Ossa}            & $\mathbf{+0.81 \pm 0.03}$ & $\mathbf{+0.41 \pm 0.03}$ & $\mathbf{+0.39 \pm 0.03}$ & $\mathbf{+0.37 \pm 0.02}$ \\
\textsc{Ossa}-\textsc{noW}        & $+0.60 \pm 0.05$ & $+0.13 \pm 0.02$ & $+0.09 \pm 0.02$ & $-0.03 \pm 0.01$ \\
\textsc{SC-Degree}       & $+0.69 \pm 0.06$ & $+0.02 \pm 0.03$ & $-0.01 \pm 0.01$ & $-0.02 \pm 0.02$ \\
\textsc{Joint-SCxFC}     & $+0.67 \pm 0.06$ & $-0.06 \pm 0.03$ & $-0.03 \pm 0.02$ & $-0.07 \pm 0.01$ \\
\textsc{Communicability} & $+0.59 \pm 0.05$ & $-0.04 \pm 0.01$ & $+0.00 \pm 0.01$ & $-0.04 \pm 0.01$ \\
\textsc{FC-Degree}       & $+0.06 \pm 0.09$ & $-0.23 \pm 0.05$ & $-0.21 \pm 0.02$ & $-0.14 \pm 0.02$ \\
\textsc{BrainGNN}        & $+0.42 \pm 0.09$ & $+0.03 \pm 0.03$ & $+0.01 \pm 0.04$ & $+0.03 \pm 0.03$ \\
\textsc{BrainNetCNN}     & $+0.28 \pm 0.08$ & $-0.08 \pm 0.05$ & $+0.04 \pm 0.04$ & $+0.00 \pm 0.03$ \\
\textsc{SDI}             & $+0.38 \pm 0.08$ & $-0.01 \pm 0.03$ & $-0.11 \pm 0.02$ & $-0.06 \pm 0.02$ \\
\textsc{Random}          & $+0.38 \pm 0.09$ & $+0.01 \pm 0.06$ & $-0.06 \pm 0.04$ & $-0.02 \pm 0.03$ \\
\bottomrule
\end{tabular}
\end{table}

\subsection{ABCD baseline $W$-fitting protocol}
\label{app:abcd-baseline-W}
For each baseline support $\widehat{\mathcal V}^m_q$, the per-method alignment $W^m_q$ is obtained by running the \textsc{Ossa} $W$-block of Algorithm~\ref{alg:single-ossa} on training subjects with $s = \one_{\widehat{\mathcal V}^m_q}$ held fixed, using three initializations (including the identity), up to $200$ $W$-update iterations per initialization, and monotone backtracking. The run with the largest training objective is retained. At $\kappa=3$, the $M_{3,\mathrm{aln}}$ random-null $W_r$ is fit using a lighter schedule of up to $50$ iterations from a single random orthogonal initialization to keep the $1000$-draw null tractable per (seed, community) unit. The deep-learning baselines (\textsc{BrainGNN}, \textsc{BrainNetCNN}) are fit on a $3 \times 3$ (learning rate, weight decay) grid selected by validation loss. 

\paragraph{Deep-learning baseline details.}
\textsc{BrainGNN} and \textsc{BrainNetCNN} are reimplemented in NumPy with
hand-derived gradients and trained per (split-seed, community, $\kappa$)
on the same train/val/test split as \textsc{Ossa}.

\begin{itemize}
\item \textbf{Input modality.} \textsc{BrainGNN}: per-community training
  mean $\bar D=\tfrac1n\sum_i A_i\!\circ\!B_i$. \textsc{BrainNetCNN}: per-subject SC submatrix $B_i$.
\item \textbf{Prediction target.} \textsc{BrainGNN}: maximize attention-pooled
  alignment energy $a^\top\!\bar D\,a$ ($a=\sigma(\theta)$). \textsc{BrainNetCNN}:
  regress the $z$-scored scalar $y_i=\langle A_i,B_i\rangle_F$.
\item \textbf{Loss.} \textsc{BrainGNN}: $-a^\top\!\bar D_{\text{tr}}a$.
  \textsc{BrainNetCNN}: mean squared error.
\item \textbf{Architecture.} \textsc{BrainGNN}: attention vector
  $\theta\in\mathbb R^p$, $a=\sigma(\theta)$, init $N(0,0.1^2)$.
  \textsc{BrainNetCNN}: 3-stage E2E$\to$E2N$\to$N2G with
  $W_{\text{row}},W_{\text{col}},W_2\!\in\!\mathbb R^{p\times p}$,
  $W_3\!\in\!\mathbb R^{p}$, init i.i.d.\ $N(0,1/p)$.
\item \textbf{Optimizer.} Adam ($\beta_1{=}0.9$, $\beta_2{=}0.999$,
  $\varepsilon{=}10^{-8}$), full-batch over the community.
\item \textbf{Learning-rate grid.} $\eta\in\{10^{-2},10^{-3},10^{-4}\}$.
\item \textbf{Weight-decay grid.} $\lambda_{\text{wd}}\in\{0,10^{-4},10^{-3}\}$.
\item \textbf{Batch size.} Full batch ($n_{\text{tr}}$ subjects).
\item \textbf{Epochs.} \textsc{BrainGNN}: $200$.
  \textsc{BrainNetCNN}: $20$.
\item \textbf{Restarts and early stopping.} $16$ restarts per
  (seed,$q$,$\kappa$); each draws $(\eta,\lambda_{\text{wd}})$ from the
  $3\!\times\!3$ grid by a seeded RNG. No within-restart early stopping. The best restart is selected on validation.
\item \textbf{Validation criterion.} \textsc{BrainGNN}:
  $-a^\top\!\bar D_{\text{val}}a$. \textsc{BrainNetCNN}: validation MSE.
  
\item \textbf{Saliency.} \textsc{BrainGNN}: $a_v=\sigma(\theta_v)$.
  \textsc{BrainNetCNN}: closed-form gating gradient
  $\mathrm{sal}_v=\tfrac1n\sum_i|(W_3^\top h^{(2)}_i-y_i)(W_3)_v h^{(2)}_{i,v}|$
  at $m\!=\!\mathbf 1$.
\item \textbf{Saliency $\to$ $\kappa$-node support.}
  $\widehat{\mathcal V}=\mathrm{Top}_\kappa(\cdot)$, ties broken by
  ascending node index. The hard support is then passed through the same
  $W$-refit and held-out metric pipeline as every other baseline.
\end{itemize}

\begin{table}[htbp]
\centering
\caption{Additional size-matched baseline comparisons on ABCD. Held-out $M_{1,\mathrm{aln}}$ is reported as mean $\pm$ s.e.\ across $30$ seed-community units. W/T/L denotes \textsc{Ossa} wins, ties, and losses against each baseline. These counts are descriptive because repeated splits reuse subjects.}
\label{tab:abcd-additional-baselines}
\tablesetup
\begin{tabular}{lcc}
\toprule
Method & $M_{1,\mathrm{aln}}$ & \textsc{Ossa} W/T/L \\
\midrule
\textsc{Ossa} & $\mathbf{10.55 \pm 0.85}$ & --- \\
NMF & $9.96 \pm 0.82$ & $28/1/1$ \\
Sparse PLS & $9.22 \pm 1.10$ & $17/11/2$ \\
Path efficiency & $8.98 \pm 0.71$ & $20/10/0$ \\
Mean first passage time & $8.32 \pm 0.60$ & $29/0/1$ \\
FAQ graph matching & $7.57 \pm 0.70$ & $26/4/0$ \\
SC-to-FC ridge prediction & $4.06 \pm 0.49$ & $30/0/0$ \\
Random & $0.18 \pm 0.04$ & $30/0/0$ \\
\bottomrule
\end{tabular}
\end{table}

\begin{table}[htbp]
\centering
\caption{Subject-level paired bootstrap comparisons on held-out subjects from one split. $\Delta M_{1,\mathrm{aln}}$ denotes \textsc{Ossa} minus the comparator. The 95\% confidence intervals use $10{,}000$ bootstrap replicates with the same resampled subject indices for all methods. Training-fitted supports and the community partition are held fixed, so the intervals exclude support-selection variability.}
\label{tab:abcd-subject-bootstrap}
\tablesetup
\begin{tabular}{lcc}
\toprule
Comparator & Mean $\Delta M_{1,\mathrm{aln}}$ & 95\% CI \\
\midrule
NMF & $+0.468$ & $[0.400,0.534]$ \\
\textsc{Communicability} & $+0.802$ & $[0.715,0.893]$ \\
\textsc{SC-Degree} & $+1.803$ & $[1.716,1.892]$ \\
\textsc{Joint-SCxFC} & $+1.874$ & $[1.785,1.967]$ \\
\textsc{FC-Degree} & $+2.829$ & $[2.748,2.911]$ \\
\textsc{Random} & $+7.427$ & $[7.292,7.568]$ \\
\bottomrule
\end{tabular}
\end{table}

\subsection{ABCD multi-clique $\lambda$-sensitivity}
\label{app:abcd-multi}

\paragraph{Diagnostic criterion.}
The \(K\)-support fit is not used as the default estimator in ABCD; it is a diagnostic for whether one FC community contains several dominant aligned structural cores. Since the \textsc{Ossa} objective is a trace alignment criterion, the primary held-out diagnostic is \(M_{1,\mathrm{aln}}\), evaluated on each fitted \(\kappa\)-node support after the same training-side \(W\)-refit used in Table~3. The correlation metrics \(M_{2,\mathrm{aln}}\) and \(M_{3,\mathrm{aln}}\) gain are retained as scale-free diagnostics of aligned FC--SC patterns after rotation, with degree-matched correction applied only to \(M_{3,\mathrm{aln}}\) gain; they complement but do not replace the objective optimized by \textsc{Ossa}.

\begin{figure}[t]
\centering
\includegraphics[width=\textwidth]{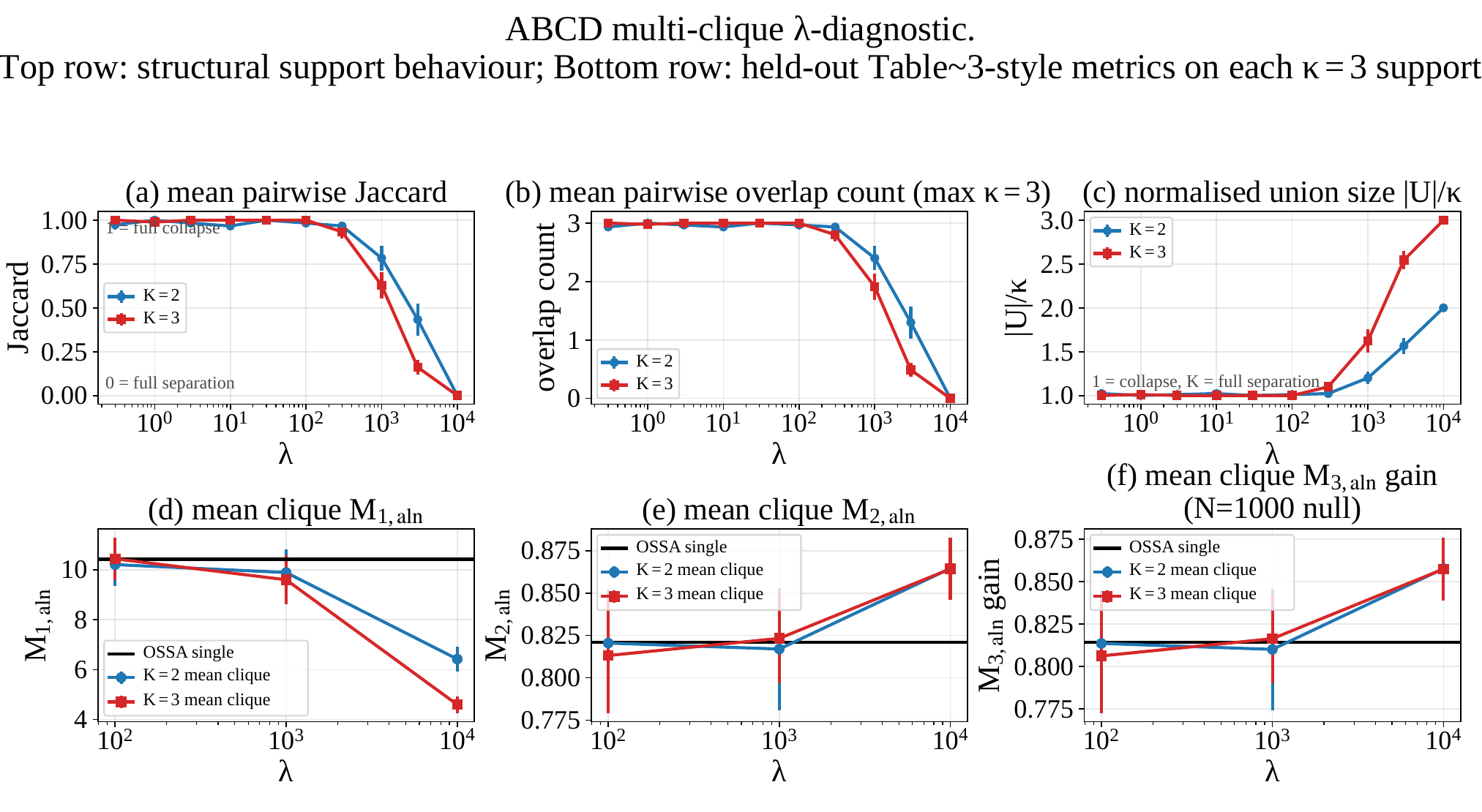}
\caption{
ABCD multi-clique \(\lambda\)-diagnostic. 
Top row: support geometry as \(\lambda\) varies. At small \(\lambda\), the \(K\) supports collapse onto the same \(\kappa\)-node set. At large \(\lambda\), the overlap penalty forces separation. 
Bottom row: held-out metrics evaluated on each \(\kappa=3\) support after the Table 3-style training-side \(W\)-refit. 
\(M_{1,\mathrm{aln}}\) is the primary held-out \textsc{Ossa} objective and decreases after forced separation. 
\(M_{2,\mathrm{aln}}\) and \(M_{3,\mathrm{aln}}\) gain are scale-free correlation diagnostics and increase at large \(\lambda\), indicating secondary correlation-structured motifs rather than multiple dominant high-\(M_1\) alignment cores.
}
\label{fig:abcd_multiclique_lambda_singlecore}
\end{figure}

\paragraph{Support geometry.}
Figure~\ref{fig:abcd_multiclique_lambda_singlecore} shows a collapse-to-separation transition. For \(\lambda \le 10^2\), the fitted supports nearly coincide: for \(K=2\), the mean pairwise Jaccard is \(0.983\) and \(|U|/\kappa=1.011\); for \(K=3\), the supports fully collapse with Jaccard \(1.000\) and \(|U|/\kappa=1.000\). Increasing \(\lambda\) first produces partial separation and then full disjointness. At \(\lambda=10^4\), the pairwise Jaccard is \(0\), with \(|U|/\kappa=2.000\) for \(K=2\) and \(3.000\) for \(K=3\). Thus the optimizer can separate the supports when forced, but the path does not reveal a stable intermediate multi-core regime.

\paragraph{Held-out metrics.}
The held-out metrics separate the primary objective from secondary correlation structure. At \(\lambda=10^2\), where the supports collapse, the mean clique-level \(M_{1,\mathrm{aln}}\) is essentially the canonical \textsc{Ossa} value: \(10.20 \pm 0.85\) for \(K=2\) and \(10.43 \pm 0.85\) for \(K=3\), compared with \(10.42 \pm 0.80\) for canonical \textsc{Ossa}. Once separation is forced, \(M_{1,\mathrm{aln}}\) decreases: at \(\lambda=10^4\), it is \(6.42 \pm 0.51\) for \(K=2\) and \(4.60 \pm 0.34\) for \(K=3\). In contrast, \(M_{2,\mathrm{aln}}\) and \(M_{3,\mathrm{aln}}\) gain rise at \(\lambda=10^4\), reaching approximately \(0.86\). We interpret this as evidence for secondary scale-free FC-SC correlation motifs, not as evidence for several dominant \textsc{Ossa} cores, because these supports no longer remain competitive under the primary alignment-magnitude objective. Therefore, under the community construction and \(\lambda\) grid used here, ABCD supports a single dominant high-\(M_1\) aligned core at \(\kappa=3\), with multi-clique \textsc{Ossa} serving as a diagnostic extension.

\subsection{Multi-support extensions of ABCD baselines}
\label{app:abcd-multisupport-baselines}

\paragraph{Protocol.}
This auxiliary analysis checks whether the single-core conclusion is an artefact of comparing the \(K\)-support \textsc{Ossa} diagnostic only with single-support baselines. We extend the non-neural support selectors in Table~\ref{tab:abcd-main} to \(K\in\{2,3\}\) supports at the same budget \(\kappa=3\). For ranking-based selectors, the \(K\) supports are consecutive disjoint \(\kappa\)-node blocks from the training-only node ranking. Thus \(V^{(1)}\) contains the top \(\kappa\) nodes, \(V^{(2)}\) contains the next \(\kappa\) nodes after removing \(V^{(1)}\), and so on. \textsc{Random}-\(K\) samples \(K\) disjoint \(\kappa\)-node blocks uniformly within each community and averages over \(R=200\) draws per seed-community unit. \textsc{Ossa}-\(K\) and \textsc{Ossa}-noW-\(K\) are evaluated on the same overlap-penalty grid \(\lambda\in\{10^2,10^3,10^4\}\). Each fitted \(\kappa\)-node support is then evaluated by the same fixed-support training-side \(W\)-refit and held-out metrics used in Table~\ref{tab:abcd-main}. We report mean-clique metrics, averaging over the \(K\) supports within each seed-community unit, so that a larger \(K\) is judged by per-support yield rather than by a mechanically larger union.

\begin{figure}[t]
\centering
\includegraphics[width=\linewidth]{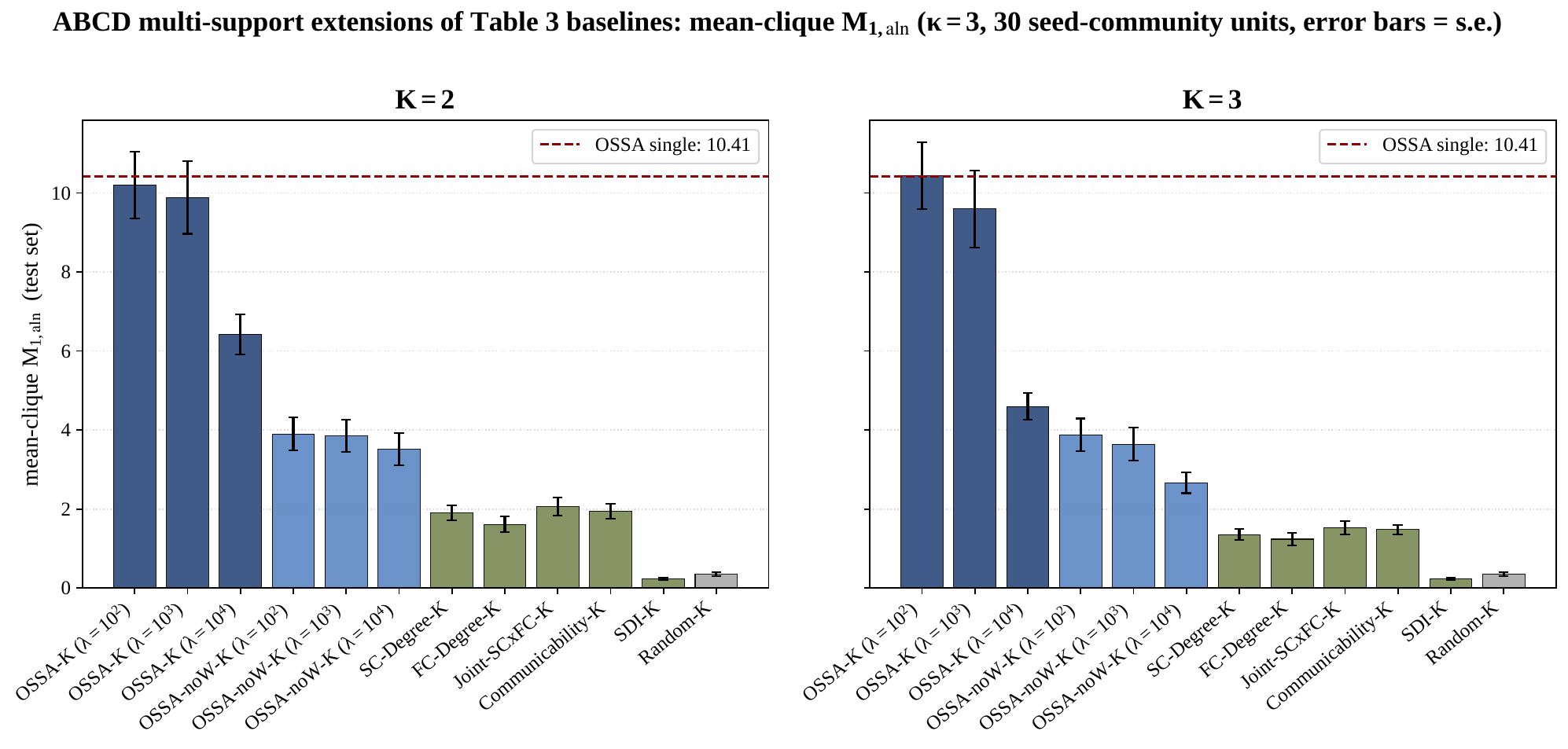}
\caption{
ABCD multi-support extensions of the Table~\ref{tab:abcd-main} baselines, evaluated by the primary held-out objective \(M_{1,\mathrm{aln}}\). Bars report mean-clique \(M_{1,\mathrm{aln}}\), averaging over the \(K\) fitted \(\kappa=3\) supports within each seed-community unit; error bars are standard errors over the 30 seed-community units. The dashed line shows the single-support \textsc{Ossa} reference used in this multi-support diagnostic. \textsc{Ossa}-\(K\) remains close to this value only when supports collapse or partially overlap; forced separation lowers the primary alignment magnitude. Ranking-based multi-support extensions are substantially lower.
}
\label{fig:abcd-multisupport-m1}
\end{figure}

\paragraph{Primary held-out alignment.}
Figure~\ref{fig:abcd-multisupport-m1} shows that extending the baselines to several supports does not produce multiple competitive high-\(M_1\) cliques. \textsc{Ossa}-\(K\) remains close to the canonical single-support value only in collapsed or partially collapsed regimes. At \(\lambda=10^2\), the \(K=2\) fit has mean-clique \(M_{1,\mathrm{aln}}=10.20\pm0.85\), but its supports nearly coincide (\(|U|/\kappa=1.01\), mean pairwise Jaccard \(0.98\)); the \(K=3\) fit fully collapses and gives \(M_{1,\mathrm{aln}}=10.43\pm0.85\). When the overlap penalty forces disjoint supports at \(\lambda=10^4\), mean-clique \(M_{1,\mathrm{aln}}\) drops to \(6.42\pm0.51\) for \(K=2\) and \(4.60\pm0.34\) for \(K=3\). The ranking-based \(K\)-support baselines are lower throughout: their best mean-clique \(M_{1,\mathrm{aln}}\) is \(2.06\pm0.23\) for \(K=2\) and \(1.53\pm0.17\) for \(K=3\). Thus forcing multiple disjoint supports does not appear beneficial as a dimension-reduction step in ABCD.

\begin{figure}[t]
\centering
\includegraphics[width=\linewidth]{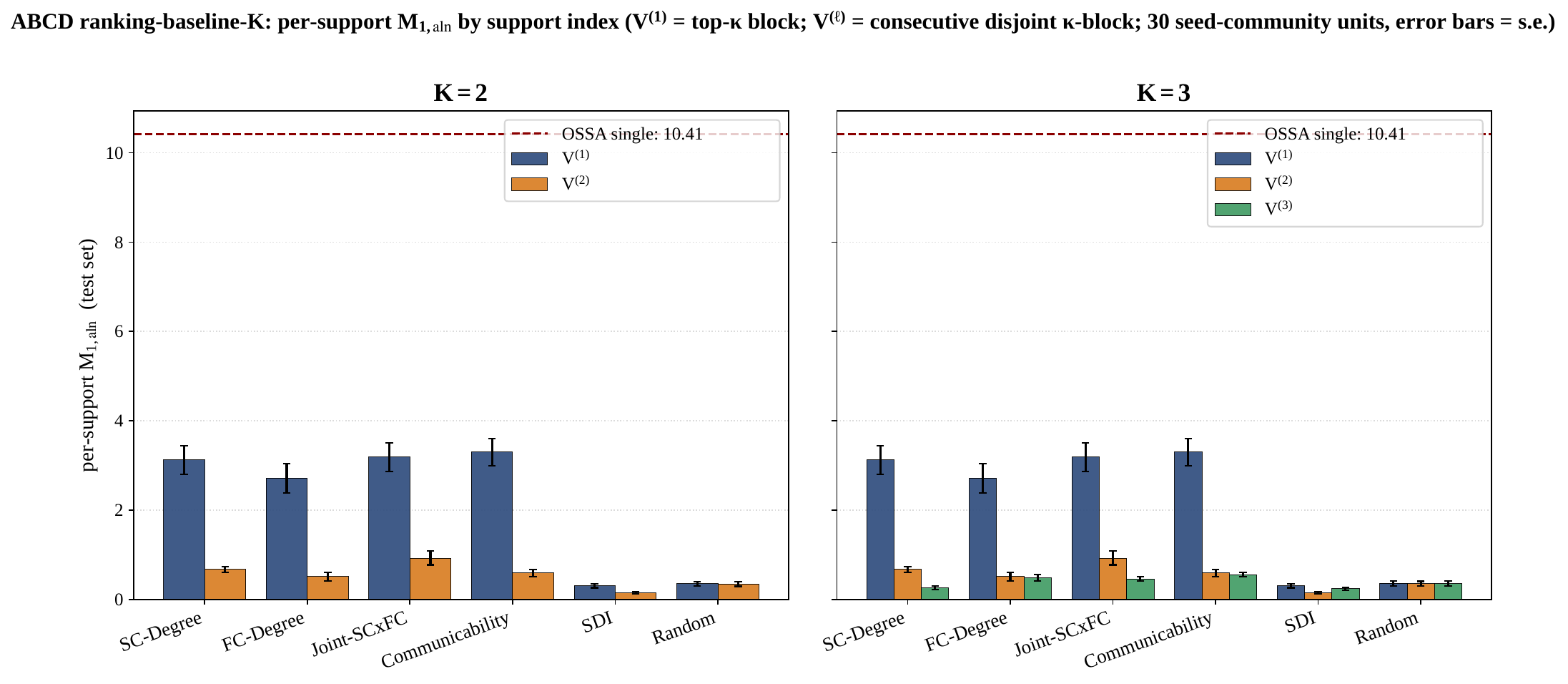}
\caption{
Per-support degradation for ranking-based baseline extensions with \(K\in\{2,3\}\). Each method forms \(K\) consecutive disjoint \(\kappa=3\) blocks from its training-only ranking. \(V^{(1)}\) corresponds to the original single-support baseline; for \textsc{SC-Degree}, \textsc{FC-Degree}, \textsc{Joint-SCxFC} and \textsc{Communicability}, the later supports have much smaller held-out \(M_{1,\mathrm{aln}}\). The dashed line shows the single-support \textsc{Ossa} reference used in this multi-support diagnostic.
}
\label{fig:abcd-multisupport-by-support}
\end{figure}

\paragraph{Support-index degradation.}
For ranking-based extensions, the support index makes the same pattern more explicit. Figure~\ref{fig:abcd-multisupport-by-support} shows that, for \textsc{SC-Degree}, \textsc{FC-Degree}, \textsc{Joint-SCxFC} and \textsc{Communicability}, \(V^{(1)}\), the first \(\kappa\)-node block, corresponds to the original single-support baseline, whereas the later supports have much smaller held-out \(M_{1,\mathrm{aln}}\). This degradation is expected when the FC-derived communities are already small: after the first high-signal support has been removed, the remaining disjoint blocks do not provide comparable high-\(M_1\) structural cores.

\paragraph{Scale-free diagnostics.}
The scale-free diagnostics tell a complementary story rather than changing the primary conclusion. As in Appendix~\ref{app:abcd-multi}, high-\(\lambda\) separated \textsc{Ossa}-\(K\) fits can increase \(M_{2,\mathrm{aln}}\) or \(M_{3,\mathrm{aln}}\) gain, but these metrics measure within-support FC-SC correlation after scale normalization. They do not replace the trace-alignment magnitude optimized by \textsc{Ossa}. We therefore interpret the high-\(\lambda\) scale-free gains as secondary correlation-structured motifs, not as evidence for several dominant high-\(M_1\) aligned cores.

\subsection{Structural connectivity within consensus supports}
\label{app:abcd-consensus-sc}

Table~\ref{tab:abcd-consensus-sc} in the main text reports the subject-averaged SC edge weights and their degree-matched reference ratios. Each consensus support forms a within-lobe triangle. The temporal and occipital supports lie in the right hemisphere, whereas the frontal support includes the left medial orbitofrontal region and bilateral frontal poles.

\begingroup
\raggedbottom
\subsection{Empirical diagnostics of the fitted alignment}
\label{app:abcd-alignment-diagnostics}

At $\kappa=3$, the normalized squared projector distance $\|\widehat W P_{\widehat V}\widehat W^\top-P_{\widehat V}\|_F^2/(2\kappa)$ has community-wise means ranging from $0.54$ to $0.67$ over ten split seeds. Reference means based on $1{,}000$ Haar-random orthogonal matrices of each community's dimension range from $0.80$ to $0.90$. All $30$ seed--community units lie below the $2.5$th percentile of their corresponding Haar reference distribution.

For regional loadings $\ell_a=\sum_{j\in\widehat V}\widehat W_{aj}^{\,2}$, we quantify concentration using the participation ratio $\mathrm{PR}=(\sum_a\ell_a)^2/\sum_a\ell_a^2$, which measures the effective number of contributing regions. Community-wise mean PR values over ten split seeds are approximately $5$--$10$, compared with $10$--$19$ under the Haar reference. The corresponding mean peak regional loadings, $\max_a\ell_a$, are approximately $0.59$--$0.69$.

We measured the fraction of $\|\widehat W P_{\widehat V}\widehat W^\top\|_F^2$ contributed by matrix entries corresponding to contralateral homologous regions. Across the three communities, its means over ten split seeds are approximately $3.2$--$8.9$ times the corresponding Haar-reference means.

For each community, we compared the fitted projectors $\widehat W P_{\widehat V}\widehat W^\top$ across all $45$ pairs of the ten split seeds using squared Frobenius distance divided by $2\kappa$. Mean distances were $0.0002$, $0.113$ and $0.0001$ for the frontal, temporal and occipital communities, respectively. The observed stability is consistent with reproducibility of the spectral structure of mean training FC and does not by itself establish biological specificity.

\par\newpage
\endgroup

\subsection{Sensitivity to community construction}
\label{app:abcd-community-sensitivity}

\begin{table}[H]
\centering
\caption{Sensitivity to community construction using the same training mean $|FC|$. Node recovery summarizes retention of the original nine selected regions under each construction. Held-out $M_1$ values are not directly comparable across $Q$, because community size changes the number of candidate edges.}
\label{tab:abcd-community-sensitivity}
\tablesetup
\begin{tabular}{lccc}
\toprule
Community construction & Sizes & Node recovery & Held-out $M_1$ \\
\midrule
$Q=2$ & $52,\ 16$ & $0.50$ & $13.83$ \\
$Q=3$, same partition refit (control) & $30,\ 23,\ 15$ & $0.89$ & $10.30$ \\
$Q=4$ & $17,\ 18,\ 14,\ 19$ & $0.56$ & $7.74$ \\
Spectral & $22,\ 15,\ 31$ & $0.61$ & $10.71$ \\
Ward hierarchical & $6,\ 20,\ 42$ & $0.61$ & $10.04$ \\
Louvain & $24,\ 22,\ 22$ & $0.67$ & $8.85$ \\
Newman leading eigenvector & $28,\ 17,\ 23$ & $0.44$ & $9.66$ \\
\bottomrule
\end{tabular}
\end{table}

\section{ABCD second-visit replication details}
\label{app:abcd-secondvisit}

This appendix documents the second-visit replication summarised in
Section~\ref{subsec:real} on the ABCD Year-2 follow-up (``second-visit'')
release. The protocol is identical to the baseline (Year-1) evaluation: no
hyperparameter is retuned, no seed is reshuffled, and the same per-method
$W$-fit budget, multi-clique grid, and aggregation rules are reused.

\paragraph{Data and pipeline.}
The ABCD second-visit release contributes $1{,}909$ subjects parcellated by
the same 68-region Desikan-Killiany atlas as the baseline \citep{desikan2006automated}. The canonical
preprocessing pipeline (transposition to $(N, p, p)$, exact symmetrization,
diagonal enforcement, Fisher-$z$ then non-negative clipping for FC, and
$\log(1+\cdot)$ for SC) and the cohort-median $5\sigma$ QC rule are applied
without modification. QC drops $12$ subjects (none with simultaneous FC and
SC flags) and leaves $n_{\mathrm{ABCD2}} = 1{,}897$ subjects. The
$60/20/20$ deterministic splits over the same $10$ seeds yield
$n_{\mathrm{train}} = 1{,}138$ and
$n_{\mathrm{val}} = n_{\mathrm{test}} \in \{379, 380\}$ per seed. The
one-time test-evaluation rule of the baseline protocol is reused: each
$(\text{method}, \text{metric}, \text{split})$ tuple is touched exactly once
on the second-visit cohort.

\begin{figure}[t]
  \centering
  \includegraphics[width=\linewidth]{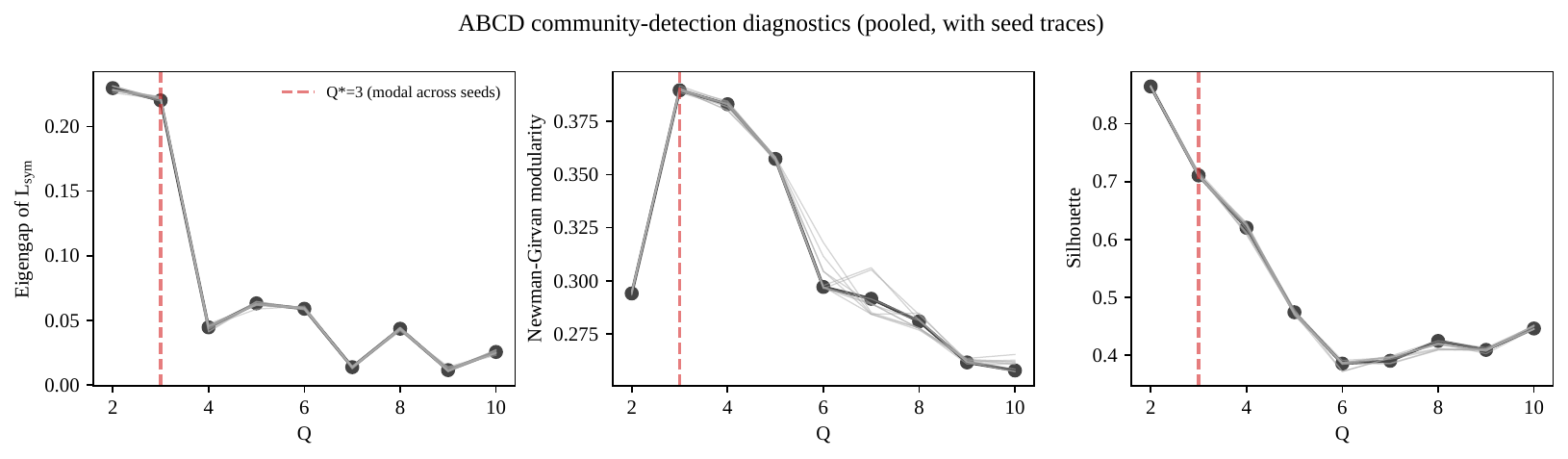}
  \caption{ABCD second-visit community-detection diagnostics. Light gray
  traces are per-seed; the dark line is the pooled training average. The
  eigengap of $L_{\mathrm{sym}}$, Newman-Girvan modularity, and silhouette
  all peak at $Q^*=3$, with no seed disagreeing, identical to the
  baseline cohort.}
  \label{fig:abcd2-community}
\end{figure}

\paragraph{Community partition.}
Spectral diagnostics on the training absolute-FC mean again select
$Q^*=3$ in every one of the $10$ split seeds. The pooled training average
likewise yields $Q_{\mathrm{pool}}=3$. Cross-seed ARI on the modal-$Q$
partitions is $0.99 \pm 0.01$ (min $0.96$ across the 45 seed pairs),
indicating near-perfect agreement up to label permutation
(Figure~\ref{fig:abcd2-community}). The pooled community sizes $(31, 23, 14)$
are within one node of the baseline pooled sizes $(30, 23, 15)$, supporting
that the same three-community organisation is recovered at the second
visit.

\begin{figure}[t]
  \centering
  \includegraphics[width=\linewidth]{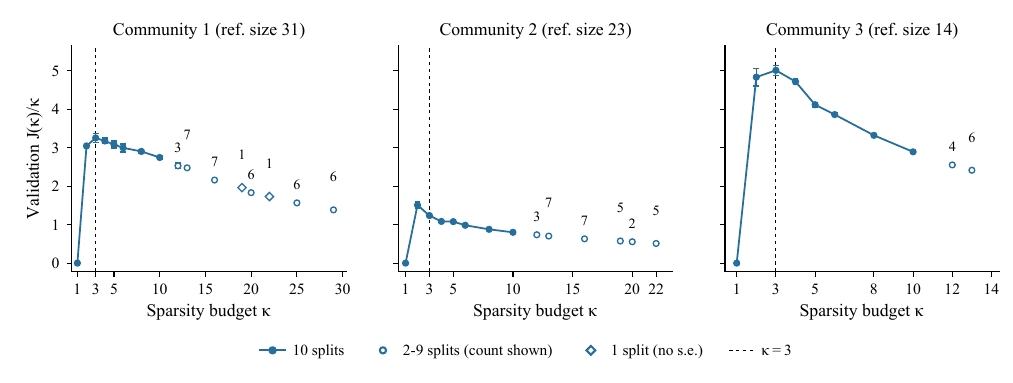}
  \caption{ABCD second-visit sparsity sweep after matching communities across splits to the pooled partition by maximum node overlap. Points show means and error bars show s.e.; counts below 10 are labeled, and single-split results have no error bar. Larger budgets may use different split subsets. The \(\kappa=1\) point is a trivial zero-objective floor. For tested budgets \(\kappa\ge3\), mean validation \(J(\kappa)/\kappa\) decreases in all three communities.}
  \label{fig:abcd2-kappa}
\end{figure}

\paragraph{Per-community budget.}
The per-community budget $\kappa = 3$ from the baseline cohort is reused without retuning. Figure~\ref{fig:abcd2-kappa} summarizes validation $J(\kappa)/\kappa$ after matching communities across split seeds to the pooled reference partition by maximum node overlap. Among the tested budgets $\kappa\geq3$, the mean is highest at $\kappa=3$ in each community, consistent with the baseline results.

\paragraph{Canonical \textsc{Ossa} evaluation.}
The summary appears in Section~\ref{subsec:real} (``Second-visit replication''); this appendix provides per-method numbers and per-community decomposition (Table~\ref{tab:abcd-secondvisit-main}). At $\kappa = 3$ over $30$ (seed, community) units, \textsc{Ossa} attains $M_{1,\mathrm{aln}} = 10.16 \pm 0.87$, with the ablation \textsc{Ossa}-\textsc{noW} ($3.23 \pm 0.30$) and the classical baseline \textsc{SC-Degree} ($3.02 \pm 0.31$) clustering at roughly one third of \textsc{Ossa}'s value. The \(M_{2,\mathrm{raw}}\to M_{2,\mathrm{aln}}\) shift again shows that high-degree or SC-summary baselines can look competitive in raw coordinates (e.g., \textsc{Communicability}: \(+0.672\), \textsc{SC-Degree}: \(+0.652\)), whereas \textsc{Ossa} leads after rotation-aware evaluation (\(+0.811\) vs. next-best \(+0.672\)). The same conclusion holds on the SC-degree-quartile-matched null gain $M_{3,\mathrm{aln}}$ ($+0.81$ vs.\ next-best $+0.67$). After matching each split's community labels to the pooled partition by maximum node overlap, \textsc{Ossa} leads on \(M_{1,\mathrm{aln}}\) in all three communities. For pooled community sizes \((31,23,14)\), its scores are \(10.17\pm0.26\), \(4.32\pm0.04\), and \(15.98\pm0.07\), respectively (mean \(\pm\) s.e. across ten splits), versus highest baseline means of \(3.66\), \(1.11\), and \(5.00\).

\begin{table}[t]
\centering
\caption{ABCD second-visit canonical \textsc{Ossa} evaluation, mean $\pm$ s.e.\ across $30$ (seed, community) units, $\kappa = 3$, on $n_{\mathrm{ABCD2}} = 1{,}897$ subjects. Conventions match Table~\ref{tab:abcd-main}: $M_{3,\mathrm{aln}}$ gain is the difference between the method's $M_{2,\mathrm{aln}}$ and the mean of $N = 1000$ SC-degree-quartile-matched random nulls; $M_3\,p$ is mean (median) over the $30$ units. The headline summary appears in Section~\ref{subsec:real} (``Second-visit replication''), with this appendix providing per-method detail.}
\label{tab:abcd-secondvisit-main}
\tablesetup
\begin{tabular}{lccccc}
\toprule
Method & $M_{1,\mathrm{aln}}$ & $M_{2,\mathrm{raw}}$ & $M_{2,\mathrm{aln}}$ &
$M_{3,\mathrm{aln}}$ gain & $M_3\,p$ \\
\midrule
\textsc{Ossa}     & $\mathbf{10.16 \pm 0.87}$ & $+0.494$ & $\mathbf{+0.811}$ & $\mathbf{+0.81}$ & $0.073\,(0.042)$ \\
\textsc{Ossa}-\textsc{noW}        & $3.23 \pm 0.30$  & $+0.514$ & $+0.539$ & $+0.54$ & $0.211\,(0.166)$ \\
\textsc{Communicability} & $3.19 \pm 0.28$  & $\mathbf{+0.672}$ & $+0.672$ & $+0.67$ & $0.145\,(0.083)$ \\
\textsc{Joint-SCxFC}     & $3.11 \pm 0.31$  & $+0.623$ & $+0.623$ & $+0.62$ & $0.173\,(0.084)$ \\
\textsc{SC-Degree}       & $3.02 \pm 0.31$  & $+0.652$ & $+0.652$ & $+0.65$ & $0.158\,(0.044)$ \\
\textsc{FC-Degree}       & $2.27 \pm 0.31$  & $-0.124$ & $-0.124$ & $-0.13$ & $0.549\,(0.633)$ \\
\textsc{BrainGNN}        & $0.64 \pm 0.17$  & $+0.285$ & $+0.324$ & $+0.32$ & $0.324\,(0.291)$ \\
\textsc{Random}          & $0.51 \pm 0.14$  & $+0.260$ & $+0.248$ & $+0.25$ & $0.370\,(0.406)$ \\
\textsc{BrainNetCNN}     & $0.37 \pm 0.12$  & $+0.328$ & $+0.432$ & $+0.43$ & $0.275\,(0.207)$ \\
\textsc{SDI}             & $0.07 \pm 0.02$  & $+0.257$ & $+0.322$ & $+0.32$ & $0.322\,(0.315)$ \\
\bottomrule
\end{tabular}
\end{table}

\paragraph{Cross-visit support replication}

To evaluate whether the selected structural supports remain stable across longitudinal
neurodevelopment, we compare the canonical $\kappa=3$ supports identified from the
baseline and Year-2 ABCD releases. For each split seed and matched FC community,
we compute a cross-visit support replication score
\[
\mathrm{CVSR}
=
\frac{
|\widehat V_{\mathrm{base}}
\cap
\widehat V_{\mathrm{Y2}}|
}{
|\widehat V_{\mathrm{base}}
\cup
\widehat V_{\mathrm{Y2}}|
},
\]
corresponding to the Jaccard overlap between the baseline and second-visit supports.
We additionally report the overlap count, overlap fraction, exact replication rate,
and a degree-matched null-corrected replication score (nCVSR). Community labels
between visits are aligned by maximum-overlap matching prior to evaluation.

Table~\ref{tab:abcd-crossvisit} shows that OSSA remains highly stable across visits, with
CVSR $=0.928\pm0.033$ and mean overlap $2.86/3$ nodes. Importantly,
this stability is achieved simultaneously with the substantially larger held-out
alignment scores reported in Tables~\ref{tab:abcd-main} and~\ref{tab:abcd-secondvisit-main}. Several graph-summary baselines,
including \textsc{SC-Degree}, \textsc{FC-Degree}, \textsc{Communicability}, and \textsc{Joint-SCxFC},
also exhibit high cross-visit support consistency, suggesting that portions of the
dominant structural organization are preserved longitudinally within FC communities.
In contrast, the deep-learning baselines exhibit substantially weaker support
replication and near-null corrected stability.

\begin{table}[t]
\centering
\caption{
Cross-visit support replication between the ABCD baseline and Year-2 follow-up releases
for canonical $\kappa=3$ supports. Results are averaged over 30 (seed, community) units.
CVSR denotes the support-set Jaccard overlap between visits. nCVSR is the null-corrected
replication score relative to size-matched or SC-degree-quartile-matched random supports.
}
\label{tab:abcd-crossvisit}
\resizebox{\textwidth}{!}{
\begin{tabular}{lcccccccc}
\toprule
Method
& Overlap
& Fraction
& CVSR
& \multicolumn{2}{c}{nCVSR}
& Exact
& $\geq 2$
& Null $p$ \\
\cmidrule(lr){5-6}
& count
& ($/\kappa$)
& (Jaccard)
& size-matched
& degree-matched
& rate
& rate
& median \\
\midrule
OSSA
& 2.86$\pm$0.08
& 0.95$\pm$0.03
& 0.928$\pm$0.033
& 0.921$\pm$0.034
& 0.935$\pm$0.031
& 0.85
& 1.00
& 0.022 \\
JOINT-SCxFC
& 3.00$\pm$0.00
& 1.00$\pm$0.00
& 1.000$\pm$0.000
& 1.000$\pm$0.000
& 1.000$\pm$0.000
& 1.00
& 1.00
& 0.019 \\
FC-DEGREE
& 2.87$\pm$0.06
& 0.96$\pm$0.02
& 0.933$\pm$0.032
& 0.929$\pm$0.034
& 0.910$\pm$0.043
& 0.87
& 1.00
& 0.019 \\
COMMUNICABILITY
& 2.83$\pm$0.07
& 0.94$\pm$0.02
& 0.917$\pm$0.035
& 0.903$\pm$0.040
& 0.877$\pm$0.051
& 0.83
& 1.00
& 0.017 \\
SC-DEGREE
& 2.83$\pm$0.07
& 0.94$\pm$0.02
& 0.917$\pm$0.035
& 0.909$\pm$0.038
& 0.893$\pm$0.044
& 0.83
& 1.00
& 0.035 \\
RANDOM
& 1.73$\pm$0.20
& 0.58$\pm$0.07
& 0.503$\pm$0.072
& 0.455$\pm$0.079
& 0.430$\pm$0.080
& 0.33
& 0.57
& 0.205 \\
SDI
& 1.73$\pm$0.11
& 0.58$\pm$0.04
& 0.433$\pm$0.038
& 0.370$\pm$0.045
& 0.370$\pm$0.040
& 0.07
& 0.67
& 0.058 \\
OSSA-NOW
& 1.50$\pm$0.21
& 0.50$\pm$0.07
& 0.430$\pm$0.073
& 0.361$\pm$0.082
& 0.349$\pm$0.084
& 0.30
& 0.40
& 0.498 \\
BRAINGNN
& 0.97$\pm$0.21
& 0.32$\pm$0.07
& 0.270$\pm$0.067
& 0.194$\pm$0.073
& 0.201$\pm$0.072
& 0.17
& 0.27
& 0.813 \\
BRAINNETCNN
& 0.47$\pm$0.12
& 0.16$\pm$0.04
& 0.103$\pm$0.029
& 0.015$\pm$0.031
& 0.020$\pm$0.030
& 0.00
& 0.10
& 1.000 \\
\bottomrule
\end{tabular}
}
\end{table}

\begin{figure}[t]
\centering
\includegraphics[width=\linewidth]{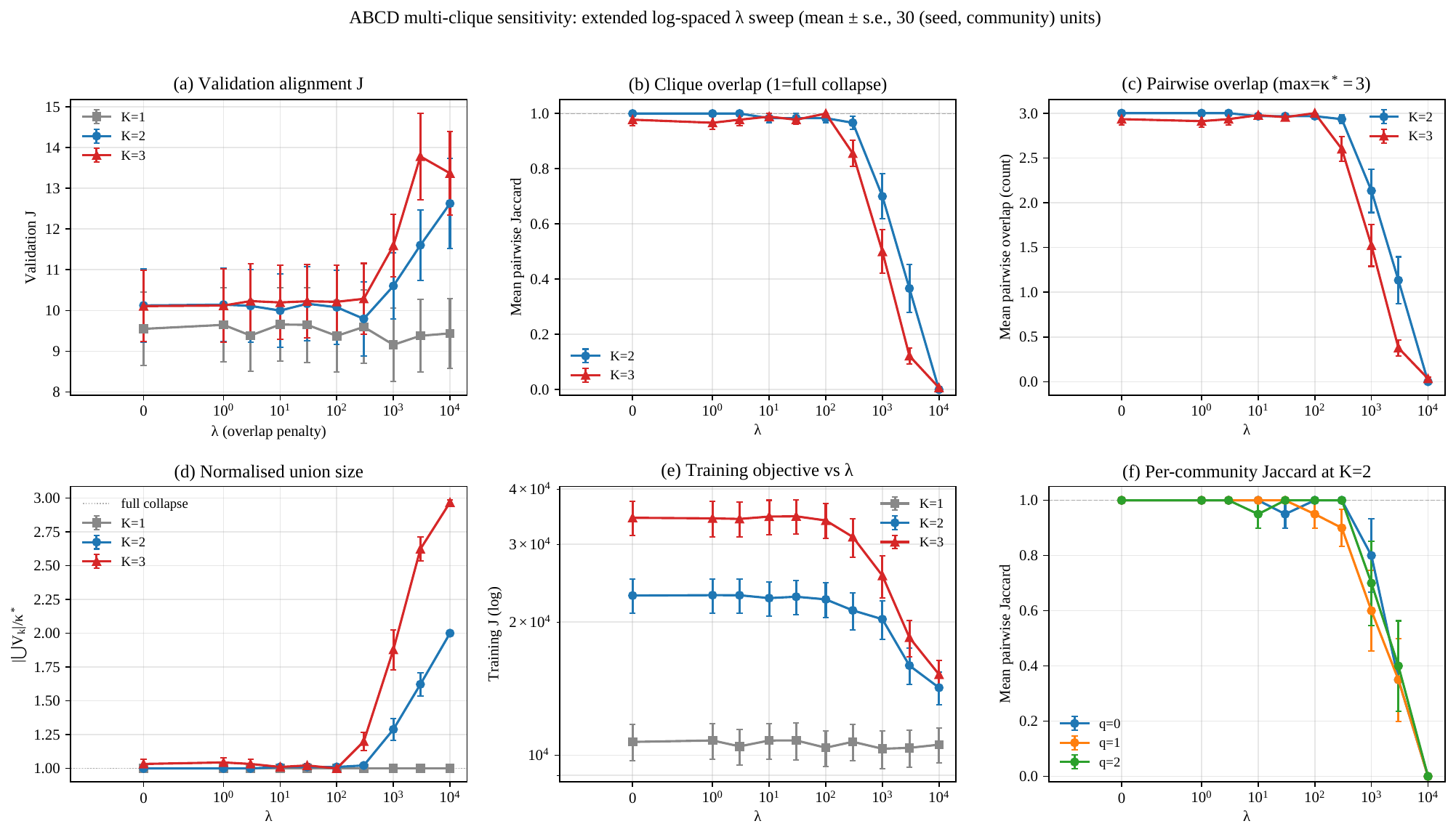}
\caption{ABCD second-visit multi-clique $\lambda$-sensitivity on a
log-spaced grid (mean $\pm$ s.e.\ over $30$ (seed, community) units;
$x$-axis on a symmetric-log scale to display $\lambda=0$). Panels show: (a) Validation $J$ evaluated on the union
of supports; (b) Mean pairwise Jaccard ($1$ = full collapse); (c) Mean
pairwise overlap count (maximum $\kappa=3$); (d) Normalised union size
$|\bigcup_\ell \widehat{\mathcal V}^{(\ell)}|/\kappa$ ($1$ = full
collapse, $K$ = full separation); (e) Training $J$ on log scale; (f)
Per-community pairwise Jaccard at $K=2$. The single-core verdict
(collapse at small $\lambda$, no per-node yield gain at large $\lambda$)
holds in every community, replicating the baseline-cohort conclusion.}
\label{fig:abcd2-multi}
\end{figure}

\paragraph{Multi-clique sensitivity ($\lambda$-diagnostic).}
We replay the same $K \in \{1, 2, 3\}$ and log-spaced $\lambda$-sweep at $\kappa = 3$ on the second-visit cohort over the same $30$ (seed, community) units (Figure~\ref{fig:abcd2-multi}); the summary appears in Section~\ref{subsec:real} (``Second-visit replication''). The diagnostic returns the same single-core signature as the baseline cohort: at $\lambda \leq 10^2$ the $K = 2$ and $K = 3$ fits collapse onto a single dominant support (mean pairwise Jaccard $\geq 0.98$, normalised union $\leq 1.03$); at $\lambda \in [10^3, 10^4]$ a sharp transition forces them into geometric expansion of the union ($K = 2$: $1.0 \to 2.0$; $K = 3$: $1.0 \to 2.97$), with no intermediate $\lambda$ yielding stable partial separation.

\section{Compute resources}
\label{app:compute}

All experiments were run on a high-performance computing cluster. CPU jobs used Intel Xeon 6342 processors, with 4 CPU cores and 16 GB of RAM allocated per job (4 GB per core). GPU resources were used only for the deep-learning baselines, which ran on an RTX 8000 GPU. Runtime was measured from loading the preprocessed SC/FC matrices to saving the final metrics and figures. Raw MRI preprocessing is excluded. Table~\ref{tab:compute-resources} reports peak memory usage (GB) and wall-clock time (minutes) for the main experiment groups.

\section{Limitations and broader impact.}
\label{app:broader-impact}
Our theory establishes local statistical guarantees for canonical OSSA, including the existence of an empirical local maximizer near the population target. 
Note that the implemented block ascent method is a finite multi-restart local solver, and we do not claim that its output lies in the specified neighborhood.
Extending guarantees to the multi-support objective is future work. Empirically, the results are limited to the Desikan--Killiany atlas, ABCD, predefined FC communities, and small support budgets, so the anatomical supports should be treated as cohort- and pipeline-dependent hypotheses until replicated across parcellations, acquisition protocols, and clinical populations. \textsc{Ossa} may support more interpretable studies of brain development and disease-related network organization, but connectome-derived supports should not be used for individual diagnosis, screening, or group profiling without external validation and uncertainty reporting.

The community-wise formulation can also be applied to predefined overlapping functional communities. Each community would retain its own alignment and support parameters, so the same objective could be fitted independently within each community, with a shared region eligible for selection in more than one. However, support selections across communities would remain uncoupled, and an SC edge included in multiple community subproblems could contribute fully to each corresponding objective. Such independent fits therefore do not describe how a shared region's structural role is distributed across its functional memberships. Whether the relevant biological structure is best described by independent community-specific cores or by a decomposition of shared structural roles across systems remains an open question that the current formulation cannot resolve. Jointly modeling these shared roles is a direction for future work.

\paragraph{ABCD data use and ethics.}
The ABCD data were obtained through the official ABCD data access mechanism and used in accordance with the applicable data-use agreement. ABCD participants were enrolled under the study's institutional review and consent/assent procedures, and our analysis used de-identified connectome-derived data only. We do not attempt individual identification and do not release subject-level neuroimaging data.

\begin{table}[t]
\centering
\caption{Compute resources for the main reported experiments.}
\label{tab:compute-resources}
\setlength{\tabcolsep}{4pt}
\renewcommand{\arraystretch}{1.08}

\begin{tabular}{
m{0.40\linewidth}
>{\centering\arraybackslash}m{0.16\linewidth}
>{\centering\arraybackslash}m{0.16\linewidth}
>{\centering\arraybackslash}m{0.16\linewidth}
}
\toprule
Experiment group & Worker type & Peak memory (GB) & Wall-clock time (min) \\
\midrule

Simulation experiments
& CPU
& 0.87
& 13 \\

ABCD main evaluation
& CPU
& 1.23
& 57 \\

Random-null/\(\kappa\)-robustness/multi-clique diagnostics
& CPU
& 0.62
& 5 \\

Deep-learning baselines
& GPU
& 1.23
& 45 \\

ABCD second-visit replication
& CPU
& 1.06
& 48 \\

\bottomrule
\end{tabular}
\end{table}

\end{document}